\documentclass[onecolumn,authoryear]{els-mrw} 

\usepackage{amsmath,amssymb,amsfonts,amsthm,makeidx,graphicx}
\usepackage{txfonts}
\usepackage{helvet}
\usepackage{wrapfig}
\usepackage{tabularx}
\usepackage{colortbl}

\def\aj{{AJ}}                   % Astronomical Journal
\def\actaa{\ref@jnl{Acta Astron.}}      % Acta Astronomica
\def\araa{\ref@jnl{ARA\&A}}             % Annual Review of Astron and Astrophys
\def\apj{\ref@jnl{ApJ}}                 % Astrophysical Journal
\def\apjl{\ref@jnl{ApJ}}                % Astrophysical Journal, Letters
\def\apjs{\ref@jnl{ApJS}}               % Astrophysical Journal, Supplement
\def\ao{\ref@jnl{Appl.~Opt.}}           % Applied Optics
\def\apss{\ref@jnl{Ap\&SS}}             % Astrophysics and Space Science
\def\aap{\ref@jnl{A\&A}}                % Astronomy and Astrophysics
\def\aapr{\ref@jnl{A\&A~Rev.}}          % Astronomy and Astrophysics Reviews
\def\aaps{\ref@jnl{A\&AS}}              % Astronomy and Astrophysics, Supplement
\def\azh{\ref@jnl{AZh}}                 % Astronomicheskii Zhurnal
\def\baas{\ref@jnl{BAAS}}               % Bulletin of the AAS
\def\bac{\ref@jnl{Bull. astr. Inst. Czechosl.}}
\def\caa{\ref@jnl{Chinese Astron. Astrophys.}}
\def\cjaa{\ref@jnl{Chinese J. Astron. Astrophys.}}
\def\icarus{\ref@jnl{Icarus}}           % Icarus
\def\jcap{\ref@jnl{J. Cosmology Astropart. Phys.}}
\def\jrasc{\ref@jnl{JRASC}}             % Journal of the RAS of Canada
\def\memras{\ref@jnl{MmRAS}}            % Memoirs of the RAS
\def\mnras{\ref@jnl{MNRAS}}             % Monthly Notices of the RAS
\def\na{\ref@jnl{New A}}                % New Astronomy
\def\nar{\ref@jnl{New A Rev.}}          % New Astronomy Review
\def\pra{\ref@jnl{Phys.~Rev.~A}}        % Physical Review A: General Physics
\def\prb{\ref@jnl{Phys.~Rev.~B}}        % Physical Review B: Solid State
\def\prc{\ref@jnl{Phys.~Rev.~C}}        % Physical Review C
\def\prd{{Phys.~Rev.~D}}        % Physical Review D
\def\pre{\ref@jnl{Phys.~Rev.~E}}        % Physical Review E
\def\prl{{Phys.~Rev.~Lett.}}    % Physical Review Letters
\def\pasa{\ref@jnl{PASA}}               % Publications of the Astron. Soc. of Australia
\def\pasp{\ref@jnl{PASP}}               % Publications of the ASP
\def\pasj{\ref@jnl{PASJ}}               % Publications of the ASJ
\def\rmxaa{\ref@jnl{Rev. Mexicana Astron. Astrofis.}}%
\def\qjras{\ref@jnl{QJRAS}}             % Quarterly Journal of the RAS
\def\skytel{\ref@jnl{S\&T}}             % Sky and Telescope
\def\solphys{\ref@jnl{Sol.~Phys.}}      % Solar Physics
\def\sovast{\ref@jnl{Soviet~Ast.}}      % Soviet Astronomy
\def\ssr{\ref@jnl{Space~Sci.~Rev.}}     % Space Science Reviews
\def\zap{\ref@jnl{ZAp}}                 % Zeitschrift fuer Astrophysik
\def\nat{{Nature}}              % Nature
\def\iaucirc{\ref@jnl{IAU~Circ.}}       % IAU Cirulars
\def\aplett{\ref@jnl{Astrophys.~Lett.}} % Astrophysics Letters
\def\apspr{\ref@jnl{Astrophys.~Space~Phys.~Res.}}
\def\bain{\ref@jnl{Bull.~Astron.~Inst.~Netherlands}} 
\def\fcp{\ref@jnl{Fund.~Cosmic~Phys.}}  % Fundamental Cosmic Physics
\def\gca{\ref@jnl{Geochim.~Cosmochim.~Acta}}   % Geochimica Cosmochimica Acta
\def\grl{\ref@jnl{Geophys.~Res.~Lett.}} % Geophysics Research Letters
\def\jcp{\ref@jnl{J.~Chem.~Phys.}}      % Journal of Chemical Physics
\def\jgr{\ref@jnl{J.~Geophys.~Res.}}    % Journal of Geophysics Research
\def\jqsrt{\ref@jnl{J.~Quant.~Spec.~Radiat.~Transf.}}
\def\memsai{\ref@jnl{Mem.~Soc.~Astron.~Italiana}}
\def\nphysa{\ref@jnl{Nucl.~Phys.~A}}   % Nuclear Physics A
\def\physrep{\ref@jnl{Phys.~Rep.}}   % Physics Reports
\def\physscr{\ref@jnl{Phys.~Scr}}   % Physica Scripta
\def\planss{\ref@jnl{Planet.~Space~Sci.}}   % Planetary Space Science
\def\procspie{\ref@jnl{Proc.~SPIE}}   % Proceedings of the SPIE
\makeatother

\begin{document}

\chapter{Gamma-ray Bursts}\label{chap1}

\author[1,2]{Andrew J. Levan}%
%\author[2]{Second Author}%

%\author[1,2]{Third Author}%

\address[1]{\orgname{Radboud University,}, \orgdiv{Department of Astrophysics/IMAPP}, \orgaddress{6525 AJ Nijmegen, The Netherlands}}
\address[2]{\orgname{University of Warwick}, \orgdiv{Department of Physics}, \orgaddress{Coventry, CV4 7AL, United Kingdom}}

\articletag{Chapter Article tagline: update of previous edition,, reprint..}

\maketitle

%\begin{glossary}[Glossary]
%\term{Europe} the model is a coherent view of capital markets data that allows users to interact with the content in a consistent manner.
%
%\term{Primates} regardless of the source. Essentially, of sources. Properly deployed.
%
%\end{glossary}

\begin{glossary}[Nomenclature]
\begin{tabular}{@{}lp{34pc}@{}}
Gamma-ray burst & A GRB is a brief flash of high energy radiation (peaking between 1-1000 keV) lasting from a few milliseconds to a few thousand seconds. \\
SN &Supernova: A transient signature associated with the collapse or explosion of a star, typically powered by newly created radioactive nickel, by interactions of the outflow with the surrounding medium, or both. \\
KN  & Kilonova: A transient associated with the merger of two neutron stars or a neutron star and a black hole. Powered by the radioactive decay of heavy elements synthesized in the merger. \\
Prompt emission & The short-lived emission that characterizes a gamma-ray burst, detected primarily at high energies with a duration from a fraction of a second to hours (typically tens of seconds).  \\
Afterglow & The multi-wavelength emission seen after a gamma-ray burst, caused by the interaction of the outgoing shockwave with the material around the progenitor star. \\ 
Synchrotron radiation  & Radiation created by the gyration of relativistic electrons around magnetic field lines. 
 \\
Gravitational Wave & Ripples in space-time that result in minute changes in the distance between two points and are caused by the non-symmetric acceleration of masses. Have been detected so far from the mergers of compact objects including neutron stars and black holes. \\
$r-$process & The route to the formation of heavy elements by the capture of neutrons at a rapid rate so that heavy isotopes are built from the capture more rapidly than they can decay. Occurs in neutron star mergers and possibly also in other astrophysical sites.  \\
%Scintillator &A type of gamma-ray detector in which gamma-rays excite electrons in the detector, and their subsequent decay produces a visible light signal.  \\
%Coded Mask &An alternative GRB detection method in which an pattern of lead tiles is placed over the detector so that sources with different spatial locations cast unique shadows on the underlying $\gamma$-ray detector.  Most well localised GRBs are now found with coded mask detectors, notably the Burst Alert Telescope (BAT) on the Neil Gehrels Swift Observatory. \\
%Lobster Eye optic &A route to focussing X-ray photons by reflections in a micro-channel plate that creates a sensitive, wide-field imager that has a short focal length and is relatively lightweight. Now used on satellites including SVOM???? and the Einstein Probe.  \\

\end{tabular}
\end{glossary}

\begin{abstract}[Abstract]
Gamma-ray bursts are flashes of high-energy radiation lasting from a fraction of a second to several hours. Military satellites made the first detections of GRBs in the late 1960s. The $\gamma$-ray emission forms from shocks in a relativistic jet launched from a compact central engine. In addition to the emission of $\gamma$-rays, the interaction of the jet with the surrounding medium yields afterglow emission that can be observed across the electromagnetic spectrum. Redshift measurements from these afterglows place GRBs from the local to the distant Universe. The central engines of GRBs are thought to be either a hyperaccreting black hole or a highly magnetized neutron star (magnetar). There is now strong observational evidence that this central engine is created either in the core collapse of a rapidly rotating massive star or via the merger of two compact objects (neutron stars or a neutron star with a black hole). The combination of stellar scale events with extreme energies and luminosities makes GRBs powerful probes of the extreme physics involved in their production and of other areas of astrophysics and cosmology. These include as the electromagnetic counterparts of gravitational wave sources, the production and acceleration of relativistic jets, the synthesis of heavy elements, the study of the interstellar and intergalactic medium, and the identification of the collapse of early generations of stars.

\end{abstract}

\begin{BoxTypeA}[]{Key points}
\begin{itemize}
\item  A gamma-ray burst is the emission of radiation from a relativistically expanding outflow from a stellar scale central engine produced in a one-off event. 
\item Gamma-ray bursts are detected normally as flashes of $\gamma-$rays and X-rays (typically peaking in the 1-1000 keV range) and show highly complex lightcurves with variability on millisecond timescales and durations up to several hours. There are at least two populations of GRBs typically defined as being of short-duration ($<2$\, s) and long-duration ($>2$\,s).  Over 10,000 GRBs have been detected to date by a large number of satellites. 
\item GRBs are cosmological in origin, with a range of redshifts from $0.1 < z < 10$ in the observed samples. The high distances imply that GRBs are the most powerful explosions known, releasing total energies (if released equally in all directions) that can exceed $10^{55}$ erg ($\sim 5 $ M$_{\odot} c^2$).
\item The combination of rapid variability that implies a small source size and the extreme energies enforces that GRBs must be relativistically beamed in our direction, likely with bulk Lorentz factors in the range $\Gamma=10-1000$, and jet opening angles of a few degrees. The GRB emission then arises from the interaction of shocks within (internal to) this expanding fireball.
\item In addition to this prompt emission a multiwavelength afterglow is created by the interaction of the outgoing jet with the external medium. This afterglow is the route through which GRBs are precisely localized on the sky, how their distances are measured, and how their host galaxies are identified. 
\item In-depth follow-up of afterglows has revealed that some GRBs, predominantly of the long duration class, are associated with high-velocity supernovae devoid of features of hydrogen and helium, implying that they occur due to the collapse of a rapidly rotating massive star that has lost its outer envelope during its stellar (or binary) evolution. 
\item Another subset of GRBs, predominantly of the short-GRB category have been associated with the merger of two compact objects thanks to the detection of faint, fast transients powered by the decay of newly synthesized heavy elements (kilonovae), and, in one case, the direction detection of gravitational waves. 
\item The combination of properties makes GRBs powerful probes of extreme physics and cosmology, including routes to probing the first stars and the process of cosmic reionisation, and as route to study jet formation and particle acceleration. 
\end{itemize}
\end{BoxTypeA}

%\section{Introduction}
%Gamma-ray bursts are the most energetic explosions known in nature and can release as much energy in their few-second duration as the Sun over its $\sim 10$ billion year lifetime. Although initially discovered by chance their study has evolved rapidly. Identifying their origin rapidly became one of the central questions in astrophysics throughout the 1980s and 1990s before answers to this question began to flow, in particular following the precise localization of bursts on the sky thanks to the discoveries of their afterglows. Today we know that GRBs are cosmological explosions that arise from at least two distinct channels -- the collapse of massive stars or the merger of two compact objects. However, many open questions

\section{Introduction}\label{chap1:sec1}

\begin{wrapfigure}{R}{0.5\textwidth}
  \vspace{-0.9cm}
\begin{center}
\includegraphics[width=0.48\textwidth]{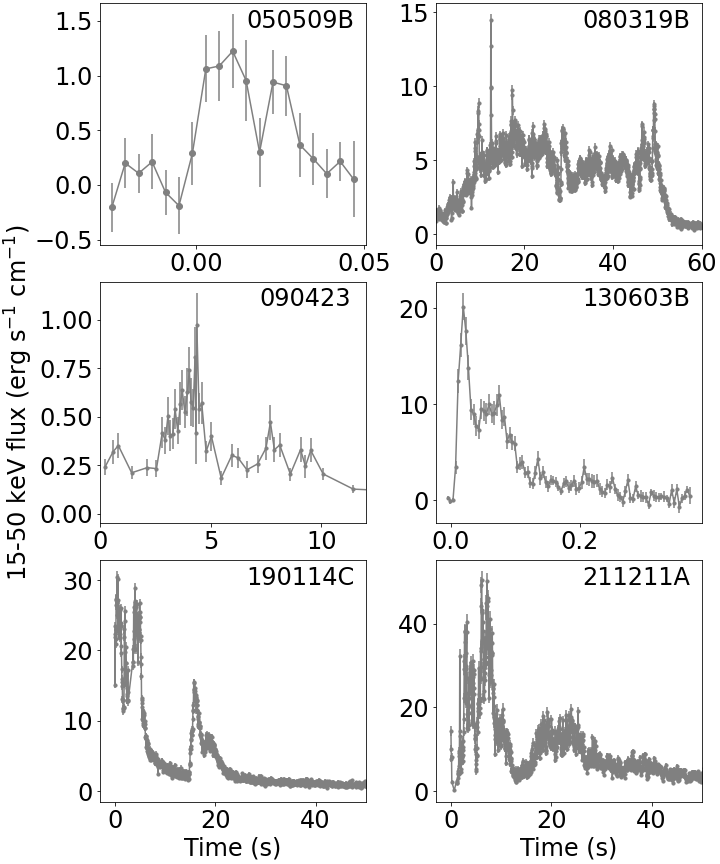} 
\end{center}
\vspace{-0.4cm}
    \caption{A sample of gamma-ray burst lightcurves observed by the {\em Swift} BAT. These are all events which have led to substantial breakthroughs in our understanding of the GRB phenomena (see Table~\ref{tab:bursts}).}
   \label{fig:lcs}
\end{wrapfigure}

Gamma-ray bursts were first discovered by both US and Russian/USSR satellites launched to police the nuclear test-ban treaty of 1963. This treaty prohibited the tests of atomic weapons in space, a location where monitoring for compliance required sensitivity to the millisecond gamma-ray emission associated with a nuclear detonation. No clear nuclear tests were detected, but the US Vela and Russian Konus satellites recorded short outbursts unrelated to apparent cosmic bodies (the Sun, Moon, Earth or Solar System objects). The first GRBs were reported to the world in 1973 and consisted of 16 events discovered by the Vela satellites \citep{kleb73}. Many missions have detected subsequent GRBs, and the total bursts recorded exceed 10,000 as of 1 Jan 2024. GRB naming is straightforward, consisting of GRB year-month-day. The first GRB detected was on 2 July 1967, GRB 670702. The first burst in a given day is assigned as the "A" burst, and subsequently ``B" etc\footnote{Before 2010, when bursts were detected infrequently, the first burst of a day does not have an "A" appended.}.

GRBs can last from milliseconds %\citep{bhat92} 
to hours 
%\citep{levan14} 
and exhibit a very wide range of lightcurve shapes (see Figure~\ref{fig:lcs}), including events that show substantial overlapping spiky structure and smooth long-lived events. A considerable fraction also exhibit the so-called Fast Rise Exponential Decay (FRED) shape. Because of the complex lightcurve behaviour, the most straightforward measure of a burst duration is the timescale over which some fraction of the emission is detected. The most commonly used diagnostic is $T_{90}$, which is the time scale over which 90\% of the total emitted $\gamma$-ray energy (more formally known as the gamma-ray fluence in erg cm$^{-2}$) is recorded. 
The spectra of GRBs peak in the keV-MeV region of the spectrum \citep{band93}, and typically more energetic bursts have spectra that peak at higher energies \citep{amati02}. This spectrum also evolves during the GRB, typically (but not always) peaking at the highest energies when the GRB is brightest and moving to lower energies when the GRB is fainter. 

%These spectra are often fit with a smoothly broken power-law, also known as a Band function in which the peak of the spectrum ($E_p$) is defined in energy space (i.e. by multiplying the energy density by the energy, or $\E F_{E} = \nu F_{\nu} = \lambda F_{\lambda}$). More recently these spectra have also been fit with physically motivated synchrotron radiation models \cite{}. Characteristic GRB lightcurves and spectra are shown in Figure~\ref{}. 

\section{A brief history}

\begin{wrapfigure}{R}{0.5\textwidth}
  \vspace{-0.9cm}
\begin{center}
\includegraphics[width=0.48\textwidth]{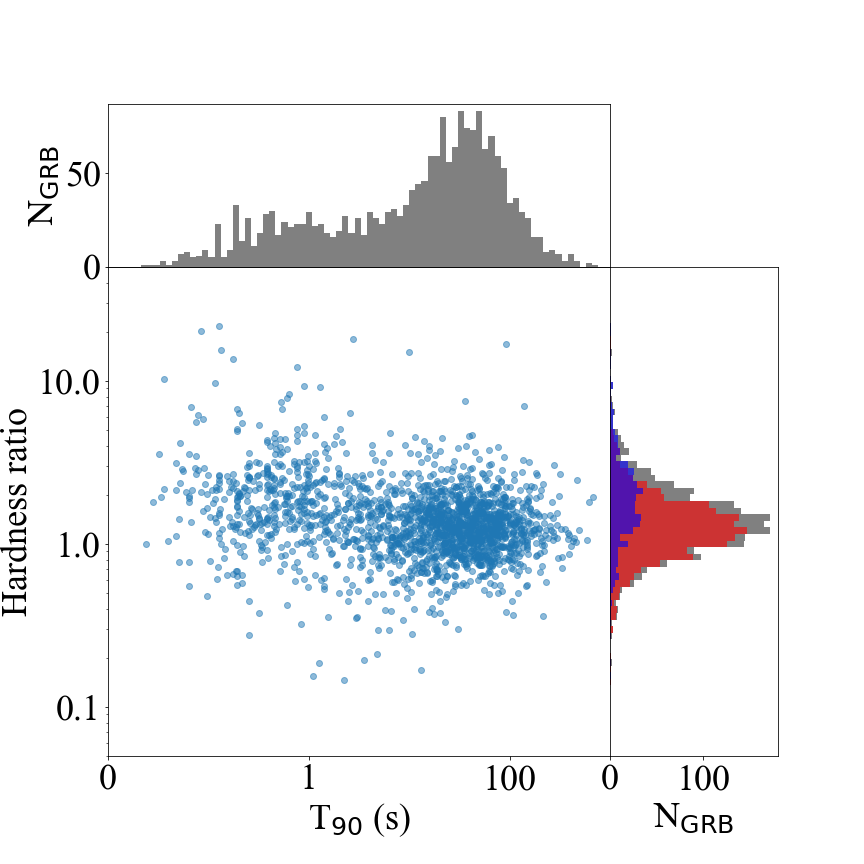} 
\end{center}
\vspace{-0.4cm}
    \caption{The hardness-duration diagram for GRBs observed by the BATSE instrument \citep{Kouveliotou+93}, which was the first to obtain a large sample of both long and short-GRBs 
    and conclusively demonstrate differences in duration and spectral hardness. The histograms show the collapsed distributions in each axis, while in the hardness panel the red histogram represents the bursts with $T_{90} >2$s and the purple $T_{90} <2$s. }
   \label{fig:sh}
\end{wrapfigure}

The discovery of GRBs led to many suggestions for their origin, ranging in distance from within the solar system to origins in distant galaxies. %By the 1990s, the list of models comprised $>100$ possibilities \citep{nemiroff94}. 
The distances to GRBs were the subject of intense discussion. The uncertainty even led to a ``Great Debate" undertaken in 1995, which was intended to celebrate the 75th anniversary of a similar debate between Curtis and Shapely on the distance scale of the Universe. Throughout the 1970s-1990s, the lack of precise positions precluded the identification of the sources of GRBs at other wavelengths and substantially limited progress towards their origin. The insight that was gained was based entirely on the prompt emission. These observations do provide several core results into the nature of GRBs, in particular:

%However, several core results did emerge. In particular these include. 

\begin{itemize}
\item
GRBs arise in two populations of duration, often known as short-GRBs and long-GRBs (Figure~\ref{fig:sh}). Short GRBs are spectrally harder (emit more high energy photons) than long GRBs \citep[e.g.][]{Kouveliotou+93}. 
\item
GRBs are distributed isotropically on the sky with no preference for the Galactic plane or other structures \citep{meegan92} %,Briggs96}. 
\item
GRBs show a distribution in number-fluence space, the so-called $\log{N} - \log{S}$ distribtution which tracks the expectation for sources uniformly distributed in a Euclidean universe (in particular a slope of $-3/2$), but deviates away from at faint fluence levels \citep{fishman94}. 
\item GRB emission shows very short timescale variability, this variability at the millisecond %\citep{bhat92} 
level places strong constraints on the size of the emitting region $c \Delta t \approx 100$km (for non-relativistic sources). 
\item For GRBs at cosmological distances the inferred photon densities are such that the system should be optically thick to pair-production, and hence we should not observe photons with energies beyond $0.511$ MeV unless the emission is relativistically beamed \citep{cav78}. 
\end{itemize}

However, there were limits to what could be learned from the prompt emission alone. Substantially, these constraints arose because of the inaccuracy of the 
celestial positions available from only $\gamma$-ray detections. This meant it was impossible to determine the distances to the bursts, constrain their energetics, or associate them with either Galactic stellar systems or extragalactic host galaxies. The realisation that cosmological GRBs must be beamed led to the development of a 
progenitor-independent model for multiwavelength emission -- the fireball model \citep[e.g.][]{mez93}.
%,mez97,piran99}. 
Here, a compact central engine drives the source, producing a relativistic outflow. This central engine creates a series of outflowing ``shells" with extremely high Lorentz factor, although each has a slightly different velocity. Because of these differences in velocity, there are {\em internal} shocks between the outgoing shells. These shocks accelerate electrons to high velocity, resulting in the observed $\gamma$-ray emission (see section~\ref{sec:prompt}).

The relativistic outflow is also not expanding into a perfect vacuum but into some medium surrounding the progenitor. This creates an {\em external} shock between the outflow and the surrounding medium. Particles accelerated at this shock also yield a synchrotron spectrum but with a much lower characteristic energy and a much longer emission duration. This emission is the GRB afterglow (see section~\ref{sec:afterglows}). Searches throughout the 1990s focussed on the prospect of identifying GRB afterglows. However, it was not until the launch of the joint Italian Dutch BeppoSAX mission that these searches bore fruit, thanks mainly to the smaller error boxes and the subsequent capability to obtain both X-ray and optical observations. These observations revealed a first X-ray \citep{costa97} and optical \citep{vanparadijs97} afterglow discovery to the burst GRB 970228. This afterglow revolution led to a suite of discoveries both from individual bursts (a sample of which are highlighted in table~\ref{tab:bursts}) and from the ensemble properties of the GRBs, their afterglows and host galaxies. In particular, these insights include securing the cosmological origin of GRBs via the measurement of redshifts \citep{metzger97}, with over 500 GRBs now having redshift measurements with a median redshift of $z \sim 2$ \citep{jakobsson06}.  From the redshifts, it was possible to secure the GRB energetics and demonstrate extreme energy production of up to $10^{55}$ erg \citep{burns23}, which may well be unphysical were it not moderated by factors of 100 or more because it is directed into only a small portion of the sky (e.g., \cite{frail01} see section~\ref{sec:energetics}). Studying the evolving afterglows was also the route by which it was possible to secure the progenitors of first the long-duration GRB population as the collapse of massive stars (GRB 980425, 030329 \citep{Galama+98,hjorth03,stanek03} see section~\ref{sec:sn}) and then the population of short GRBs as compact object mergers (\citep{tanvir13,berger13,Abbott+17a} see section~\ref{sec:mergers}). However, the latter was also significantly aided by the co-incident detection of a gravitational wave signal \citep{Abbott+17a}). While recent results cast doubt on this relatively simple distinction between long- and short-GRB progenitors \citep{rastinejad22,troja22,levan24a}, these two progenitor channels remain the only two (out of the hundreds proposed) that have been robustly identified as occurring in nature. 

Studying the nature of GRBs themselves only represents part of the activities of the field in the past decades. Comparable effort has been deployed to utilize GRBs as probes, both using their bright emission as a backlight to illuminate lines of sight to the distant Universe and making use of their extreme energetics to probe the physics in their creation in conditions that are far beyond those attainable in any ground-based laboratory (see section~\ref{sec:extreme}).

\begin{table}[!h]
\caption{Summary of important GRBs with the results from them.}%%%Table caption goes here
\label{tab:bursts}
\begin{tabular}{|l|l|l|}%%%The number of columns has to be defined here
\noalign{\global\arrayrulewidth=1mm}
\hline
GRB & Summary & core references \\ 
\hline

\noalign{\global\arrayrulewidth=0.1mm}
%\arrayrulecolor{gray}
%910503 & Prompt & \cite{nowak94} \\
670702	& The first detected GRB, first reported by in 1973 & \cite{kleb73} \\
\hline
790305B		& An extremely bright GRB, now known to represent a burst from a soft   &  \cite{mazets79,cline80} \\ 
& gamma-repeater (magnetar) in the Large Magellanic Cloud. & \cite{evans80} \\
\hline
970228  & First GRB with an X-ray and optical afterglow, and hence accurate position & \cite{costa97,vanparadijs97}\\
\hline
970508  & First GRB with a measured redshift, $z=0.835$, showing GRBs are cosmological & \cite{metzger97} \\
\hline
980425 & Unusual long GRB in the local Universe (40 Mpc), associated with novel & \cite{Galama+98}  \\  & broad-lined  hydrogen poor (type Ic) supernova & \\
\hline
%990123 & The first burst to be observed in optical light during the prompt phase, with a very bright early afterglow  \\
030329 &  Burst which secured the relation between cosmological long GRBs and & \cite{hjorth03} \\ &  broad-lined type Ic supernovae & \cite{stanek03} \\
\hline
050509B & First short duration GRB an afterglow, also with an ancient host galaxy & \cite{gehrels05,bloom06} \\
\hline
%050904 & \\
%060614 & A long \\
080319B & A gamma-ray burst with an afterglow so bright it could have been seen with & \cite{racusin08} \\ &  the naked eye despite a redshift of $z=0.937$ & \cite{bloom09} \\
\hline
090423 & The most distant GRB with a secure redshift, $z=8.23$ & \cite{tanvir09,salvaterra09} \\
\hline
130603B & The first short-GRB with a suggested kilonova created in a compact object & \cite{tanvir13,berger13} \\
&  merger & \\
\hline
%090429B &  \\
%130427A & \\
170817A & A short duration GRB associated with a gravitational wave & \cite{Abbott+17a,Goldstein+2017}  \\
& detected binary neutron star GW170817  & \cite{Coulter+17} \\
\hline 
190114C & A very bright GRB detected from ground based Cherenkov detectors & \cite{magic19} \\ & in the TeV regime & \\
\hline
211211A & A long GRB with an associated kilonova indicating more complex & \cite{rastinejad22,troja22} \\ & progenitor models than previously suggested  & \cite{yang22} \\
\hline
221009A & The brightest (observed) and most energetic  GRB of all time,  & \cite{burns23,lesage23} \\
& likely the brightest GRB in c. 10000 years &  \cite{malesani23,ravasio24} \\
%230307A & \\
\hline
\end{tabular}
\vspace*{-4pt}
\end{table}%%%End of the table

\section{Gamma-ray burst detectors} 
The wavelength of gamma rays is comparable to the atomic separation of atoms in glass, so gamma rays cannot be focussed in the same way as optical light and require a different detection method. The effects of the atmosphere mean that detection must, in general, be done in space. 
%However, advances in technology mean that very high energy gamma-rays can now be detected via ground-based Cherenkov telescopes, which image the flash of optical photons created as an incoming very high energy gamma-ray interacts because it is travelling faster than the local speed of light in the atmosphere \citep[e.g.][]{magic19}.

Many gamma-ray burst detectors are scintillators with no intrinsic spatial resolution. They instead record flashes of light created by the incidence of the $\gamma-$ray photon on the detector material. The first GRB detectors were scintillators, and many GRB missions continue to rely on them. This includes the missions that have identified the most GRBs, the Burst and Transient Alert Explorer (BATSE) and the Fermi satellite's Gamma-ray Burst Monitor (GBM). While the scintillators have no spatial resolution, burst positions can be recovered if several are placed on the spacecraft and the relative rates in detectors with different locations and orientations are recorded. Such positions usually have substantial uncertainties of several degrees. Positions from different detectors can enable far smaller sky localisations using triangulation via time of flight measurements to widely spaced detectors. This method underpins the InterPlanetary Network (IPN), which uses a series of often small $\gamma$-ray detectors mounted on probes throughout the solar system, including Mars Oddesey, BeppiColumbo and even the Voyager probes, to obtain precise positions. 

More recently, positions have been obtained using coded mask detectors. In this scenario, a series of blocking tiles (typically made of lead) are arranged over the detector in a random pattern. When illuminated by a cosmic $\gamma$-ray source, these tiles case a unique pattern of shadows, which can be interpreted to provide a position for the burst relative to the detector. Such positions typically have arcminute localization. The first well-localized bursts were found in the coded mask Wide Field Camera (WFC) on the BeppoSAX mission, and the largest sample of $>1500$ accurately positioned bursts have been detected by the Burst Alert Telescope (BAT) onboard the Neil Gehrels Swift Observatory (hereafter {\em Swift}). 

Increasingly, attempts are being made to locate GRBs or GRB-like phenomena outside the $\gamma$-ray regime. This includes searches for very high energy emission with Cherenkov telescopes and softer, wide-field X-ray detectors such as those working on wide-field imaging with Lobster Eye technology %\citep{yuan22}, 
or even identifying GRB afterglow emission in the optical, in the absence of a prompt GRB detection %\citep[e.g][]{cenko13}. 
These techniques can also provide positions for GRBs with arcminute precision. However, since the detection is outside the spectral window commonly used for GRBs, the relationship between the discovered transient emission and GRBs is not always apparent.

\begin{figure*}
\vspace{-0.1cm}
\begin{center}
\includegraphics[width=0.8\textwidth]{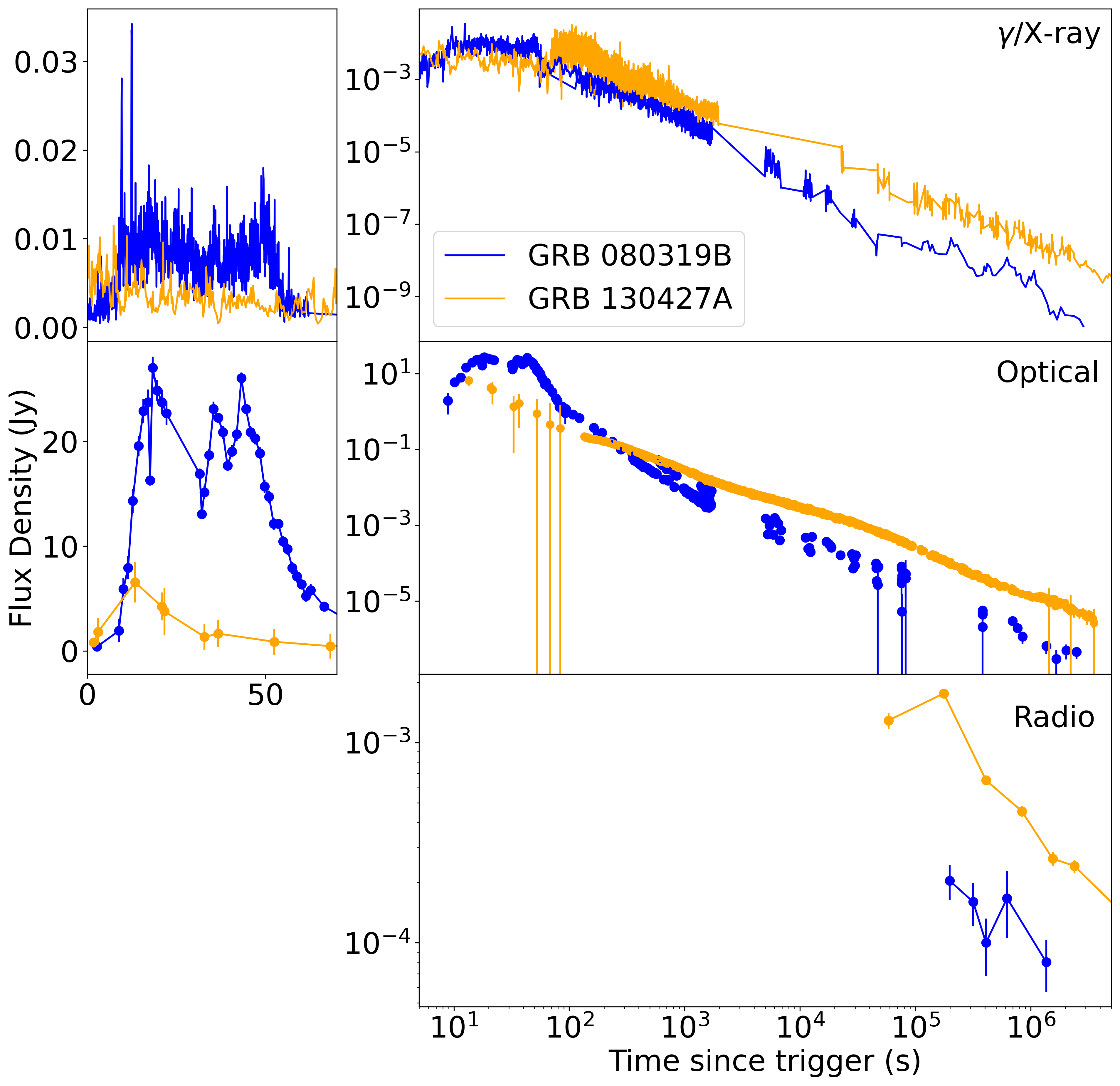} 
\end{center}
\vspace{-0.4cm}
    \caption{The observed properties of a GRB, demonstrated with the bright bursts GRB 080319B \citep{racusin08} and GRB 130427A \citep{perley14}. The burst begins with the prompt emission, detected primarily in $\gamma$-rays, but on rare occasions also seen in optical light. This is followed by a much longer-lived afterglow which dominates the emission beyond a few hundred seconds in each case. The optical and radio light is dominated by the afterglow emission, but the observed frequencies lie on different parts of the synchrotron spectrum, which can yield varied temporal behavior for the different wavelength regimes. Ultimately, mapping the evolving multi-wavelength afterglow enables the full properties of the evolving spectrum to be mapped and hence the details of the explosion to be probed.  }
   \label{fig:afterglow}
\end{figure*}

\section{GRB prompt emission}
\label{sec:prompt}
GRBs are almost always detected by their prompt emission, and most often via this prompt emission seen in the $\gamma$-ray regime. While, in principle, many different physical systems can create outbursts of $\gamma$-ray emission, the term GRB is normally only used for events that are extragalactic (or most likely extragalactic) and do not repeat. Other classes of $\gamma-$ray emitting object include soft gamma-repeaters \citep[e.g.][]{kouveliotou98}, flare stars \citep[e.g.][]{osten10}, X-ray binaries \citep[e.g.][]{zanin16}, and unusual tidal disruption events \citep[e.g.][]{bloom11}. These are now normally distinguished from GRBs, although it is likely that large GRB catalogs do contain some of these events. 

The prompt emission is thought to be caused by the interaction of shocks within the GRB jet (see Figure~\ref{fig:progenitors}), although the exact physical origin of the emission remains a matter of debate. The simplest model posits direct collisions between shells emitted with slightly different Lorentz factors \citep{daigne} are responsible for $\gamma$-ray production. 
Alternatively, it may arise from so-called collisionless shocks, where the shocks between shells are the result of interactions in the ionized plasma rather than collisions between the particles in the outflows \citep{waxman06}. There are also models in which the jet is magnetically powered, and magnetic reconnection -- the re-arrangement of the magnetic field and release of excess energy in photons -- creates $\gamma$-rays \citep{kumar15}. Increasingly, more complex models that utilize elements of these different physical processes are also gaining success in explaining the GRBs. This includes the Internal-Collision-induced MAgnetic Reconnection and Turbulence (ICMART) model \citep[e.g.][]{zhang14} where small discrete pockets are created and the GRB itself consists of the sum of emission from many of these pockets, providing a plausible explanation for the highly varied GRB lightcurves. 

Prompt GRB emission typically has a rapid onset, but the morphology of the resulting lightcurve can be extremely varied, including events that exhibit large-scale variability on a wide range of timescales and are highly spiky. Other events show multiple distinct emission episodes spread out over the burst duration.
%, although the most common number of pulses is one \citep{maccary24}. 
Some bursts show so-called precursor emission in which a fainter event proceeds the main emission and the source returns to quiescence (or at least to the level of detector sensitivity) before the main outburst \citep[e.g.][]{coppin20}. Several example GRB lightcurves that demonstrate this wide range of variability are shown in Figure~\ref{fig:lcs}.

Typically, a GRB begins with the emission of predominantly the more energetic gamma-ray photons (also often referred to as ``hard" photons); as the burst proceeds, it is common, but not ubiquitous, for there to be increasingly lower energy (or ``soft") photons emitted. This is known in the field as hard-to-soft evolution. Most bursts show this evolution, although there are a handful of events, including some recently found by the Einstein Probe, in which the X-ray emission leads the harder $\gamma$-rays \citep{liu24}.

At any given time the spectral shape of the prompt emission, if expressed in the $E F_E = E^2 N_{\gamma}$ regime\footnote{Note that multiplying the flux density by the energy or the photon number by $E^2$ provides a monochromatic flux or energy, for example in erg $s^{-1}$ cm$^{2}$. Hence when expressed in this form the spectral peak corresponds to where most of the energy is emitted.}
shows a rise to a peak followed by a subsequent fall-off. In this sense GRBs are accurately named, they emit most of their energy in the $\gamma$-ray regime. The spectrum is commonly fit with a smoothly broken power-law known as a Band spectrum \citep{band93}, which is characterized as a smoothly broken power-law connecting the regimes either side of the spectral peak. Bursts can exhibit a wide range of peak energies from $<10$ keV to $>1000$ keV, and this led to some attempts to classify GRBs based on the relative fluxes in different bands including X-ray Flashes, X-ray rich GRBs and GRBs \citep[e.g.][]{sakamoto05}. Such classifications are rarely used these days, although there is recently more interest in the Fast X-ray Transients (FXTs) which may be related to GRBs and XRFs \citep{qv23,levan24b}. 

It should be noted that the Band function is an empirical fit to the observed spectral energy distribution of GRB prompt emission. It is not a physical model for the emission processes. Recently, authors have also achieved significant success in using more physically motivated models to explain the prompt emission. One of these is to use 
synchrotron models \citep[e.g.][]{ravasio19}, and in several bursts, the spectral slopes before and after the break appear to match well with the expectation of synchrotron theory. Alternatively, the expanding jet may be well described by a thermal model (suitably blueshifted), and while at first sight, the spectra appear non-thermal these models can also provide a very good description of the observations of some GRBs \citep[e.g][]{ryde05}.  

Although GRB spectra have predominantly been defined as featureless, showing only continuum slopes, the opportunities afforded by the brightest burst of all time, GRB 221009A, provided a surprising result that the prompt emission contained a time-evolving emission line at $\sim 10$ MeV, most likely as a result of pair-production within the GRB-jet \citep{ravasio24}. 

GRB prompt emission is ultimately visible up to energies of at least tens of MeV. There have been detections of emission from GRBs in the GeV regime, for example, with the {\em Fermi}-LAT, and more recently even with ground-based $\gamma$-ray detectors in the TeV regime \citep{magic19}, with several thousand very high energy photons detected from GRB 221009A \citep{lhaaso23}. However, despite the extreme energies, it appears more likely that these photons are related to inverse Compton emission from the afterglow and not the prompt component. 
%Although the peak of the GRB spectrum is often reported as a single value based on the integrated light collected from the GRB spectral evolution means that such measurements should be viewed with caution. 

%, there is substantial spectral variation within a given burst. Typically bursts exhibit a "harder when brighter" structure, in which the more intense pulses are characterised by a higher $E_p$, although this behaviour is not ubqitous. There is also frequently an observed hard to soft evolution in a GRB in which the burst is harder at early times and then evolves to softer emission later. However, there are a handful of events, most recently explified by detections with the Einstein Probe, in which soft X-ray emission apparently preceeds the hard $\gamma$-ray emission \citep{}. 

\section{GRB afterglows} 
\label{sec:afterglows}

\subsection{Afterglow production}

While prompt emission is thought to be produced by the interaction between shocks generated in the central engine of the gamma-ray burst. Afterglow emission is created by the interaction of the outgoing shock with the surrounding medium. This critical difference yields emission that decays more slowly than the prompt emission and is visible across the electromagnetic spectrum, with afterglow emission visible as a (typically fading) source in the hours to weeks after the burst (see Figure~\ref{fig:afterglow} for typical afterglow evolution). The highest energy detections of afterglow emission are in the TeV region% \citep[e.g.][]{magic19,lhaaso23}, 
while the lowest lie in the very low-frequency radio regime. %\citep[e.g.][]{vanderhorst08}. 

The principle of afterglow emission is that the outgoing shock sweeps up the material in the surrounding medium. It is moving ultra relativistically at first with Lorentz factors of several hundred but begins to slow down as it entrains the matter. 
The details of the emitted radiation can vary according to an array of macro- and microphysical parameters, which result in potentially different observational signatures. However, the general features are those of a synchrotron spectrum \citep{Sari+98}.
The electrons in the shock do not all achieve the same energy but rather have a spectrum of energies, often expressed as Lorentz factors ($\gamma_e$, to distinguish from the bulk Lorentz factor of the outflow $\Gamma$), so that the number of electrons of given energy $N_{\gamma_e} \propto \gamma_e^{-p}$, where $p$ is the so-called electron index and typically lies in the region $2 < p < 3$, but under certain conditions can exceed these bounds. 
These electrons then 
radiate to produce a characteristic spectral shape that consists of a series of joined power-laws with different spectral shapes depending on the regime in which the accelerated electrons find themselves \citep[e.g.][]{Sari+98}. Briefly, at low frequency, the photons emitted from the shock are reabsorbed, and so the slope below the so-called self-absorption frequency $(\nu_a)$ is expected to scale as $F_{\nu} \propto \nu^2$, above this frequency it scales as $F_{\nu} \propto \nu^{1/3}$ up to a peak frequency which typically lies in regimes not readily accessible to observations. Above this peak, the spectrum typically evolves as $F_{\nu} \propto \nu^{-(p-1)/2}$ up to a final break where electrons can effectively cool (the cooling break) at which point it steepens to $F_{\nu} \propto \nu^{-p/2}$.

As the burst evolves the afterglow becomes fainter, and the spectral breaks move to a lower frequency. In the optical and X-ray, the afterglows typically show a monotonic decline (see Figure~\ref{fig:afterglow}), but the shape of the moving spectrum means that 
while the overall (bolometric) flux from the afterglow falls the motion of the spectral breaks can result in a brightening afterglow in certain regimes. For example, the radio afterglows of GRBs often brighten over the first few days of observations. 

A final critical feature of the temporal evolution of the afterglow is that the GRB outflow is both geometrically narrow and relativistically beamed (moving toward the observer at a high Lorentz factor) into an outflow with an opening angle of $\theta_j \sim 2-10$ degrees \citep[e.g.][]{frail01}. During the early part of the jet expansion the head of the jet is not all in causal contact, and so the behavior is comparable to that in the case of an isotropic expansion. At a critical Lorentz factor, where the $\Gamma \sim (1 / \theta_j$), both sides of the jet enter causal contact, leading to a steepening of the afterglow decay, the so-called jet-break, which also provides a route to measuring the collimation and total energy of a GRB. 

In practice, the details of the afterglow evolution depend on the structure of the jet itself, which may be top-hat (same energy at all locations in the jet and zero outside) or structured (where the energy is typically highest closest to the jet-axis, but evolves to lower energy as one moves off-axis, \cite[e.g.][]{kumar03}). 

Many observations now broadly confirm this picture of the multi-wavelength afterglows. At frequencies above the peak frequency (e.g. in the X-ray and optical regime) the bursts typically fade as $\sim t^{-1}$ for hours to days after the burst before undergoing a steepening to $\sim t^{-2}$ because of the jet break. Multi-wavelength observations from radio to X-ray can successfully map the evolving spectral energy distribution and confirm the broad picture of synchrotron radiation. 

However, there are exceptions to the clean behavior that can be observed, some of which are well understood, and others are not. Two particular physical processes modify the behavior of the afterglows. The first is prolonged activity in the central engine (see section~\ref{sec:engines}). Typically the observed duration of a burst is the duration that the engine is active {\em after} it has pierced whatever encases it (e.g. the envelope of a massive star). GRB durations are normally short. However, a substantial fraction of bursts have evidence for prolonged central engine activity at a lower level than during the prompt emission. This is most evident in well-sampled X-ray lightcurves which in addition to the monotonic afterglow decay can also show plateau or flaring activity indicative of this engine \citep[e.g.][]{zhang06}. Indeed, such flares can also be seen on occasion at other wavelengths. 

The second modification arises from the so-called reverse shock emission. This is where a shock wave propagates back into the expanding jet, creating an additional shock that may introduce measurable emission with a different spectral shape to the standard forward shock emission. The name refers to the motion of the shock in the frame of the forward shock. It is important to understand that the shock is not genuinely moving inwards, towards the progenitor. Several reverse shock signatures have now been seen, most readily in the radio band \citep[e.g.][]{laskar13}. 

\subsection{The importance of afterglows}
While the physics of afterglow production and evolution is a major topic in its own right, an important value of afterglows is that they provide accurate positions for the bursts on the sky. These accurate positions can then be targetted by much more sensitive multi-wavelength observations that provide a complete view of the GRB and its environment. It is only through afterglows that GRB distances, energetics, and progenitors have become accessible. 

After the location, the most important 
diagnostic available from a GRB afterglow is the burst redshift. In particular, the GRB represents a bright backlight that shines through its host galaxy. In doing so, light at the specific wavelengths of atomic transitions is absorbed creating a fingerprint of those elements in the afterglow spectrum of the GRB, typically measured in optical light. A comparison of these line positions to their rest-frame ones enables a measurement of the burst redshift and hence distance (with a cosmological model), as well as providing diagnostics regarding the make-up of the GRB host galaxy (see section~\ref{sec:probes}).

\begin{figure*}
\vspace{-0.1cm}
\begin{center}
\includegraphics[width=1.0\textwidth]{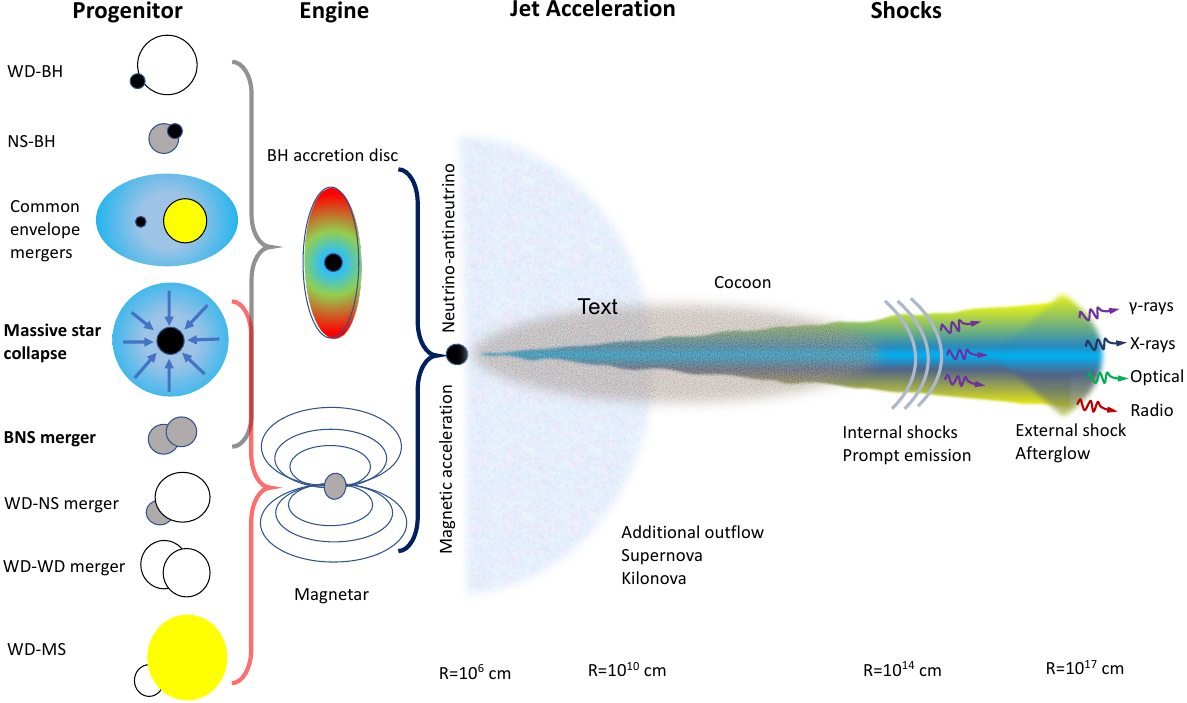} 
\end{center}
\vspace{-0.4cm}
    \caption{The physical processes underlying the production of a GRB. The system begins with the formation of a progenitor. Or the many progenitors suggested in the literature several remain viable, although only the collapse of massive stars and the merger of two neutron stars have been observationally confirmed. This progenitor proceeds to produce a central engine that is either an accreting black hole or a highly magnetised neutron star. The central engine powers the jet, and interactions both inside (prompt) and external to the jet (afterglow) are responsible for the observed emission.}
   \label{fig:progenitors}
\end{figure*}

%\section{GRB host galaxies} 
%\label{sec:hosts}
%GRBs are found in host galaxies across cosmic history. These host galaxies provide information about the GRBs themselves and are also of use for studying various other areas such as the cosmic evolution of metals, the luminosity function of galaxies with cosmic time, or contributions of faint galaxies to the budget of ionising photons in the distant Universe. 

%GRBs which arise from stellar collapse are found in star forming galaxies, and these galaxies are typically of low luminosity and small size \cite{}. Where available the measured metallicities of the galaxies are also low and these observations form the core support for models in which core collapse GRBs are formed in stars of low metallicity. It is unclear if the metallicity threshold is a hard cut-off or a gradual decline in the rate of GRBs with increasing metallicity, and estimates of the metallicity at which the GRB rate becomes suppressed vary from approximately solar \cite{perley16} to 0.3 solar or less \cite{graham13}. This is notably still somewhat higher than the metallicity in stellar models in which single stars can retain sufficient angular momentum to form centrifugally supported accretion discs during collapse. 

\section{Energetics}
\label{sec:energetics}

Prior to the identification of GRB redshifts one of the strongest arguments for a Galactic origin was that the energetics at cosmological distances were extreme. It is common to calculate the energy of an event under the assumption that it is emitting isotropically, although this is not the case for GRBs. As such one can define an isotropic energy\footnote{A common source of confusion is that $E_{iso}$ contains a $(1+z)$ term while $L_{iso}$ does not. This is because the fluence measured is equal to the flux multiplied by the {\em observer} time, not the rest frame time. It is also important to note that while in principle $E_{iso}$ is a bolometric measurement (i.e. the total energy) it is also common to see it measured in only a specific observed energy range, such as the 15-150 keV of the {\em Swift} BAT.} $(E_{iso} = 4 \pi d_L^2 S_{obs} / (1+z))$ and an isotropic peak luminosity  ($L_{iso} = 4 \pi d_L^2 F_{obs}$). For long-duration GRBs, the energies observed span the range $10^{48} < E_{iso}  < 10^{55}$ erg \citep{amati02}, with local low luminosity GRBs at the low energy end, and the brightest of all time, GRB 221009A covering the upper envelope (even though it also had a redshift of only $z=0.151$). The short duration GRBs span isotropic energies in the range $10^{46} < E_{iso}  < 10^{52}$ erg \citep{fong+15}, although most have $E_{iso}  > 10^{48}$ erg and the faintest burst is the exceptional GRB 170817A, associated with the gravitational wave event GW170817 with $E_{iso} = 5 \times 10^{46}$ erg \citep{Goldstein+2017}.  

These very large energetics correspond to a significant fraction of a solar rest mass but are ameliorated by relativistic beaming. Measurements of achromatic jet breaks in long-GRB afterglows suggest typical opening angles of a few degrees and corresponding beaming fractions of $f_b = (1- \cos{\theta_j})$. These reduce the total energy requirements by factors of 15-1500 for angles between $20 > \theta_j > 2$ degrees. For long GRBs the beaming corrected energies are much more narrowly peaked around 
$E_{iso} = 10^{51}$ erg s$^{-1}$ \citep[e.g.][]{frail01}, implying that typically the more energetic bursts have narrower jets. For short GRBs determining the presence of jet-breaks has been more challenging because of the typically fainter afterglows, although evidence suggests that the jets are typically wider (on average 15-20 degrees), although the lower fluence also results in lower isotropic energy releases, with a mean $E_{iso} \sim 2 \times 10^{51}$ erg \citep{fong+15}.

In addition to the distribution of energies, there has been substantial work investigating if the GRB energetics correlate with any other properties of the burst or its environment. 
This is done both to gain insight into the GRBs themselves, but has also been motivated by a desire to find energetic correlations with small intrinsic scatter which may enable distance measurements and hence make GRBs valuable as standard candles. The most well-known of these relations are the relations between the energy (or luminosity) and the peak of the $\nu F_{\nu}$ spectrum $(E_p)$, also known as the Amati \citep{amati02} and Yonetoku \citep{yonetoku04} relations, while it is also apparent that the beaming corrected energies of the bursts also correlate with the spectral peak \citep{ghirlanda04}.  Other relations have also been searched for including between afterglow decay rate and luminosity, and the duration of apparent plateaux \citep[e.g][]{dainotti10}. However, the robustness of these relations, and if they can ultimately enable reliable cosmological inference is currently unclear. 

\section{Central Engines} 
\label{sec:engines}
Compactness arguments point to the source of GRBs being very small (scales of tens to hundreds of km). To achieve the energies observed for GRBs places stringent constraints on the energy densities in this compact region, and leads to the realization that GRBs are engine-driven explosions, in which the action of the compact source is responsible for the GRB creation. There are rather few physical systems that can create such strong central engines, and only two are considered seriously as GRB engines. These are hyper-accreting stellar mass black holes or very rapidly spinning, highly magnetized neutron stars (magnetars).

%However, it important to note that while the nature of the engines has limited range, there are multiple routes to the creation of such engines, such that a given engine does not point directly to the progenitor system of the GRB. Indeed, there are a large number of routes to the production of both hyperaccreting black holes and magnetars (see Figure~\ref{fig:progenitors} and
%\cite{fryer,levan06,beniamini}).

\subsection{Black hole engines}
The classical central engine for a GRB is that of a newly formed black hole surrounded by a hyper-accreting disc, perhaps accreting at rates $(\dot{m})$ of several solar masses per second and so creating a burst whose luminosity can 
be extremely high ($L \sim \eta \dot{m} c^2$), where $\eta \sim 0.1$ is the efficiency. The resulting engine converts the infalling matter into energy available to power the jet, perhaps by magnetic processes, such as the Blandford Zjanek mechanism \citep[e.g][]{kom09}. Alternatively, the jet may be powered via the annihilation of neutrinos and anti-neutrinos produced from the accretion disc which is sufficiently hot both to produce copious neutrinos and to reduce the mean-free path of these neutrinos such that a significant fraction will interact just above the disc \citep[e.g.][]{birkl07}. 
The result is that a relatively efficient process (e.g. 10\%) can power an outflow that continues to be powered while accretion continues onto the central engine. 

The creation of black hole engines in stars is non-trivial since the rotation of the star must be sufficiently large that not all material accretes directly onto the black hole during formation. This leads to models for black hole engines that require significant rotation, either in the form of a rapidly rotating single star (i.e. one that has not lost significant mass and angular momentum to stellar winds) or via binary interactions \citep[e.g.][]{fryer}. 

\subsection{Magnetar engines} 
Magnetar central engines have only been considered for GRBs in the last 20-years \citep{Metzger+08_magnetarEE}. In this case, the energy that powers the GRB arises not from accretion but from the tapping of the rotational energy of the magnetar. This has a maximum since neutron stars cannot spin arbitrarily rapidly without disrupting (exceeding the so-called break-up rotational speed). Hence the total energy available from a magnetar central engine is $E_{tot} = {1 \over 2} I \omega^2$, where $I$ is the moment of inertia and $\omega$ the rotational velocity. Magnetars thought to power GRBs must have initial spin periods in the millisecond regime, and are often known as millisecond magnetars. In contrast to black hole engines the total energy from a magnetar-driven explosion cannot exceed its rotational energy although uncertainties in the efficiency in which this energy is converted to radiation, and in the beaming factors mean there are few GRBs where magnetars can be excluded as engines. In principle, magnetar central engines are often invoked in a variety of extreme transients including GRBs and superluminous supernovae, with the difference in the observed signal (i.e. GRB vs supernova) related to the spin-down time of the magnetar and the structure of the progenitor star \citep{Metzger+15}.

Importantly, the nature of the engine provides rather limited information about the nature of the progenitor star. There are a wide range of routes through which the different types of engines can be formed. A comprehensive list of mechanisms for black hole accretion disc formation was given in \cite{fryer}, and some of these routes are indicated in Figure~\ref{fig:progenitors}. Similarly, methods for the creation of magnetars include the collapse of massive stars \citep{ben94}, the merger of white dwarfs \citep{levan06} or the merger of neutron stars \citep{Abbott+17a}. Massive star origins are apparent in the Milky Way, but other routes may operate as well. Indeed, the rate of formation of magnetars in the Milky Way is several orders of magnitude higher than the rate of production of GRBs \citep[e.g.][]{beniamini}, even when allowing for the uncertain beaming corrections (it is a significant fraction of the core-collapse supernovae rate). If GRBs are created from magnetars then these magnetars must have different properties at birth than those in the bulk of the magnetar population. 

\section{Progenitors}
\label{sec:progenitrs}
\subsection{Massive star progenitors}
\label{sec:sn}

\begin{figure*}
\vspace{-0.1cm}
\begin{center}
\includegraphics[width=1.0\textwidth]{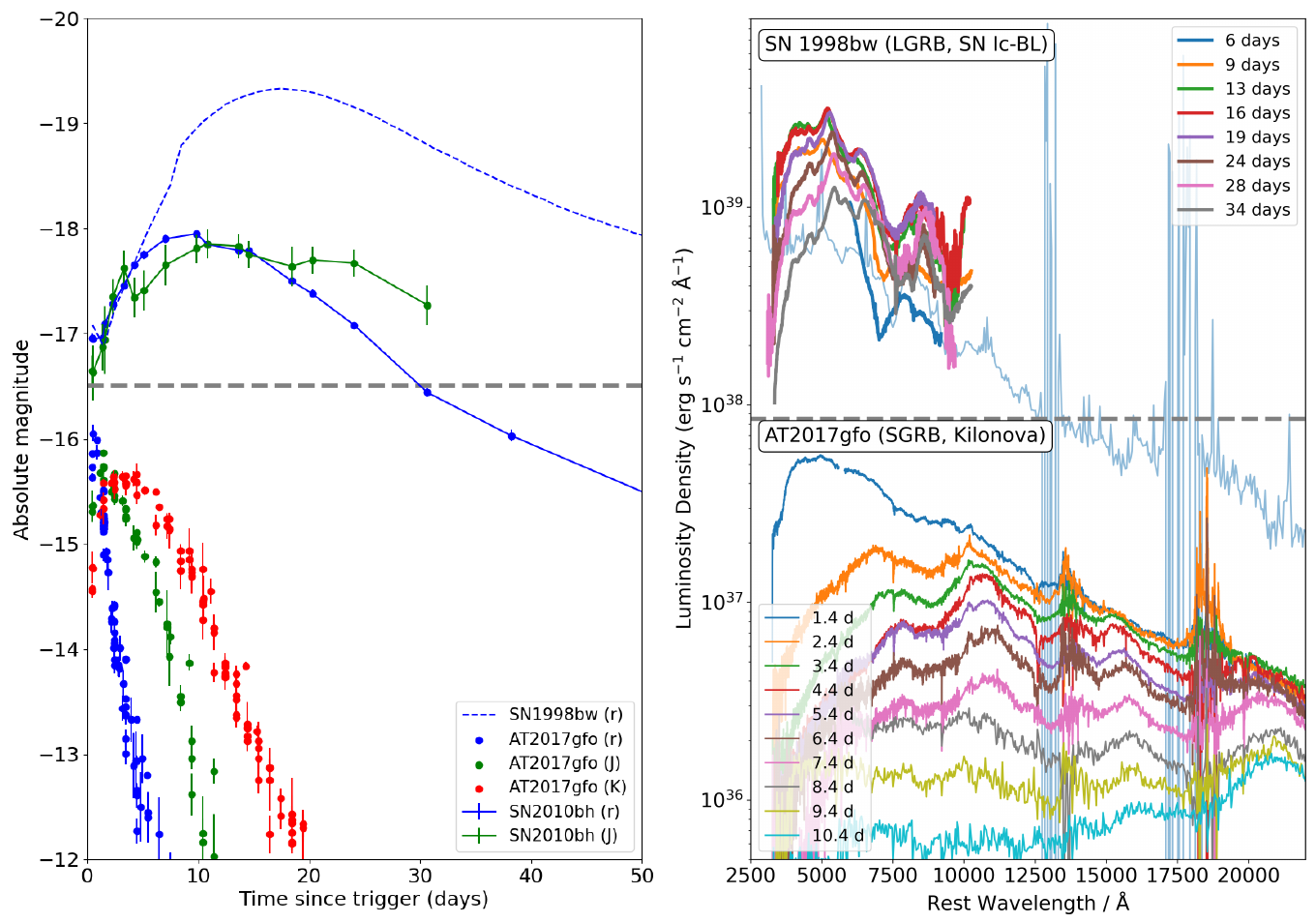} 
\end{center}
\vspace{-0.4cm}
    \caption{The observational differences between supernovae and kilonova associated with GRBs showing the lightcurves (left) and spectra (right). We compare the best-sampled kilonova AT2017gfo with well-sampled GRB supernovae, in particular, SN~1998bw, although we also show the X-shooter spectrum of SN~2010bh to demonstrate the IR behavior of broad-lined SN Ic. Supernova lightcurves are approximately 100 times brighter at peak than kilonova and reach an optical peak on timescales of 10-20 days compared to $<1$ day for kilonova. However, supernovae remain dominated by optical light for most of their evolution, while the spectra of kilonovae show a clear shift into the infrared and peak beyond one micron on timescales of only a few days.}
   \label{fig:sn_kn}
\end{figure*}

A subset of gamma-ray bursts arise from the collapse of massive stars \citep{Galama+98,hjorth03}, and traditionally this has been viewed as the origin of the long-duration GRB population. Initial evidence for this arose from the highly star-forming host galaxies, and the absence of passive galaxies from the long-GRB host population \citep{bloom02,fruchter06}. More robust evidence arises 
from the presence of unusual supernovae whose spectra are devoid of both hydrogen and helium (known as type Ic supernovae) and that exhibit very high expansion velocities of 20,000-30,000 km s$^{-1}$,  resulting in significant doppler broadening of their spectral lines.  The first event with a clear supernova signature is GRB 980425, associated with 
SN~1998bw \citep{Galama+98}. However, this burst was extremely unusual arising from a distance of only 40 Mpc (still the closest supernova GRB more than 25 years later) and with extremely low luminosity. The supernova luminosity was also unusual, reaching a peak absolute magnitude of $M_V \sim -19.3$, approximately 2 magnitudes brighter than the mean for core-collapse supernova events \cite{frohmaier21}.  Following its discovery intensive searches in the afterglows of more distant and more energetic GRBs revealed telltale supernova signatures in the form of red bumps which slowed, or even reversed the afterglow decay on timescales of 10-15 $(1+z)$ days and had similar luminosity to SN~1998bw (see Figure~\ref{fig:sn_kn}). Eventually, robust spectroscopic evidence for the supernova connection in cosmological GRBs came from the nearby GRB 030329 ($z=0.17$, \cite{hjorth03,stanek03}), where detailed spectroscopic follow-up revealed a supernova extremely similar to SN 1998bw, cementing the association. 

There are now more than 50 cases where supernovae have been claimed in GRBs based on bumps in photometric lightcurves and 30 cases where there is reasonable spectroscopic evidence for them. In the vast majority of these cases, the GRBs are of the long-duration variety, leading to the link between long-duration GRBs and massive stars. However, in one case there is now a supernova signature in a short-duration GRB, indicating a more complex selection function \cite{rossi21}. However, in general, it seems that the overwhelming majority of supernova GRBs are from the population of bursts with $T_{90} > 2$s. 

While the supernova connection demonstrates that the progenitors of these GRBs are massive stars, it does not answer the detailed questions about their progenitors. What are the progenitor masses? Are the stars formed via single-star routes (at least sometimes) or are binary interactions essential? Is the lack of hydrogen in the supernova spectra due to loss in stellar winds, common envelopes, mass transfer, or even via burning in chemically homogeneous evolution? While substantial progress toward the origins of ``normal" core-collapse supernovae has been made via the direct detection of progenitors, this seems unlikely to be the case for GRBs since even the closest events are beyond the distances where progenitor identification is straightforward. Hence, understanding the more detailed properties of the progenitor stars must be done indirectly. So far this can be approached from two distinct avenues. The first is to attempt to understand the properties of the star at explosion by reconstructing the details of the supernova, in particular in terms of the total ejecta mass and the nickel mass created \citep[e.g.][]{modjaz18}. These results typically confirm that the supernovae in GRBs have higher energies (as evidenced by their velocities) and ejecta more nickel than typical supernovae (GRB-SNe have $\sim 0.5$ M$_{\odot}$ of nickel on average, compared to $\sim 0.1$ for other stripped-envelope supernovae \citep[e.g.][]{anderson19}, although GRB-supernovae are less obviously outliers to the population of broad lined SNe Ic that are seen without associated GRBs. These high ejecta masses and energies support the inference that GRBs arise from the explosion of relatively massive stellar cores.

The second channel is to use the environments of the GRBs as a proxy for the stars that form them. This has also been done very successfully, for example, %using the variations in the H$\alpha$ equivalent width as a route to measuring the age (and hence potentially zero age main sequence mass) of the progenitors, 
using the locations of the GRBs in comparison with supernovae to argue that their progenitors are more massive \citep{fruchter06}, and using the properties of the hosts as a whole to demonstrate that GRBs with supernovae form preferentially in low metallicity environments \citep[e.g.][]{fruchter06,graham13}. For nearby GRBs substantial progress has been made using integral field spectrographs to obtain spatially resolved observations, while for the more distant (but larger) samples either integrated spectroscopy or even imaging observations have been the standard \citep[e.g.][]{perley16}. 

Hence these combined diagnostics paint a thorough, if still incomplete picture of the progenitor of a (normally long-duration) supernova GRB; it is a massive star (likely $>20$M$_{\odot}$ at low metallicity (with a threshold in the region 0.3-1 Z$_{\odot}$) that is likely rapidly rotating. The metallicity thresholds are substantially above the thresholds where single massive stars are expected to create GRB-like explosions and this suggests that, at least in the local Universe, that binary interactions are likely important in creating GRB progenitors.

\begin{figure*}
\vspace{-0.1cm}
\begin{center}
\includegraphics[width=1.0\textwidth]{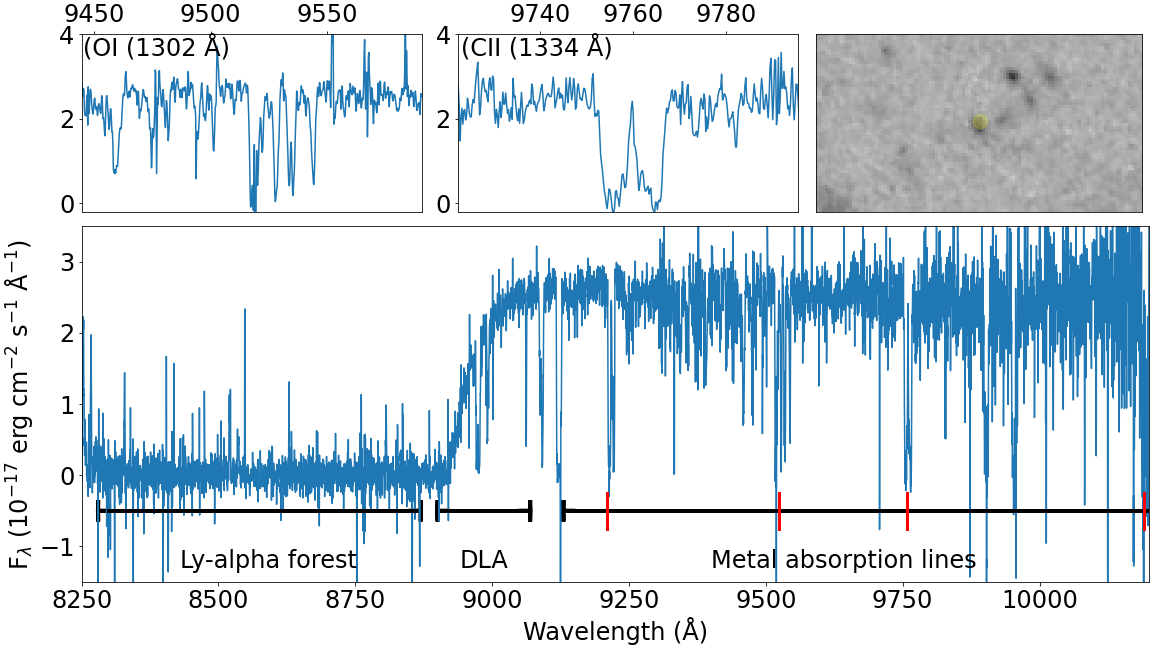} 
\end{center}
\vspace{-0.4cm}
    \caption{Spectroscopy of the afterglow of GRB 210905A at $z=6.29$ \citep{saccardi23}. The intrinsic afterglow emission is a power-law however the imprint of both the interstellar medium in the host galaxy and the intergalactic medium is clearly visible in the spectrum. In particular, at wavelengths shorter than Ly-$\alpha$ (1216$\AA$) the spectrum is highly absorbed by many neutral hydrogen clouds in the intergalactic medium (the Ly-$\alpha$ forest). At the wavelength of Ly-$\alpha$ in the host galaxy, there is substantial absorption with a high column density $\sim 10^{21}$ atoms cm$^{-2}$, and then redward of Ly$-\alpha$ most of the light from the afterglow escapes, but the imprint of specific metals in the host galaxy can clearly be seen. In this case, there are different absorbers in the host with a range of different velocities (see top left and center insets). Spectra such as this one allow much detail to be discerned about the host galaxy. Finally, the galaxy can also be studied in emission with facilities such as the Hubble Space Telescope (top right). }
\label{fig:spec}
\end{figure*}

\subsection{Compact binary progenitors} 
\label{sec:mergers}
Linking the mergers of compact objects to at least some gamma-ray bursts has been more challenging than establishing the supernova connection. Compact object mergers have traditionally been associated with the short population of GRBs, partly because of the evidence of supernovae in many long bursts. However, evidence for this origin was more difficult to come by. It was obtained first by studies of the host galaxies and environments of the short bursts which were shown to include ancient elliptical galaxies without any sign of ongoing star formation \citep[e.g.][]{gehrels05}. Furthermore, the offsets of the bursts around their host galaxies were systemically higher than for the long GRBs \citep[e.g.][]{FongBerger13}, consistent with a combination of kicks to the binaries during the formation of each neutron star which propels the progenitor from its birth-site at velocities of up to several hundred kilometers per second. 

Evidence for compact object progenitors was further strengthened by the identification of kilonova signatures in the afterglows of GRBs. Kilonovae were first postulated as faint, fast transients powered by the decay of heavy elements synthesized in a merger\footnote{Other names in the literature include mini-SN and macronova, but all refer to the same physical phenomena.}. Because the ejecta masses are low and the velocities high, kilonovae are expected to be visible for much shorter periods than supernovae (days rather than weeks, see Figure~\ref{fig:sn_kn}). Furthermore, the make-up of kilonova ejecta is very different from supernova. In particular, because kilonovae are powered by the production of the heaviest elements (e.g. strontium, gold, platinum, uranium) created in the rapid capture of neutrons ($r-$process) they contain elements with complex electron orbitals that create a very high opacity to outgoing light. As a consequence, kilonovae are expected to become red, and peak in the near, or even mid-infrared days after the merger \cite{barneskasen13}. 

In the short GRB 130603B, observations with the {\em Hubble Space Telescope} show that by 10 days the optical/IR  transient reddened substantially from early times, reaching a peak absolute magnitude of $\sim -16$ before fading rapidly \citep{tanvir13,berger13}. Following this initial discovery, similar faint bumps were reported in further bursts \citep[][and references therein]{Gompertz+18}. Although the sampling of these counterparts is sparse they can readily be interpreted as arising from kilonovae emission. 

Conclusive evidence for compact object mergers was finally found in 2017 with the detection of the binary neutron star merger GW170817 in gravitational waves \citep{Abbott+17a}, followed 1.7 seconds later by the detection of the short GRB 170817A by both the {\em Fermi} and {\em INTEGRAL} satellies \citep{Goldstein+2017,Savchenko+17} This event was subsequently localized to the galaxy NGC 4993 at 40 Mpc distance \citep{coulter17}, implying that GRB 170817A was extremely underluminous in $\gamma$-rays compared to most short GRBs. However, further observations in the X-ray and radio \citep{Troja+17,Hallinan+2017}, and in particular VLBI observations \citep{Mooley+18,Ghirlanda+19} suggested that this GRB was likely viewed $\sim 25$ degrees from its jet-axis, and given the jet structure, this explains the relative $\gamma$-ray faintness as well as the unusual afterglow behavior that rises for the first $\sim 100$ days before reaching a broad peak and subsequently decaying. 

In addition to GRB and its afterglow the optical counterpart of GW170817A was dominated by its kilonova emission and is by far the best-sampled kilonova both in terms of its lightcurve \citep[e.g.][]{Arcavi+17,Chornock+17,Drout+17, Lipunov+17,Tanvir+17,Valenti+17} and spectroscopically \citep[e.g.][]{Pian+17,Shappee+17,Smartt+17}.  

The discovery and characterization of the kilonova population began in the short-GRB population, lending support to a relatively clean distinction between the progenitors of short and long-GRBs. However, more recently two kilonovae have been identified in the afterglows of long-duration GRBs \citep{rastinejad22,troja22}. These kilonovae, identified by the characteristic rapid redening and in one case via the likely identification of $r-$process production via JWST spectroscopy \citep{levan24a} demonstrate that a potentially significant fraction of nearby long-GRBs could arise from mergers.

\subsection{Durations, spectra and progenitors, the definition of a GRB}
\label{sec:definition}
The observed dichotomy in GRB duration splits them into long- and short-duration bursts (Figure~\ref{fig:sh}). It is logical to associate this distinction with two different progenitor populations. However, recent observations cast doubt on this simple association. While supernovae are very rare in short-GRBs, kilonovae may contribute substantially (if sub-dominantly) to long-GRBs \citep{levan24a}.
 Hence, the duration of a burst is not a strong indication of its nature. Some authors have suggested replacing the terms long- and short- with Type II and Type I GRBs \citep{zhang09}, although the use of this nomenclature varies in the field. It is clear now that duration should not be used directly as a proxy for progenitors, and that both merger-GRBs and collapsar-GRBs can be present at a wide range of duration. 

Furthermore, increasingly GRB-like phenomena are being identified outside of the $\gamma$-ray regime. For several years wide-field optical surveys have been identifying rapidly changing transients with power-law spectra and from cosmological distances. The properties of these events are consistent with GRB afterglows in which a prompt $\gamma$-ray signature was not seen \citep[e.g.][]{cenko13}. The absence of $\gamma$-ray emission from these events may be due either to physical effects, such as bursts viewed off-axis, or with high baryon loading that precludes $\gamma$-ray production, or because of observational limitations. However, such events do show supernova signatures and are otherwise similar to GRBs. 

In a similar vein, there are also populations of short-lived transients found at other wavelengths whose relation to GRBs remains unclear. For example, the population of Fast X-ray Transients that has been building from archival observations with {\em Chandra}, {\em Swift} and {\em XMM-Newton} \citep{qv23}. In most cases, these events do not have $\gamma$-ray counterparts, but also would not be observable to current $\gamma$-ray detectors for the typical X-ray to $\gamma$-ray ratios seen in GRBs. They are typically of much longer duration than GRBs, although such observational differences could be ascribed to selection effects \citep{levan24b}. More recently, the Einstein Probe has observed several Fast X-ray Transients (typically of shorter duration) associated with GRBs, suggesting a causal link that is further strengthened by the presence of broad-lined type Ic supernovae, even where no $\gamma-$ray signal is observed \citep[e.g.][]{vandalen24}. The conclusion would appear to be that at least a substantial fraction of FXTs are related to the same physical mechanisms as GRBs. 

This in turn leads to the challenge of defining a GRB. Simply describing it as a flash of $\gamma$-rays appears unsatisfactory since many cosmic sources emit such flashes but are not GRBs. Indeed, since the line between X-rays and gamma-rays is arbitrary and set by us, setting a line between sources emitting in the X-ray or $\gamma$-ray regime does not make sense since while such detections are often made by different technology, they are not necessarily indicative of fundamentally different underlying emission. Alternatively, ascribing GRBs to solely to a given physical mechanism is also unsatisfactory because they are known to represent multiple different progenitors. However, there are features of all of the events that appear to be GRBs that can unify. We therefore suggest that a good definition for a GRB is: 
{\bf A gamma-ray burst is the emission of radiation from a relativistically expanding outflow from a stellar scale central engine produced in a one-off event}. This definition unifies disparate observational signatures of the same physical mechanisms and suggests that many objects not detected in the $\gamma$-ray regime may be GRBs. However, it is also crucial to note that this definition requires the detection of the relativistic ejecta -- a kilonova, or broad-lined type Ic event is {\em not} a GRB (even if it may have hosted one) because while the progenitor is correct the emission mechanism is not. 

\subsection{Rates}
One of the critical questions in the various progenitor scenarios is how frequently they occur in nature. GRBs, of all varieties, are intrinsically rare events, with those observable to current instrumentation having a rate of only a few per day across the Universe. However, the actual volumetric rates are the subject of significant uncertainty due to rate evolution, beaming, small sample sizes, and difficulties in determining the detectability of different bursts as a function of distance (or redshift) \citep{Littlejohns+13}. In Table~\ref{tab:rates} we report current estimates of the rates of GRBs of different types, as well as a comparison to the expectations for 
different progenitor models and massive stars.

\begin{table}[!h]
\caption{Rates of GRBs and related observational and physical phenomena. Note that the GRB rates are as observed, and should be corrected for beaming, increasing the progenitor rates by a factor of $\sim 10-100$.}%%%Table caption goes here
\label{tab:rates}
\begin{center}
\begin{tabular}{|lll|}%%%The number of columns has to be defined here
%\noalign{\global\arrayrulewidth=1mm}
\hline
Events & Volumetric rate (Gpc$^{-3}$ yr$^{-1}$) & References  \\ 
\hline
LGRBs ($L>10^{50}$ erg s$^{-1}$) & $0.8 \pm 0.1$ & \cite{sun15} \\
LGRBs ($L> 5 \times 10^{46}$) & 164$^{+98}_{-65}$ & \cite{sun15} \\
SGRBs & 1-3 & \cite{sun15} \\
\hline
Core collapse SNe & $(9.10_{-1.27}^{+1.56}) \times 10^{4}$ & \cite{frohmaier21}\\
Stripped Envelope SNe  &  $(2.41_{-0.64}^{+0.81}) \times 10^{4}$ & \cite{frohmaier21} \\
%Type Ia supernovae \\
Superluminous supernovae & $35_{-13}^{+25}$ & \cite{frohmaier21} \\
Kilonovae & $<4000$ & \cite{andreoni20} \\
Fast X-ray Transients ($L >10^{44}$ erg s$^{-1}$) & 2000-5000 & \cite{qv23} \\
\hline
BNS mergers & 10-1700&  \cite{gwtc3}\\
BH-NS mergers  &  8-140 & \cite{gwtc3}\\
BBH mergers & 16-61  & \cite{gwtc3} \\
%Magnetar formation & \\
\hline

%\noalign{\global\arrayrulewidth=0.1mm}
%\arrayrulecolor{gray}
%910503 & Prompt & \cite{nowak94} \\

\hline
\end{tabular}
\end{center}
\vspace*{-4pt}
\end{table}%%%End of the table

\section{Probes of Extreme Astrophysics and Cosmology}
\label{sec:probes}

\subsection{GRBs as probes of distant galaxies and the intergalactic medium}
In addition to understanding the nature of gamma-ray bursts themselves, the GRBs provide a bright backlight that can be used to address questions in several other areas of astrophysics. Firstly, the extreme luminosities of GRBs make them visible across the observable universe and so they can be traced across a wide range of redshift \citep[e.g.][]{jakobsson06}. There are even suggestions that the first generations of very massive stars could have driven powerful and very long-duration GRBs \citep{toma16}, although their true masses, and if they drive GRB-like events or explode without leaving a remnant as a pair-instability supernova remains uncertain. Nonetheless, the prospect of using GRBs as probes of the distant Universe continues to be a major driver in the field, and a core goal of many of the proposed next-generation GRB detection missions including THESEUS (ESA \cite{theseus}), GAMOW (NASA \cite{gamow}) and HIZ-GUDAM (JAXA \cite{hizgundam})\footnote{At the time of writing all of these missions are proposals, and none have been selected for flight}.   

Beyond their simple presence, the GRB and its afterglow, shine through first the interstellar medium in the host galaxy and then the intergalactic medium between the burst and the observer, and in doing so illuminate material \citep[e.g.][]{Selsing+19}. It is from this that we can measure the redshifts of distant GRBs through the combination of narrow metal lines in their host galaxies\footnote{Here, we use astronomers parlance that metals are elements heavier than H and He, and the most common elements seen in GRB afterglow spectra include C, N, O, Si, S, Mg, Fe} and from much broader Damped Lyman Alpha absorption (DLAs) (see Figure~\ref{fig:spec}). These spectra can provide direct measurements of the metallicities of cold gas within distant galaxies, even for galaxies that are too faint for detection in deep surveys with, for example, {\em HST} and {\em JWST}.  In doing so they can provide pictures of distant galaxies that are more physically complete than available by other methods. The redshifts alone also provide potential information since many long-duration GRBs arise from the collapse of massive stars which have short lives and hence GRBs can trace star formation \citep{tanvir12}, or at least low metallicity star formation should a strong metallicity bias be in effect \citep{graham13}. 

A particular goal is to use GRBs to probe the so-called reionization era, during which the photons produced by early luminous objects (exactly which luminous objects remains a subject of discussion) ionized an intergalactic medium which had been neutral since the formation of the Cosmic Microwave Background. This is possible both because massive star progenitors of GRBs may well be a dominant source of ionizing photons in the Universe, and because measurements of GRB afterglows can allow inference of the fraction of escaping radiation along a given line of sight to a massive star \citep[e.g.][]{tanvir19}.

A final cosmological utility arises from the detection of GRBs at extremely high energy (for example in the TeV regime). Such detections have now been made in a handful of cases. Importantly, because they are copious photon emitters it is likely that at least some GRBs produce high fluxes of such high-energy photons. Unlike optical/IR photons these very high-energy photons do not travel to us largely unimpeded but are scattered by the Extragalactic Background Light (EBL). This EBL is the combined optical/IR background light from all the stars in the Universe, and so offers potentially profound cosmological insight. The emission of GRBs, in particular, if observed with sufficiently high signal-to-noise offers the possibility of comparing the {\em observed} emission with that expected under extrapolation of reasonable physical models to determine the optical depth to high energy photons as a function of distance(redshift) or line of sight (sky position) \citep{magic19,lhaaso23}. Such observations remain in their infancy but should become more routine and diagnostic in the era of the next generation of detectors such as the Cherenkov Telescope Array (CTA).

\subsection{GRB and extreme physics}
\label{sec:extreme}
As well as their cosmological utility, GRBs are also widely touted as probes of extreme physics. Perhaps most directly they provide insight into the production and acceleration of jets and the motion of material at ultra-relativistic velocity. In this regard, they are highly complementary to studies of jets in X-ray binaries and around supermassive black holes. The detailed studies of both emission production in prompt emission \citep[e.g.][]{ryde05,ravasio19,ravasio24} and in afterglows \citep[e.g.][]{racusin08,laskar13} directly feed into studies of these issues.

A further straightforward test is that because the prompt emission occurs both in a short time and is visible across several decades of energy is to use GRBs to test the so-called Lorentz Invariance Violation -- more specifically the concept that the speed of light is invariant. Some versions of quantum gravity postulate a dispersion in the speed of light, for example as a function of photon energy. Over the long path lengths from GRBs at cosmological distances, even small differences in speeds become measurable as a delay in the arrival time of photons of different energies. Although there are astrophysical effects that result in spectral evolution in GRBs, it is still possible to use the observations to place stringent constraints on variations in the speed of light of order 1 part in 10$^{12}$ \citep[e.g.][]{lhasso23b}. 

Finally, GRBs and multi-messenger astronomy are intimately linked and are a promising route for enlarging samples of events observed beyond the electromagnetic spectrum. In particular, while many possible signatures are considered for the electromagnetic counterparts to gravitational wave sources, perhaps most notably including kilonovae whose emission can be seen independently of viewing angle, GRBs offer potential as they are visible to much larger distances than kilonovae. This increased horizon is particularly important for next-generation (so-called third generation) gravitational wave detectors such as Cosmic Explorer and the Einstein Telescope whose cosmological reach is such that they can observe 
merging neutron star binaries to $z \sim 3$, and black hole neutron star systems at even higher redshifts. The lack of further GW-EM sources besides GW170817/GRB170817A/AT2017gfo suggests that mergers with EM counterparts are rare within the current horizons of gravitational wave detectors. Building meaningful samples of multi-messenger events is likely to require these third-generation detectors in concert with GRBs. Furthermore, particle acceleration in the jets of GRBs is a promising site for the production of high-energy neutrinos that have so far only been associated with extreme active galactic nuclei \citep[Blazars,][]{icecube18}. While these models are still uncertain, and the detection of neutrinos coincident with GRBs and GW events will be challenging, even in the next decade, such a detection would represent a remarkable opportunity to build an extremely detailed understanding of the anatomy of a merging binary that includes the masses of the merging components, their equations of state, how and where mass is ejected in the merger, which elements are synthesized, by what mechanisms are particles accelerated, and how relativistic jets form.

\section{Summary}
In the 50+ years since their discovery GRBs have moved from an astrophysical curiosity, through an era when their origins were one of the central questions in astronomy to the current epoch in which GRB identification and follow-up has become routine, but where new surprises continue to be found, and where GRBs are deployed as probes of an increasingly diverse range of pressing questions in contemporary physics, astrophysics and cosmology. While the progenitors of at least some GRBs are unambiguously clear, what drives the observed properties of the GRB (e.g. duration) is still an open question. How GRBs relate to extreme transients identified in other regimes is also unclear -- are many of the fast-fading optical transients now uncovered in synoptic sky surveys really just undiscovered GRBs? How are the Fast X-ray Transients, discovered in archival observations with X-ray satellites and now in real-time with the Einstein Probe related to GRB phenomena? Are more progenitor channels still to be identified, such as WD-IMBH disruptions as ultra-long GRBs, or the collapse of first-generation, population III stars? 

The challenge of GRB detection is that it must, predominantly, be done from space. The detection and localization of GRB signals rely on the presence of a suitable satellite, and different missions are optimized for different forms of GRB detection. For example, the {\em Fermi} satellite provides exceptional spectral and timing information but poor localization, {\em Swift} provides excellent and rapid localization, as well as multi-wavelength follow-up, but is more limited in the information it provides for the prompt emission. 
The challenge going forward is to ensure that the community has access to all the information currently available, and ideally more besides (e.g. better sky coverage, fainter bursts, prompt redshift information). Many axes for improvement can allow GRBs to reach their potential to answer the questions outlined above, and these are the subject of many new enterprises that seek to fully exploit GRBs as probes of (astro)physics and cosmology across the visible Universe.

%\seealso{article title article title}

\bibliographystyle{Harvard}
\begin{thebibliography*}{115}
\providecommand{\bibtype}[1]{}
\providecommand{\natexlab}[1]{#1}
{\catcode`\|=0\catcode`\#=12\catcode`\@=11\catcode`\\=12
|immediate|write|@auxout{\expandafter\ifx\csname
  natexlab\endcsname\relax\gdef\natexlab#1{#1}\fi}}
\renewcommand{\url}[1]{{\tt #1}}
\providecommand{\urlprefix}{URL }
\expandafter\ifx\csname urlstyle\endcsname\relax
  \providecommand{\doi}[1]{doi:\discretionary{}{}{}#1}\else
  \providecommand{\doi}{doi:\discretionary{}{}{}\begingroup
  \urlstyle{rm}\Url}\fi
\providecommand{\bibinfo}[2]{#2}
\providecommand{\eprint}[2][]{\url{#2}}

\bibtype{Article}%
\bibitem[{Abbott} et al.(2017)]{Abbott+17a}
\bibinfo{author}{{Abbott} BP}, \bibinfo{author}{{Abbott} R},
  \bibinfo{author}{{Abbott} TD}, \bibinfo{author}{{Acernese} F},
  \bibinfo{author}{{Ackley} K}, \bibinfo{author}{{Adams} C},
  \bibinfo{author}{{Adams} T}, \bibinfo{author}{{Addesso} P},
  \bibinfo{author}{{Adhikari} RX}, \bibinfo{author}{{Adya} VB},
  \bibinfo{author}{{Affeldt} C}, \bibinfo{author}{{Afrough} M},
  \bibinfo{author}{{Agarwal} B}, \bibinfo{author}{{Agathos} M},
  \bibinfo{author}{{Agatsuma} K}, \bibinfo{author}{{Aggarwal} N},
  \bibinfo{author}{{Aguiar} OD}, \bibinfo{author}{{Aiello} L} and
  \bibinfo{author}{et~al.} (\bibinfo{year}{2017}), \bibinfo{month}{Oct.}
\bibinfo{title}{{GW170817: Observation of Gravitational Waves from a Binary
  Neutron Star Inspiral}}.
\bibinfo{journal}{{\em \prl}} \bibinfo{volume}{119} (\bibinfo{number}{16}),
  \bibinfo{eid}{161101}. \bibinfo{doi}{\doi{10.1103/PhysRevLett.119.161101}}.
\eprint{1710.05832}.

\bibtype{Article}%
\bibitem[{Abbott} et al.(2023)]{gwtc3}
\bibinfo{author}{{Abbott} R}, \bibinfo{author}{{Abbott} TD},
  \bibinfo{author}{{Acernese} F}, \bibinfo{author}{{Ackley} K},
  \bibinfo{author}{{Adams} C}, \bibinfo{author}{{Adhikari} N},
  \bibinfo{author}{{Adhikari} RX}, \bibinfo{author}{{Adya} VB},
  \bibinfo{author}{{Affeldt} C}, \bibinfo{author}{{Agarwal} D},
  \bibinfo{author}{{Agathos} M}, \bibinfo{author}{{Agatsuma} K},
  \bibinfo{author}{{Aggarwal} N}, \bibinfo{author}{{Aguiar} OD},
  \bibinfo{author}{{Aiello} L}, \bibinfo{author}{{Ain} A},
  \bibinfo{author}{{Ajith} P}, \bibinfo{author}{{Akutsu} T},
  \bibinfo{author}{{de Alarc{\'o}n} PF}, \bibinfo{author}{{Akcay} S},
  \bibinfo{author}{{Albanesi} S}, \bibinfo{author}{{Allocca} A},
  \bibinfo{author}{{Altin} PA}, \bibinfo{author}{{Amato} A},
  \bibinfo{author}{{Anand} C}, \bibinfo{author}{{Anand} S},
  \bibinfo{author}{{Ananyeva} A}, \bibinfo{author}{{Anderson} SB},
  \bibinfo{author}{{Anderson} WG}, \bibinfo{author}{{Ando} M},
  \bibinfo{author}{{Andrade} T}, \bibinfo{author}{{Andres} N},
  \bibinfo{author}{{Andri{\'c}} T}, \bibinfo{author}{{Angelova} SV},
  \bibinfo{author}{{Ansoldi} S}, \bibinfo{author}{{Antelis} JM},
  \bibinfo{author}{{Antier} S}, \bibinfo{author}{{Antonini} F},
  \bibinfo{author}{{Appert} S}, \bibinfo{author}{{Arai} K},
  \bibinfo{author}{{Arai} K}, \bibinfo{author}{{Arai} Y},
  \bibinfo{author}{{Araki} S}, \bibinfo{author}{{Araya} A},
  \bibinfo{author}{{Araya} MC}, \bibinfo{author}{{Areeda} JS},
  \bibinfo{author}{{Ar{\`e}ne} M}, \bibinfo{author}{{Aritomi} N},
  \bibinfo{author}{{Arnaud} N}, \bibinfo{author}{{Arogeti} M},
  \bibinfo{author}{{Aronson} SM}, \bibinfo{author}{{Arun} KG},
  \bibinfo{author}{{Asada} H}, \bibinfo{author}{{Asali} Y},
  \bibinfo{author}{{Ashton} G}, \bibinfo{author}{{Aso} Y},
  \bibinfo{author}{{Assiduo} M}, \bibinfo{author}{{Aston} SM},
  \bibinfo{author}{{Astone} P}, \bibinfo{author}{{Aubin} F},
  \bibinfo{author}{{Austin} C}, \bibinfo{author}{{Babak} S},
  \bibinfo{author}{{Badaracco} F}, \bibinfo{author}{{Bader} MKM},
  \bibinfo{author}{{Badger} C}, \bibinfo{author}{{Bae} S},
  \bibinfo{author}{{Bae} Y}, \bibinfo{author}{{Baer} AM},
  \bibinfo{author}{{Bagnasco} S}, \bibinfo{author}{{Bai} Y},
  \bibinfo{author}{{Baiotti} L}, \bibinfo{author}{{Baird} J},
  \bibinfo{author}{{Bajpai} R}, \bibinfo{author}{{Ball} M},
  \bibinfo{author}{{Ballardin} G}, \bibinfo{author}{{Ballmer} SW},
  \bibinfo{author}{{Balsamo} A}, \bibinfo{author}{{Baltus} G},
  \bibinfo{author}{{Banagiri} S}, \bibinfo{author}{{Bankar} D},
  \bibinfo{author}{{Barayoga} JC}, \bibinfo{author}{{Barbieri} C},
  \bibinfo{author}{{Barish} BC}, \bibinfo{author}{{Barker} D},
  \bibinfo{author}{{Barneo} P}, \bibinfo{author}{{Barone} F},
  \bibinfo{author}{{Barr} B}, \bibinfo{author}{{Barsotti} L},
  \bibinfo{author}{{Barsuglia} M}, \bibinfo{author}{{Barta} D},
  \bibinfo{author}{{Bartlett} J}, \bibinfo{author}{{Barton} MA},
  \bibinfo{author}{{Bartos} I}, \bibinfo{author}{{Bassiri} R},
  \bibinfo{author}{{Basti} A}, \bibinfo{author}{{Bawaj} M},
  \bibinfo{author}{{Bayley} JC}, \bibinfo{author}{{Baylor} AC},
  \bibinfo{author}{{Bazzan} M}, \bibinfo{author}{{B{\'e}csy} B},
  \bibinfo{author}{{Bedakihale} VM}, \bibinfo{author}{{Bejger} M},
  \bibinfo{author}{{Belahcene} I}, \bibinfo{author}{{Benedetto} V},
  \bibinfo{author}{{Beniwal} D}, \bibinfo{author}{{Bennett} TF},
  \bibinfo{author}{{Bentley} JD}, \bibinfo{author}{{Benyaala} M},
  \bibinfo{author}{{Bergamin} F}, \bibinfo{author}{{Berger} BK},
  \bibinfo{author}{{Bernuzzi} S}, \bibinfo{author}{{Berry} CPL},
  \bibinfo{author}{{Bersanetti} D}, \bibinfo{author}{{Bertolini} A},
  \bibinfo{author}{{Betzwieser} J}, \bibinfo{author}{{Beveridge} D},
  \bibinfo{author}{{Bhandare} R}, \bibinfo{author}{{Bhardwaj} U},
  \bibinfo{author}{{Bhattacharjee} D}, \bibinfo{author}{{Bhaumik} S},
  \bibinfo{author}{{Bilenko} IA}, \bibinfo{author}{{Billingsley} G},
  \bibinfo{author}{{Bini} S}, \bibinfo{author}{{Birney} R},
  \bibinfo{author}{{Birnholtz} O}, \bibinfo{author}{{Biscans} S},
  \bibinfo{author}{{Bischi} M}, \bibinfo{author}{{Biscoveanu} S},
  \bibinfo{author}{{Bisht} A}, \bibinfo{author}{{Biswas} B},
  \bibinfo{author}{{Bitossi} M}, \bibinfo{author}{{Bizouard} MA},
  \bibinfo{author}{{Blackburn} JK}, \bibinfo{author}{{Blair} CD},
  \bibinfo{author}{{Blair} DG}, \bibinfo{author}{{Blair} RM},
  \bibinfo{author}{{Bobba} F}, \bibinfo{author}{{Bode} N},
  \bibinfo{author}{{Boer} M}, \bibinfo{author}{{Bogaert} G},
  \bibinfo{author}{{Boldrini} M}, \bibinfo{author}{{Bonavena} LD},
  \bibinfo{author}{{Bondu} F}, \bibinfo{author}{{Bonilla} E},
  \bibinfo{author}{{Bonnand} R}, \bibinfo{author}{{Booker} P},
  \bibinfo{author}{{Boom} BA}, \bibinfo{author}{{Bork} R},
  \bibinfo{author}{{Boschi} V}, \bibinfo{author}{{Bose} N},
  \bibinfo{author}{{Bose} S}, \bibinfo{author}{{Bossilkov} V},
  \bibinfo{author}{{Boudart} V}, \bibinfo{author}{{Bouffanais} Y},
  \bibinfo{author}{{Bozzi} A}, \bibinfo{author}{{Bradaschia} C},
  \bibinfo{author}{{Brady} PR}, \bibinfo{author}{{Bramley} A},
  \bibinfo{author}{{Branch} A}, \bibinfo{author}{{Branchesi} M},
  \bibinfo{author}{{Brandt} J}, \bibinfo{author}{{Brau} JE},
  \bibinfo{author}{{Breschi} M}, \bibinfo{author}{{Briant} T},
  \bibinfo{author}{{Briggs} JH}, \bibinfo{author}{{Brillet} A},
  \bibinfo{author}{{Brinkmann} M}, \bibinfo{author}{{Brockill} P},
  \bibinfo{author}{{Brooks} AF}, \bibinfo{author}{{Brooks} J},
  \bibinfo{author}{{Brown} DD}, \bibinfo{author}{{Brunett} S},
  \bibinfo{author}{{Bruno} G}, \bibinfo{author}{{Bruntz} R},
  \bibinfo{author}{{Bryant} J}, \bibinfo{author}{{Bulik} T},
  \bibinfo{author}{{Bulten} HJ}, \bibinfo{author}{{Buonanno} A},
  \bibinfo{author}{{Buscicchio} R}, \bibinfo{author}{{Buskulic} D},
  \bibinfo{author}{{Buy} C}, \bibinfo{author}{{Byer} RL},
  \bibinfo{author}{{Cadonati} L}, \bibinfo{author}{{Cagnoli} G},
  \bibinfo{author}{{Cahillane} C}, \bibinfo{author}{{Bustillo} JC},
  \bibinfo{author}{{Callaghan} JD}, \bibinfo{author}{{Callister} TA},
  \bibinfo{author}{{Calloni} E}, \bibinfo{author}{{Cameron} J},
  \bibinfo{author}{{Camp} JB}, \bibinfo{author}{{Canepa} M},
  \bibinfo{author}{{Canevarolo} S}, \bibinfo{author}{{Cannavacciuolo} M},
  \bibinfo{author}{{Cannon} KC}, \bibinfo{author}{{Cao} H},
  \bibinfo{author}{{Cao} Z}, \bibinfo{author}{{Capocasa} E},
  \bibinfo{author}{{Capote} E}, \bibinfo{author}{{Carapella} G},
  \bibinfo{author}{{Carbognani} F}, \bibinfo{author}{{Carlin} JB},
  \bibinfo{author}{{Carney} MF}, \bibinfo{author}{{Carpinelli} M},
  \bibinfo{author}{{Carrillo} G}, \bibinfo{author}{{Carullo} G},
  \bibinfo{author}{{Carver} TL}, \bibinfo{author}{{Diaz} JC},
  \bibinfo{author}{{Casentini} C}, \bibinfo{author}{{Castaldi} G},
  \bibinfo{author}{{Caudill} S}, \bibinfo{author}{{Cavagli{\`a}} M},
  \bibinfo{author}{{Cavalier} F}, \bibinfo{author}{{Cavalieri} R},
  \bibinfo{author}{{Ceasar} M}, \bibinfo{author}{{Cella} G},
  \bibinfo{author}{{Cerd{\'a}-Dur{\'a}n} P}, \bibinfo{author}{{Cesarini} E},
  \bibinfo{author}{{Chaibi} W}, \bibinfo{author}{{Chakravarti} K},
  \bibinfo{author}{{Subrahmanya} SC}, \bibinfo{author}{{Champion} E},
  \bibinfo{author}{{Chan} CH}, \bibinfo{author}{{Chan} C},
  \bibinfo{author}{{Chan} CL}, \bibinfo{author}{{Chan} K},
  \bibinfo{author}{{Chan} M}, \bibinfo{author}{{Chandra} K},
  \bibinfo{author}{{Chanial} P}, \bibinfo{author}{{Chao} S},
  \bibinfo{author}{{Chapman-Bird} CEA}, \bibinfo{author}{{Charlton} P},
  \bibinfo{author}{{Chase} EA}, \bibinfo{author}{{Chassande-Mottin} E},
  \bibinfo{author}{{Chatterjee} C}, \bibinfo{author}{{Chatterjee} D},
  \bibinfo{author}{{Chatterjee} D}, \bibinfo{author}{{Chaturvedi} M},
  \bibinfo{author}{{Chaty} S}, \bibinfo{author}{{Chatziioannou} K},
  \bibinfo{author}{{Chen} C}, \bibinfo{author}{{Chen} HY},
  \bibinfo{author}{{Chen} J}, \bibinfo{author}{{Chen} K},
  \bibinfo{author}{{Chen} X}, \bibinfo{author}{{Chen} YB},
  \bibinfo{author}{{Chen} YR}, \bibinfo{author}{{Chen} Z},
  \bibinfo{author}{{Cheng} H}, \bibinfo{author}{{Cheong} CK},
  \bibinfo{author}{{Cheung} HY}, \bibinfo{author}{{Chia} HY},
  \bibinfo{author}{{Chiadini} F}, \bibinfo{author}{{Chiang} CY},
  \bibinfo{author}{{Chiarini} G}, \bibinfo{author}{{Chierici} R},
  \bibinfo{author}{{Chincarini} A}, \bibinfo{author}{{Chiofalo} ML},
  \bibinfo{author}{{Chiummo} A}, \bibinfo{author}{{Cho} G},
  \bibinfo{author}{{Cho} HS}, \bibinfo{author}{{Choudhary} RK},
  \bibinfo{author}{{Choudhary} S}, \bibinfo{author}{{Christensen} N},
  \bibinfo{author}{{Chu} H}, \bibinfo{author}{{Chu} Q}, \bibinfo{author}{{Chu}
  YK}, \bibinfo{author}{{Chua} S}, \bibinfo{author}{{Chung} KW},
  \bibinfo{author}{{Ciani} G}, \bibinfo{author}{{Ciecielag} P},
  \bibinfo{author}{{Cie{\'s}lar} M}, \bibinfo{author}{{Cifaldi} M},
  \bibinfo{author}{{Ciobanu} AA}, \bibinfo{author}{{Ciolfi} R},
  \bibinfo{author}{{Cipriano} F}, \bibinfo{author}{{Cirone} A},
  \bibinfo{author}{{Clara} F}, \bibinfo{author}{{Clark} EN},
  \bibinfo{author}{{Clark} JA}, \bibinfo{author}{{Clarke} L},
  \bibinfo{author}{{Clearwater} P}, \bibinfo{author}{{Clesse} S},
  \bibinfo{author}{{Cleva} F}, \bibinfo{author}{{Coccia} E},
  \bibinfo{author}{{Codazzo} E}, \bibinfo{author}{{Cohadon} PF},
  \bibinfo{author}{{Cohen} DE}, \bibinfo{author}{{Cohen} L},
  \bibinfo{author}{{Colleoni} M}, \bibinfo{author}{{Collette} CG},
  \bibinfo{author}{{Colombo} A}, \bibinfo{author}{{Colpi} M},
  \bibinfo{author}{{Compton} CM}, \bibinfo{author}{{Constancio} M},
  \bibinfo{author}{{Conti} L}, \bibinfo{author}{{Cooper} SJ},
  \bibinfo{author}{{Corban} P}, \bibinfo{author}{{Corbitt} TR},
  \bibinfo{author}{{Cordero-Carri{\'o}n} I}, \bibinfo{author}{{Corezzi} S},
  \bibinfo{author}{{Corley} KR}, \bibinfo{author}{{Cornish} N},
  \bibinfo{author}{{Corre} D}, \bibinfo{author}{{Corsi} A},
  \bibinfo{author}{{Cortese} S}, \bibinfo{author}{{Costa} CA},
  \bibinfo{author}{{Cotesta} R}, \bibinfo{author}{{Coughlin} MW},
  \bibinfo{author}{{Coulon} JP}, \bibinfo{author}{{Countryman} ST},
  \bibinfo{author}{{Cousins} B}, \bibinfo{author}{{Couvares} P},
  \bibinfo{author}{{Coward} DM}, \bibinfo{author}{{Cowart} MJ},
  \bibinfo{author}{{Coyne} DC}, \bibinfo{author}{{Coyne} R},
  \bibinfo{author}{{Creighton} JDE}, \bibinfo{author}{{Creighton} TD},
  \bibinfo{author}{{Criswell} AW}, \bibinfo{author}{{Croquette} M},
  \bibinfo{author}{{Crowder} SG}, \bibinfo{author}{{Cudell} JR},
  \bibinfo{author}{{Cullen} TJ}, \bibinfo{author}{{Cumming} A},
  \bibinfo{author}{{Cummings} R}, \bibinfo{author}{{Cunningham} L},
  \bibinfo{author}{{Cuoco} E}, \bibinfo{author}{{Cury{\l}o} M},
  \bibinfo{author}{{Dabadie} P}, \bibinfo{author}{{Canton} TD},
  \bibinfo{author}{{Dall'Osso} S}, \bibinfo{author}{{D{\'a}lya} G},
  \bibinfo{author}{{Dana} A}, \bibinfo{author}{{Daneshgaranbajastani} LM},
  \bibinfo{author}{{D'Angelo} B}, \bibinfo{author}{{Danila} B},
  \bibinfo{author}{{Danilishin} S}, \bibinfo{author}{{D'Antonio} S},
  \bibinfo{author}{{Danzmann} K}, \bibinfo{author}{{Darsow-Fromm} C},
  \bibinfo{author}{{Dasgupta} A}, \bibinfo{author}{{Datrier} LEH},
  \bibinfo{author}{{Datta} S}, \bibinfo{author}{{Dattilo} V},
  \bibinfo{author}{{Dave} I}, \bibinfo{author}{{Davier} M},
  \bibinfo{author}{{Davies} GS}, \bibinfo{author}{{Davis} D},
  \bibinfo{author}{{Davis} MC}, \bibinfo{author}{{Daw} EJ},
  \bibinfo{author}{{Dean} R}, \bibinfo{author}{{Debra} D},
  \bibinfo{author}{{Deenadayalan} M}, \bibinfo{author}{{Degallaix} J},
  \bibinfo{author}{{de Laurentis} M}, \bibinfo{author}{{Del{\'e}glise} S},
  \bibinfo{author}{{Del Favero} V}, \bibinfo{author}{{de Lillo} F},
  \bibinfo{author}{{de Lillo} N}, \bibinfo{author}{{Del Pozzo} W},
  \bibinfo{author}{{Demarchi} LM}, \bibinfo{author}{{de Matteis} F},
  \bibinfo{author}{{D'Emilio} V}, \bibinfo{author}{{Demos} N},
  \bibinfo{author}{{Dent} T}, \bibinfo{author}{{Depasse} A},
  \bibinfo{author}{{de Pietri} R}, \bibinfo{author}{{De Rosa} R},
  \bibinfo{author}{{de Rossi} C}, \bibinfo{author}{{Desalvo} R},
  \bibinfo{author}{{de Simone} R}, \bibinfo{author}{{Dhurandhar} S},
  \bibinfo{author}{{D{\'\i}az} MC}, \bibinfo{author}{{Diaz-Ortiz} M},
  \bibinfo{author}{{Didio} NA}, \bibinfo{author}{{Dietrich} T},
  \bibinfo{author}{{di Fiore} L}, \bibinfo{author}{{di Fronzo} C},
  \bibinfo{author}{{di Giorgio} C}, \bibinfo{author}{{di Giovanni} F},
  \bibinfo{author}{{di Giovanni} M}, \bibinfo{author}{{di Girolamo} T},
  \bibinfo{author}{{di Lieto} A}, \bibinfo{author}{{Ding} B},
  \bibinfo{author}{{di Pace} S}, \bibinfo{author}{{di Palma} I},
  \bibinfo{author}{{di Renzo} F}, \bibinfo{author}{{Divakarla} AK},
  \bibinfo{author}{{Dmitriev} A}, \bibinfo{author}{{Doctor} Z},
  \bibinfo{author}{{D'Onofrio} L}, \bibinfo{author}{{Donovan} F},
  \bibinfo{author}{{Dooley} KL}, \bibinfo{author}{{Doravari} S},
  \bibinfo{author}{{Dorrington} I}, \bibinfo{author}{{Drago} M},
  \bibinfo{author}{{Driggers} JC}, \bibinfo{author}{{Drori} Y},
  \bibinfo{author}{{Ducoin} JG}, \bibinfo{author}{{Dupej} P},
  \bibinfo{author}{{Durante} O}, \bibinfo{author}{{D'Urso} D},
  \bibinfo{author}{{Duverne} PA}, \bibinfo{author}{{Dwyer} SE},
  \bibinfo{author}{{Eassa} C}, \bibinfo{author}{{Easter} PJ},
  \bibinfo{author}{{Ebersold} M}, \bibinfo{author}{{Eckhardt} T},
  \bibinfo{author}{{Eddolls} G}, \bibinfo{author}{{Edelman} B},
  \bibinfo{author}{{Edo} TB}, \bibinfo{author}{{Edy} O},
  \bibinfo{author}{{Effler} A}, \bibinfo{author}{{Eguchi} S},
  \bibinfo{author}{{Eichholz} J}, \bibinfo{author}{{Eikenberry} SS},
  \bibinfo{author}{{Eisenmann} M}, \bibinfo{author}{{Eisenstein} RA},
  \bibinfo{author}{{Ejlli} A}, \bibinfo{author}{{Engelby} E},
  \bibinfo{author}{{Enomoto} Y}, \bibinfo{author}{{Errico} L},
  \bibinfo{author}{{Essick} RC}, \bibinfo{author}{{Estell{\'e}s} H},
  \bibinfo{author}{{Estevez} D}, \bibinfo{author}{{Etienne} Z},
  \bibinfo{author}{{Etzel} T}, \bibinfo{author}{{Evans} M},
  \bibinfo{author}{{Evans} TM}, \bibinfo{author}{{Ewing} BE},
  \bibinfo{author}{{Fafone} V}, \bibinfo{author}{{Fair} H},
  \bibinfo{author}{{Fairhurst} S}, \bibinfo{author}{{Farah} AM},
  \bibinfo{author}{{Farinon} S}, \bibinfo{author}{{Farr} B},
  \bibinfo{author}{{Farr} WM}, \bibinfo{author}{{Farrow} NW},
  \bibinfo{author}{{Fauchon-Jones} EJ}, \bibinfo{author}{{Favaro} G},
  \bibinfo{author}{{Favata} M}, \bibinfo{author}{{Fays} M},
  \bibinfo{author}{{Fazio} M}, \bibinfo{author}{{Feicht} J},
  \bibinfo{author}{{Fejer} MM}, \bibinfo{author}{{Fenyvesi} E},
  \bibinfo{author}{{Ferguson} DL}, \bibinfo{author}{{Fernandez-Galiana} A},
  \bibinfo{author}{{Ferrante} I}, \bibinfo{author}{{Ferreira} TA},
  \bibinfo{author}{{Fidecaro} F}, \bibinfo{author}{{Figura} P},
  \bibinfo{author}{{Fiori} I}, \bibinfo{author}{{Fishbach} M},
  \bibinfo{author}{{Fisher} RP}, \bibinfo{author}{{Fittipaldi} R},
  \bibinfo{author}{{Fiumara} V}, \bibinfo{author}{{Flaminio} R},
  \bibinfo{author}{{Floden} E}, \bibinfo{author}{{Fong} H},
  \bibinfo{author}{{Font} JA}, \bibinfo{author}{{Fornal} B},
  \bibinfo{author}{{Forsyth} PWF}, \bibinfo{author}{{Franke} A},
  \bibinfo{author}{{Frasca} S}, \bibinfo{author}{{Frasconi} F},
  \bibinfo{author}{{Frederick} C}, \bibinfo{author}{{Freed} JP},
  \bibinfo{author}{{Frei} Z}, \bibinfo{author}{{Freise} A},
  \bibinfo{author}{{Frey} R}, \bibinfo{author}{{Fritschel} P},
  \bibinfo{author}{{Frolov} VV}, \bibinfo{author}{{Fronz{\'e}} GG},
  \bibinfo{author}{{Fujii} Y}, \bibinfo{author}{{Fujikawa} Y},
  \bibinfo{author}{{Fukunaga} M}, \bibinfo{author}{{Fukushima} M},
  \bibinfo{author}{{Fulda} P}, \bibinfo{author}{{Fyffe} M},
  \bibinfo{author}{{Gabbard} HA}, \bibinfo{author}{{Gadre} BU},
  \bibinfo{author}{{Gair} JR}, \bibinfo{author}{{Gais} J},
  \bibinfo{author}{{Galaudage} S}, \bibinfo{author}{{Gamba} R},
  \bibinfo{author}{{Ganapathy} D}, \bibinfo{author}{{Ganguly} A},
  \bibinfo{author}{{Gao} D}, \bibinfo{author}{{Gaonkar} SG},
  \bibinfo{author}{{Garaventa} B}, \bibinfo{author}{{Garc{\'\i}a} F},
  \bibinfo{author}{{Garc{\'\i}a-N{\'u}{\~n}ez} C},
  \bibinfo{author}{{Garc{\'\i}a-Quir{\'o}s} C}, \bibinfo{author}{{Garufi} F},
  \bibinfo{author}{{Gateley} B}, \bibinfo{author}{{Gaudio} S},
  \bibinfo{author}{{Gayathri} V}, \bibinfo{author}{{Ge} GG},
  \bibinfo{author}{{Gemme} G}, \bibinfo{author}{{Gennai} A},
  \bibinfo{author}{{George} J}, \bibinfo{author}{{George} RN},
  \bibinfo{author}{{Gerberding} O}, \bibinfo{author}{{Gergely} L},
  \bibinfo{author}{{Gewecke} P}, \bibinfo{author}{{Ghonge} S},
  \bibinfo{author}{{Ghosh} A}, \bibinfo{author}{{Ghosh} A},
  \bibinfo{author}{{Ghosh} S}, \bibinfo{author}{{Ghosh} S},
  \bibinfo{author}{{Giacomazzo} B}, \bibinfo{author}{{Giacoppo} L},
  \bibinfo{author}{{Giaime} JA}, \bibinfo{author}{{Giardina} KD},
  \bibinfo{author}{{Gibson} DR}, \bibinfo{author}{{Gier} C},
  \bibinfo{author}{{Giesler} M}, \bibinfo{author}{{Giri} P},
  \bibinfo{author}{{Gissi} F}, \bibinfo{author}{{Glanzer} J},
  \bibinfo{author}{{Gleckl} AE}, \bibinfo{author}{{Godwin} P},
  \bibinfo{author}{{Golomb} J}, \bibinfo{author}{{Goetz} E},
  \bibinfo{author}{{Goetz} R}, \bibinfo{author}{{Gohlke} N},
  \bibinfo{author}{{Goncharov} B}, \bibinfo{author}{{Gonz{\'a}lez} G},
  \bibinfo{author}{{Gopakumar} A}, \bibinfo{author}{{Gosselin} M},
  \bibinfo{author}{{Gouaty} R}, \bibinfo{author}{{Gould} DW},
  \bibinfo{author}{{Grace} B}, \bibinfo{author}{{Grado} A},
  \bibinfo{author}{{Granata} M}, \bibinfo{author}{{Granata} V},
  \bibinfo{author}{{Grant} A}, \bibinfo{author}{{Gras} S},
  \bibinfo{author}{{Grassia} P}, \bibinfo{author}{{Gray} C},
  \bibinfo{author}{{Gray} R}, \bibinfo{author}{{Greco} G},
  \bibinfo{author}{{Green} AC}, \bibinfo{author}{{Green} R},
  \bibinfo{author}{{Gretarsson} AM}, \bibinfo{author}{{Gretarsson} EM},
  \bibinfo{author}{{Griffith} D}, \bibinfo{author}{{Griffiths} W},
  \bibinfo{author}{{Griggs} HL}, \bibinfo{author}{{Grignani} G},
  \bibinfo{author}{{Grimaldi} A}, \bibinfo{author}{{Grimm} SJ},
  \bibinfo{author}{{Grote} H}, \bibinfo{author}{{Grunewald} S},
  \bibinfo{author}{{Gruning} P}, \bibinfo{author}{{Guerra} D},
  \bibinfo{author}{{Guidi} GM}, \bibinfo{author}{{Guimaraes} AR},
  \bibinfo{author}{{Guix{\'e}} G}, \bibinfo{author}{{Gulati} HK},
  \bibinfo{author}{{Guo} HK}, \bibinfo{author}{{Guo} Y},
  \bibinfo{author}{{Gupta} A}, \bibinfo{author}{{Gupta} A},
  \bibinfo{author}{{Gupta} P}, \bibinfo{author}{{Gustafson} EK},
  \bibinfo{author}{{Gustafson} R}, \bibinfo{author}{{Guzman} F},
  \bibinfo{author}{{Ha} S}, \bibinfo{author}{{Haegel} L},
  \bibinfo{author}{{Hagiwara} A}, \bibinfo{author}{{Haino} S},
  \bibinfo{author}{{Halim} O}, \bibinfo{author}{{Hall} ED},
  \bibinfo{author}{{Hamilton} EZ}, \bibinfo{author}{{Hammond} G},
  \bibinfo{author}{{Han} WB}, \bibinfo{author}{{Haney} M},
  \bibinfo{author}{{Hanks} J}, \bibinfo{author}{{Hanna} C},
  \bibinfo{author}{{Hannam} MD}, \bibinfo{author}{{Hannuksela} O},
  \bibinfo{author}{{Hansen} H}, \bibinfo{author}{{Hansen} TJ},
  \bibinfo{author}{{Hanson} J}, \bibinfo{author}{{Harder} T},
  \bibinfo{author}{{Hardwick} T}, \bibinfo{author}{{Haris} K},
  \bibinfo{author}{{Harms} J}, \bibinfo{author}{{Harry} GM},
  \bibinfo{author}{{Harry} IW}, \bibinfo{author}{{Hartwig} D},
  \bibinfo{author}{{Hasegawa} K}, \bibinfo{author}{{Haskell} B},
  \bibinfo{author}{{Hasskew} RK}, \bibinfo{author}{{Haster} CJ},
  \bibinfo{author}{{Hattori} K}, \bibinfo{author}{{Haughian} K},
  \bibinfo{author}{{Hayakawa} H}, \bibinfo{author}{{Hayama} K},
  \bibinfo{author}{{Hayes} FJ}, \bibinfo{author}{{Healy} J},
  \bibinfo{author}{{Heidmann} A}, \bibinfo{author}{{Heidt} A},
  \bibinfo{author}{{Heintze} MC}, \bibinfo{author}{{Heinze} J},
  \bibinfo{author}{{Heinzel} J}, \bibinfo{author}{{Heitmann} H},
  \bibinfo{author}{{Hellman} F}, \bibinfo{author}{{Hello} P},
  \bibinfo{author}{{Helmling-Cornell} AF}, \bibinfo{author}{{Hemming} G},
  \bibinfo{author}{{Hendry} M}, \bibinfo{author}{{Heng} IS},
  \bibinfo{author}{{Hennes} E}, \bibinfo{author}{{Hennig} J},
  \bibinfo{author}{{Hennig} MH}, \bibinfo{author}{{Hernandez} AG},
  \bibinfo{author}{{Vivanco} FH}, \bibinfo{author}{{Heurs} M},
  \bibinfo{author}{{Hild} S}, \bibinfo{author}{{Hill} P},
  \bibinfo{author}{{Himemoto} Y}, \bibinfo{author}{{Hines} AS},
  \bibinfo{author}{{Hiranuma} Y}, \bibinfo{author}{{Hirata} N},
  \bibinfo{author}{{Hirose} E}, \bibinfo{author}{{Hochheim} S},
  \bibinfo{author}{{Hofman} D}, \bibinfo{author}{{Hohmann} JN},
  \bibinfo{author}{{Holcomb} DG}, \bibinfo{author}{{Holland} NA},
  \bibinfo{author}{{Hollows} IJ}, \bibinfo{author}{{Holmes} ZJ},
  \bibinfo{author}{{Holt} K}, \bibinfo{author}{{Holz} DE},
  \bibinfo{author}{{Hong} Z}, \bibinfo{author}{{Hopkins} P},
  \bibinfo{author}{{Hough} J}, \bibinfo{author}{{Hourihane} S},
%  \bibinfo{author}{{Howell} EJ}, \bib (\bibinfo{year}{2023}),
%  \bibinfo{month}{Jan.}
\bibinfo{title}{{Population of Merging Compact Binaries Inferred Using
  Gravitational Waves through GWTC-3}}.
\bibinfo{journal}{{\em Physical Review X}} \bibinfo{volume}{13}
  (\bibinfo{number}{1}), \bibinfo{eid}{011048}.
  \bibinfo{doi}{\doi{10.1103/PhysRevX.13.011048}}.
\eprint{2111.03634}.

\bibtype{Article}%
\bibitem[{Amati} et al.(2002)]{amati02}
\bibinfo{author}{{Amati} L}, \bibinfo{author}{{Frontera} F},
  \bibinfo{author}{{Tavani} M}, \bibinfo{author}{{in't Zand} JJM},
  \bibinfo{author}{{Antonelli} A}, \bibinfo{author}{{Costa} E},
  \bibinfo{author}{{Feroci} M}, \bibinfo{author}{{Guidorzi} C},
  \bibinfo{author}{{Heise} J}, \bibinfo{author}{{Masetti} N},
  \bibinfo{author}{{Montanari} E}, \bibinfo{author}{{Nicastro} L},
  \bibinfo{author}{{Palazzi} E}, \bibinfo{author}{{Pian} E},
  \bibinfo{author}{{Piro} L} and  \bibinfo{author}{{Soffitta} P}
  (\bibinfo{year}{2002}), \bibinfo{month}{Jul.}
\bibinfo{title}{{Intrinsic spectra and energetics of BeppoSAX Gamma-Ray Bursts
  with known redshifts}}.
\bibinfo{journal}{{\em A\&A}} \bibinfo{volume}{390}: \bibinfo{pages}{81--89}.
  \bibinfo{doi}{\doi{10.1051/0004-6361:20020722}}.
\eprint{astro-ph/0205230}.

\bibtype{Article}%
\bibitem[{Amati} et al.(2021)]{theseus}
\bibinfo{author}{{Amati} L}, \bibinfo{author}{{O'Brien} PT},
  \bibinfo{author}{{G{\"o}tz} D}, \bibinfo{author}{{Bozzo} E},
  \bibinfo{author}{{Santangelo} A}, \bibinfo{author}{{Tanvir} N},
  \bibinfo{author}{{Frontera} F}, \bibinfo{author}{{Mereghetti} S},
  \bibinfo{author}{{Osborne} JP}, \bibinfo{author}{{Blain} A},
  \bibinfo{author}{{Basa} S}, \bibinfo{author}{{Branchesi} M},
  \bibinfo{author}{{Burderi} L}, \bibinfo{author}{{Caballero-Garc{\'\i}a} M},
  \bibinfo{author}{{Castro-Tirado} AJ}, \bibinfo{author}{{Christensen} L},
  \bibinfo{author}{{Ciolfi} R}, \bibinfo{author}{{De Rosa} A},
  \bibinfo{author}{{Doroshenko} V}, \bibinfo{author}{{Ferrara} A},
  \bibinfo{author}{{Ghirlanda} G}, \bibinfo{author}{{Hanlon} L},
  \bibinfo{author}{{Heddermann} P}, \bibinfo{author}{{Hutchinson} I},
  \bibinfo{author}{{Labanti} C}, \bibinfo{author}{{Le Floch} E},
  \bibinfo{author}{{Lerman} H}, \bibinfo{author}{{Paltani} S},
  \bibinfo{author}{{Reglero} V}, \bibinfo{author}{{Rezzolla} L},
  \bibinfo{author}{{Rosati} P}, \bibinfo{author}{{Salvaterra} R},
  \bibinfo{author}{{Stratta} G}, \bibinfo{author}{{Tenzer} C} and
  \bibinfo{author}{{Theseus Consortium}} (\bibinfo{year}{2021}),
  \bibinfo{month}{Dec.}
\bibinfo{title}{{The THESEUS space mission: science goals, requirements and
  mission concept}}.
\bibinfo{journal}{{\em Experimental Astronomy}} \bibinfo{volume}{52}
  (\bibinfo{number}{3}): \bibinfo{pages}{183--218}.
  \bibinfo{doi}{\doi{10.1007/s10686-021-09807-8}}.
\eprint{2104.09531}.

\bibtype{Article}%
\bibitem[{Anderson}(2019)]{anderson19}
\bibinfo{author}{{Anderson} JP} (\bibinfo{year}{2019}), \bibinfo{month}{Aug.}
\bibinfo{title}{{A meta-analysis of core-collapse supernova $^{56}$Ni masses}}.
\bibinfo{journal}{{A\&A}} \bibinfo{volume}{628}, \bibinfo{eid}{A7}.
  \bibinfo{doi}{\doi{10.1051/0004-6361/201935027}}.
\eprint{1906.00761}.

\bibtype{Article}%
\bibitem[{Andreoni} et al.(2020)]{andreoni20}
\bibinfo{author}{{Andreoni} I}, \bibinfo{author}{{Kool} EC},
  \bibinfo{author}{{Sagu{\'e}s Carracedo} A}, \bibinfo{author}{{Kasliwal} MM},
  \bibinfo{author}{{Bulla} M}, \bibinfo{author}{{Ahumada} T},
  \bibinfo{author}{{Coughlin} MW}, \bibinfo{author}{{Anand} S},
  \bibinfo{author}{{Sollerman} J}, \bibinfo{author}{{Goobar} A},
  \bibinfo{author}{{Kaplan} DL}, \bibinfo{author}{{Loveridge} TT},
  \bibinfo{author}{{Karambelkar} V}, \bibinfo{author}{{Cooke} J},
  \bibinfo{author}{{Bagdasaryan} A}, \bibinfo{author}{{Bellm} EC},
  \bibinfo{author}{{Cenko} SB}, \bibinfo{author}{{Cook} DO},
  \bibinfo{author}{{De} K}, \bibinfo{author}{{Dekany} R},
  \bibinfo{author}{{Delacroix} A}, \bibinfo{author}{{Drake} A},
  \bibinfo{author}{{Duev} DA}, \bibinfo{author}{{Fremling} C},
  \bibinfo{author}{{Golkhou} VZ}, \bibinfo{author}{{Graham} MJ},
  \bibinfo{author}{{Hale} D}, \bibinfo{author}{{Kulkarni} SR},
  \bibinfo{author}{{Kupfer} T}, \bibinfo{author}{{Laher} RR},
  \bibinfo{author}{{Mahabal} AA}, \bibinfo{author}{{Masci} FJ},
  \bibinfo{author}{{Rusholme} B}, \bibinfo{author}{{Smith} RM},
  \bibinfo{author}{{Tzanidakis} A}, \bibinfo{author}{{Van Sistine} A} and
  \bibinfo{author}{{Yao} Y} (\bibinfo{year}{2020}), \bibinfo{month}{Dec.}
\bibinfo{title}{{Constraining the Kilonova Rate with Zwicky Transient Facility
  Searches Independent of Gravitational Wave and Short Gamma-Ray Burst
  Triggers}}.
\bibinfo{journal}{{\em ApJ}} \bibinfo{volume}{904} (\bibinfo{number}{2}),
  \bibinfo{eid}{155}. \bibinfo{doi}{\doi{10.3847/1538-4357/abbf4c}}.
\eprint{2008.00008}.

\bibtype{Article}%
\bibitem[{Arcavi} et al.(2017)]{Arcavi+17}
\bibinfo{author}{{Arcavi} I}, \bibinfo{author}{{Hosseinzadeh} G},
  \bibinfo{author}{{Howell} DA}, \bibinfo{author}{{McCully} C},
  \bibinfo{author}{{Poznanski} D}, \bibinfo{author}{{Kasen} D},
  \bibinfo{author}{{Barnes} J}, \bibinfo{author}{{Zaltzman} M},
  \bibinfo{author}{{Vasylyev} S}, \bibinfo{author}{{Maoz} D} and
  \bibinfo{author}{{Valenti} S} (\bibinfo{year}{2017}), \bibinfo{month}{Nov.}
\bibinfo{title}{{Optical emission from a kilonova following a
  gravitational-wave-detected neutron-star merger}}.
\bibinfo{journal}{{\em Nature}} \bibinfo{volume}{551} (\bibinfo{number}{7678}):
  \bibinfo{pages}{64--66}. \bibinfo{doi}{\doi{10.1038/nature24291}}.
\eprint{1710.05843}.

\bibtype{Article}%
\bibitem[{Band} et al.(1993)]{band93}
\bibinfo{author}{{Band} D}, \bibinfo{author}{{Matteson} J},
  \bibinfo{author}{{Ford} L}, \bibinfo{author}{{Schaefer} B},
  \bibinfo{author}{{Palmer} D}, \bibinfo{author}{{Teegarden} B},
  \bibinfo{author}{{Cline} T}, \bibinfo{author}{{Briggs} M},
  \bibinfo{author}{{Paciesas} W}, \bibinfo{author}{{Pendleton} G},
  \bibinfo{author}{{Fishman} G}, \bibinfo{author}{{Kouveliotou} C},
  \bibinfo{author}{{Meegan} C}, \bibinfo{author}{{Wilson} R} and
  \bibinfo{author}{{Lestrade} P} (\bibinfo{year}{1993}), \bibinfo{month}{Aug.}
\bibinfo{title}{{BATSE Observations of Gamma-Ray Burst Spectra. I. Spectral
  Diversity}}.
\bibinfo{journal}{{\em ApJ}} \bibinfo{volume}{413}: \bibinfo{pages}{281}.
  \bibinfo{doi}{\doi{10.1086/172995}}.

\bibtype{Article}%
\bibitem[{Barnes} and {Kasen}(2013)]{barneskasen13}
\bibinfo{author}{{Barnes} J} and  \bibinfo{author}{{Kasen} D}
  (\bibinfo{year}{2013}), \bibinfo{month}{Sep.}
\bibinfo{title}{{Effect of a High Opacity on the Light Curves of Radioactively
  Powered Transients from Compact Object Mergers}}.
\bibinfo{journal}{{\em ApJ}} \bibinfo{volume}{775}, \bibinfo{eid}{18}.
  \bibinfo{doi}{\doi{10.1088/0004-637X/775/1/18}}.
\eprint{1303.5787}.

\bibtype{Article}%
\bibitem[{Beniamini} et al.(2019{\natexlab{a}})]{ben94}
\bibinfo{author}{{Beniamini} P}, \bibinfo{author}{{Hotokezaka} K},
  \bibinfo{author}{{van der Horst} A} and  \bibinfo{author}{{Kouveliotou} C}
  (\bibinfo{year}{2019}{\natexlab{a}}), \bibinfo{month}{Jul.}
\bibinfo{title}{{Formation rates and evolution histories of magnetars}}.
\bibinfo{journal}{{\em MNRAS}} \bibinfo{volume}{487} (\bibinfo{number}{1}):
  \bibinfo{pages}{1426--1438}. \bibinfo{doi}{\doi{10.1093/mnras/stz1391}}.
\eprint{1903.06718}.

\bibtype{Article}%
\bibitem[{Beniamini} et al.(2019{\natexlab{b}})]{beniamini}
\bibinfo{author}{{Beniamini} P}, \bibinfo{author}{{Hotokezaka} K},
  \bibinfo{author}{{van der Horst} A} and  \bibinfo{author}{{Kouveliotou} C}
  (\bibinfo{year}{2019}{\natexlab{b}}), \bibinfo{month}{Jul.}
\bibinfo{title}{{Formation rates and evolution histories of magnetars}}.
\bibinfo{journal}{{\em MNRAS}} \bibinfo{volume}{487} (\bibinfo{number}{1}):
  \bibinfo{pages}{1426--1438}. \bibinfo{doi}{\doi{10.1093/mnras/stz1391}}.
\eprint{1903.06718}.

\bibtype{Article}%
\bibitem[{Berger} et al.(2013)]{berger13}
\bibinfo{author}{{Berger} E}, \bibinfo{author}{{Fong} W} and
  \bibinfo{author}{{Chornock} R} (\bibinfo{year}{2013}), \bibinfo{month}{Sep.}
\bibinfo{title}{{An r-process Kilonova Associated with the Short-hard GRB
  130603B}}.
\bibinfo{journal}{{\em ApJl}} \bibinfo{volume}{774}, \bibinfo{eid}{L23}.
  \bibinfo{doi}{\doi{10.1088/2041-8205/774/2/L23}}.
\eprint{1306.3960}.

\bibtype{Article}%
\bibitem[{Birkl} et al.(2007)]{birkl07}
\bibinfo{author}{{Birkl} R}, \bibinfo{author}{{Aloy} MA},
  \bibinfo{author}{{Janka} HT} and  \bibinfo{author}{{M{\"u}ller} E}
  (\bibinfo{year}{2007}), \bibinfo{month}{Feb.}
\bibinfo{title}{{Neutrino pair annihilation near accreting, stellar-mass black
  holes}}.
\bibinfo{journal}{{\em A\&A}} \bibinfo{volume}{463} (\bibinfo{number}{1}):
  \bibinfo{pages}{51--67}. \bibinfo{doi}{\doi{10.1051/0004-6361:20066293}}.
\eprint{astro-ph/0608543}.

\bibtype{Article}%
\bibitem[{Bloom} et al.(2002)]{bloom02}
\bibinfo{author}{{Bloom} JS}, \bibinfo{author}{{Kulkarni} SR} and
  \bibinfo{author}{{Djorgovski} SG} (\bibinfo{year}{2002}),
  \bibinfo{month}{Mar.}
\bibinfo{title}{{The Observed Offset Distribution of Gamma-Ray Bursts from
  Their Host Galaxies: A Robust Clue to the Nature of the Progenitors}}.
\bibinfo{journal}{{\em \aj}} \bibinfo{volume}{123} (\bibinfo{number}{3}):
  \bibinfo{pages}{1111--1148}. \bibinfo{doi}{\doi{10.1086/338893}}.
\eprint{astro-ph/0010176}.

\bibtype{Article}%
\bibitem[{Bloom} et al.(2006)]{bloom06}
\bibinfo{author}{{Bloom} JS}, \bibinfo{author}{{Prochaska} JX},
  \bibinfo{author}{{Pooley} D}, \bibinfo{author}{{Blake} CH},
  \bibinfo{author}{{Foley} RJ}, \bibinfo{author}{{Jha} S},
  \bibinfo{author}{{Ramirez-Ruiz} E}, \bibinfo{author}{{Granot} J},
  \bibinfo{author}{{Filippenko} AV}, \bibinfo{author}{{Sigurdsson} S},
  \bibinfo{author}{{Barth} AJ}, \bibinfo{author}{{Chen} HW},
  \bibinfo{author}{{Cooper} MC}, \bibinfo{author}{{Falco} EE},
  \bibinfo{author}{{Gal} RR}, \bibinfo{author}{{Gerke} BF},
  \bibinfo{author}{{Gladders} MD}, \bibinfo{author}{{Greene} JE},
  \bibinfo{author}{{Hennanwi} J}, \bibinfo{author}{{Ho} LC},
  \bibinfo{author}{{Hurley} K}, \bibinfo{author}{{Koester} BP},
  \bibinfo{author}{{Li} W}, \bibinfo{author}{{Lubin} L},
  \bibinfo{author}{{Newman} J}, \bibinfo{author}{{Perley} DA},
  \bibinfo{author}{{Squires} GK} and  \bibinfo{author}{{Wood-Vasey} WM}
  (\bibinfo{year}{2006}), \bibinfo{month}{Feb.}
\bibinfo{title}{{Closing in on a Short-Hard Burst Progenitor: Constraints from
  Early-Time Optical Imaging and Spectroscopy of a Possible Host Galaxy of GRB
  050509b}}.
\bibinfo{journal}{{\em ApJ}} \bibinfo{volume}{638}: \bibinfo{pages}{354--368}.
  \bibinfo{doi}{\doi{10.1086/498107}}.
\eprint{astro-ph/0505480}.

\bibtype{Article}%
\bibitem[{Bloom} et al.(2009)]{bloom09}
\bibinfo{author}{{Bloom} JS}, \bibinfo{author}{{Perley} DA},
  \bibinfo{author}{{Li} W}, \bibinfo{author}{{Butler} NR},
  \bibinfo{author}{{Miller} AA}, \bibinfo{author}{{Kocevski} D},
  \bibinfo{author}{{Kann} DA}, \bibinfo{author}{{Foley} RJ},
  \bibinfo{author}{{Chen} HW}, \bibinfo{author}{{Filippenko} AV},
  \bibinfo{author}{{Starr} DL}, \bibinfo{author}{{Macomber} B},
  \bibinfo{author}{{Prochaska} JX}, \bibinfo{author}{{Chornock} R},
  \bibinfo{author}{{Poznanski} D}, \bibinfo{author}{{Klose} S},
  \bibinfo{author}{{Skrutskie} MF}, \bibinfo{author}{{Lopez} S},
  \bibinfo{author}{{Hall} P}, \bibinfo{author}{{Glazebrook} K} and
  \bibinfo{author}{{Blake} CH} (\bibinfo{year}{2009}), \bibinfo{month}{Jan.}
\bibinfo{title}{{Observations of the Naked-Eye GRB 080319B: Implications of
  Nature's Brightest Explosion}}.
\bibinfo{journal}{{\em ApJ}} \bibinfo{volume}{691}: \bibinfo{pages}{723--737}.
  \bibinfo{doi}{\doi{10.1088/0004-637X/691/1/723}}.
\eprint{0803.3215}.

\bibtype{Article}%
\bibitem[{Bloom} et al.(2011)]{bloom11}
\bibinfo{author}{{Bloom} JS}, \bibinfo{author}{{Giannios} D},
  \bibinfo{author}{{Metzger} BD}, \bibinfo{author}{{Cenko} SB},
  \bibinfo{author}{{Perley} DA}, \bibinfo{author}{{Butler} NR},
  \bibinfo{author}{{Tanvir} NR}, \bibinfo{author}{{Levan} AJ},
  \bibinfo{author}{{O'Brien} PT}, \bibinfo{author}{{Strubbe} LE},
  \bibinfo{author}{{De Colle} F}, \bibinfo{author}{{Ramirez-Ruiz} E},
  \bibinfo{author}{{Lee} WH}, \bibinfo{author}{{Nayakshin} S},
  \bibinfo{author}{{Quataert} E}, \bibinfo{author}{{King} AR},
  \bibinfo{author}{{Cucchiara} A}, \bibinfo{author}{{Guillochon} J},
  \bibinfo{author}{{Bower} GC}, \bibinfo{author}{{Fruchter} AS},
  \bibinfo{author}{{Morgan} AN} and  \bibinfo{author}{{van der Horst} AJ}
  (\bibinfo{year}{2011}), \bibinfo{month}{Jul.}
\bibinfo{title}{{A Possible Relativistic Jetted Outburst from a Massive Black
  Hole Fed by a Tidally Disrupted Star}}.
\bibinfo{journal}{{\em Science}} \bibinfo{volume}{333}: \bibinfo{pages}{203}.
  \bibinfo{doi}{\doi{10.1126/science.1207150}}.
\eprint{1104.3257}.

\bibtype{Article}%
\bibitem[{Burns} et al.(2023)]{burns23}
\bibinfo{author}{{Burns} E}, \bibinfo{author}{{Svinkin} D},
  \bibinfo{author}{{Fenimore} E}, \bibinfo{author}{{Kann} DA},
  \bibinfo{author}{{Ag{\"u}{\'\i} Fern{\'a}ndez} JF},
  \bibinfo{author}{{Frederiks} D}, \bibinfo{author}{{Hamburg} R},
  \bibinfo{author}{{Lesage} S}, \bibinfo{author}{{Temiraev} Y},
  \bibinfo{author}{{Tsvetkova} A}, \bibinfo{author}{{Bissaldi} E},
  \bibinfo{author}{{Briggs} MS}, \bibinfo{author}{{Dalessi} S},
  \bibinfo{author}{{Dunwoody} R}, \bibinfo{author}{{Fletcher} C},
  \bibinfo{author}{{Goldstein} A}, \bibinfo{author}{{Hui} CM},
  \bibinfo{author}{{Hristov} BA}, \bibinfo{author}{{Kocevski} D},
  \bibinfo{author}{{Lysenko} AL}, \bibinfo{author}{{Mailyan} B},
  \bibinfo{author}{{Mangan} J}, \bibinfo{author}{{McBreen} S},
  \bibinfo{author}{{Racusin} J}, \bibinfo{author}{{Ridnaia} A},
  \bibinfo{author}{{Roberts} OJ}, \bibinfo{author}{{Ulanov} M},
  \bibinfo{author}{{Veres} P}, \bibinfo{author}{{Wilson-Hodge} CA} and
  \bibinfo{author}{{Wood} J} (\bibinfo{year}{2023}), \bibinfo{month}{Mar.}
\bibinfo{title}{{GRB 221009A: The Boat}}.
\bibinfo{journal}{{\em ApJl}} \bibinfo{volume}{946} (\bibinfo{number}{1}),
  \bibinfo{eid}{L31}. \bibinfo{doi}{\doi{10.3847/2041-8213/acc39c}}.
\eprint{2302.14037}.

\bibtype{Article}%
\bibitem[{Cao} et al.(2024)]{lhasso23b}
\bibinfo{author}{{Cao} Z}, \bibinfo{author}{{Aharonian} F},
  \bibinfo{author}{{Axikegu} Bai YX}, \bibinfo{author}{{Bao} YW},
  \bibinfo{author}{{Bastieri} D}, \bibinfo{author}{{Bi} XJ},
  \bibinfo{author}{{Bi} YJ}, \bibinfo{author}{{Bian} W},
  \bibinfo{author}{{Bukevich} AV}, \bibinfo{author}{{Cao} Q},
  \bibinfo{author}{{Cao} WY}, \bibinfo{author}{{Cao} Z},
  \bibinfo{author}{{Chang} J}, \bibinfo{author}{{Chang} JF},
  \bibinfo{author}{{Chen} AM}, \bibinfo{author}{{Chen} ES},
  \bibinfo{author}{{Chen} HX}, \bibinfo{author}{{Chen} L},
  \bibinfo{author}{{Chen} L}, \bibinfo{author}{{Chen} L},
  \bibinfo{author}{{Chen} MJ}, \bibinfo{author}{{Chen} ML},
  \bibinfo{author}{{Chen} QH}, \bibinfo{author}{{Chen} S},
  \bibinfo{author}{{Chen} SH}, \bibinfo{author}{{Chen} SZ},
  \bibinfo{author}{{Chen} TL}, \bibinfo{author}{{Chen} Y},
  \bibinfo{author}{{Cheng} N}, \bibinfo{author}{{Cheng} YD},
  \bibinfo{author}{{Cui} MY}, \bibinfo{author}{{Cui} SW},
  \bibinfo{author}{{Cui} XH}, \bibinfo{author}{{Cui} YD},
  \bibinfo{author}{{Dai} BZ}, \bibinfo{author}{{Dai} HL},
  \bibinfo{author}{{Dai} ZG}, \bibinfo{author}{{Danzengluobu} Dong XQ},
  \bibinfo{author}{{Duan} KK}, \bibinfo{author}{{Fan} JH},
  \bibinfo{author}{{Fan} YZ}, \bibinfo{author}{{Fang} J},
  \bibinfo{author}{{Fang} JH}, \bibinfo{author}{{Fang} K},
  \bibinfo{author}{{Feng} CF}, \bibinfo{author}{{Feng} H},
  \bibinfo{author}{{Feng} L}, \bibinfo{author}{{Feng} SH},
  \bibinfo{author}{{Feng} XT}, \bibinfo{author}{{Feng} Y},
  \bibinfo{author}{{Feng} YL}, \bibinfo{author}{{Gabici} S},
  \bibinfo{author}{{Gao} B}, \bibinfo{author}{{Gao} CD}, \bibinfo{author}{{Gao}
  Q}, \bibinfo{author}{{Gao} W}, \bibinfo{author}{{Gao} WK},
  \bibinfo{author}{{Ge} MM}, \bibinfo{author}{{Geng} LS},
  \bibinfo{author}{{Giacinti} G}, \bibinfo{author}{{Gong} GH},
  \bibinfo{author}{{Gou} QB}, \bibinfo{author}{{Gu} MH}, \bibinfo{author}{{Guo}
  FL}, \bibinfo{author}{{Guo} XL}, \bibinfo{author}{{Guo} YQ},
  \bibinfo{author}{{Guo} YY}, \bibinfo{author}{{Han} YA},
  \bibinfo{author}{{Hasan} M}, \bibinfo{author}{{He} HH}, \bibinfo{author}{{He}
  HN}, \bibinfo{author}{{He} JY}, \bibinfo{author}{{He} Y},
  \bibinfo{author}{{Hor} YK}, \bibinfo{author}{{Hou} BW},
  \bibinfo{author}{{Hou} C}, \bibinfo{author}{{Hou} X}, \bibinfo{author}{{Hu}
  HB}, \bibinfo{author}{{Hu} Q}, \bibinfo{author}{{Hu} SC},
  \bibinfo{author}{{Huang} DH}, \bibinfo{author}{{Huang} TQ},
  \bibinfo{author}{{Huang} WJ}, \bibinfo{author}{{Huang} XT},
  \bibinfo{author}{{Huang} XY}, \bibinfo{author}{{Huang} Y},
  \bibinfo{author}{{Ji} XL}, \bibinfo{author}{{Jia} HY}, \bibinfo{author}{{Jia}
  K}, \bibinfo{author}{{Jiang} K}, \bibinfo{author}{{Jiang} XW},
  \bibinfo{author}{{Jiang} ZJ}, \bibinfo{author}{{Jin} M},
  \bibinfo{author}{{Kang} MM}, \bibinfo{author}{{Karpikov} I},
  \bibinfo{author}{{Kuleshov} D}, \bibinfo{author}{{Kurinov} K},
  \bibinfo{author}{{Li} BB}, \bibinfo{author}{{Li} CM}, \bibinfo{author}{{Li}
  C}, \bibinfo{author}{{Li} C}, \bibinfo{author}{{Li} D}, \bibinfo{author}{{Li}
  F}, \bibinfo{author}{{Li} HB}, \bibinfo{author}{{Li} HC},
  \bibinfo{author}{{Li} J}, \bibinfo{author}{{Li} J}, \bibinfo{author}{{Li} K},
  \bibinfo{author}{{Li} SD}, \bibinfo{author}{{Li} WL}, \bibinfo{author}{{Li}
  WL}, \bibinfo{author}{{Li} XR}, \bibinfo{author}{{Li} X},
  \bibinfo{author}{{Li} YZ}, \bibinfo{author}{{Li} Z}, \bibinfo{author}{{Li}
  Z}, \bibinfo{author}{{Liang} EW}, \bibinfo{author}{{Liang} YF},
  \bibinfo{author}{{Lin} SJ}, \bibinfo{author}{{Liu} B}, \bibinfo{author}{{Liu}
  C}, \bibinfo{author}{{Liu} D}, \bibinfo{author}{{Liu} DB},
  \bibinfo{author}{{Liu} H}, \bibinfo{author}{{Liu} HD}, \bibinfo{author}{{Liu}
  J}, \bibinfo{author}{{Liu} JL}, \bibinfo{author}{{Liu} MY},
  \bibinfo{author}{{Liu} RY}, \bibinfo{author}{{Liu} SM},
  \bibinfo{author}{{Liu} W}, \bibinfo{author}{{Liu} Y}, \bibinfo{author}{{Liu}
  YN}, \bibinfo{author}{{Luo} Q}, \bibinfo{author}{{Luo} Y},
  \bibinfo{author}{{Lv} HK}, \bibinfo{author}{{Ma} BQ}, \bibinfo{author}{{Ma}
  LL}, \bibinfo{author}{{Ma} XH}, \bibinfo{author}{{Mao} JR},
  \bibinfo{author}{{Min} Z}, \bibinfo{author}{{Mitthumsiri} W},
  \bibinfo{author}{{Mu} HJ}, \bibinfo{author}{{Nan} YC},
  \bibinfo{author}{{Neronov} A}, \bibinfo{author}{{Ou} LJ},
  \bibinfo{author}{{Pattarakijwanich} P}, \bibinfo{author}{{Pei} ZY},
  \bibinfo{author}{{Qi} JC}, \bibinfo{author}{{Qi} MY}, \bibinfo{author}{{Qiao}
  BQ}, \bibinfo{author}{{Qin} JJ}, \bibinfo{author}{{Raza} A},
  \bibinfo{author}{{Ruffolo} D}, \bibinfo{author}{{S{\'a}iz} A},
  \bibinfo{author}{{Saeed} M}, \bibinfo{author}{{Semikoz} D},
  \bibinfo{author}{{Shao} L}, \bibinfo{author}{{Shchegolev} O},
  \bibinfo{author}{{Sheng} XD}, \bibinfo{author}{{Shu} FW},
  \bibinfo{author}{{Song} HC}, \bibinfo{author}{{Stenkin} YV},
  \bibinfo{author}{{Stepanov} V}, \bibinfo{author}{{Su} Y},
  \bibinfo{author}{{Sun} DX}, \bibinfo{author}{{Sun} QN},
  \bibinfo{author}{{Sun} XN}, \bibinfo{author}{{Sun} ZB},
  \bibinfo{author}{{Takata} J}, \bibinfo{author}{{Tam} PHT},
  \bibinfo{author}{{Tang} QW}, \bibinfo{author}{{Tang} R},
  \bibinfo{author}{{Tang} ZB}, \bibinfo{author}{{Tian} WW},
  \bibinfo{author}{{Wang} C}, \bibinfo{author}{{Wang} CB},
  \bibinfo{author}{{Wang} GW}, \bibinfo{author}{{Wang} HG},
  \bibinfo{author}{{Wang} HH}, \bibinfo{author}{{Wang} JC},
  \bibinfo{author}{{Wang} K}, \bibinfo{author}{{Wang} K},
  \bibinfo{author}{{Wang} LP}, \bibinfo{author}{{Wang} LY},
  \bibinfo{author}{{Wang} PH}, \bibinfo{author}{{Wang} R},
  \bibinfo{author}{{Wang} W}, \bibinfo{author}{{Wang} XG},
  \bibinfo{author}{{Wang} XY}, \bibinfo{author}{{Wang} Y},
  \bibinfo{author}{{Wang} YD}, \bibinfo{author}{{Wang} YJ},
  \bibinfo{author}{{Wang} ZH}, \bibinfo{author}{{Wang} ZX},
  \bibinfo{author}{{Wang} Z}, \bibinfo{author}{{Wang} Z},
  \bibinfo{author}{{Wei} DM}, \bibinfo{author}{{Wei} JJ},
  \bibinfo{author}{{Wei} YJ}, \bibinfo{author}{{Wen} T}, \bibinfo{author}{{Wu}
  CY}, \bibinfo{author}{{Wu} HR}, \bibinfo{author}{{Wu} QW},
  \bibinfo{author}{{Wu} S}, \bibinfo{author}{{Wu} XF}, \bibinfo{author}{{Wu}
  YS}, \bibinfo{author}{{Xi} SQ}, \bibinfo{author}{{Xia} J},
  \bibinfo{author}{{Xiang} GM}, \bibinfo{author}{{Xiao} DX},
  \bibinfo{author}{{Xiao} G}, \bibinfo{author}{{Xin} YL},
  \bibinfo{author}{{Xing} Y}, \bibinfo{author}{{Xiong} DR},
  \bibinfo{author}{{Xiong} Z}, \bibinfo{author}{{Xu} DL}, \bibinfo{author}{{Xu}
  RF}, \bibinfo{author}{{Xu} RX}, \bibinfo{author}{{Xu} WL},
  \bibinfo{author}{{Xue} L}, \bibinfo{author}{{Yan} DH}, \bibinfo{author}{{Yan}
  JZ}, \bibinfo{author}{{Yan} T}, \bibinfo{author}{{Yang} CW},
  \bibinfo{author}{{Yang} CY}, \bibinfo{author}{{Yang} F},
  \bibinfo{author}{{Yang} FF}, \bibinfo{author}{{Yang} LL},
  \bibinfo{author}{{Yang} MJ}, \bibinfo{author}{{Yang} RZ},
  \bibinfo{author}{{Yang} WX}, \bibinfo{author}{{Yao} YH},
  \bibinfo{author}{{Yao} ZG}, \bibinfo{author}{{Yin} LQ},
  \bibinfo{author}{{Yin} N}, \bibinfo{author}{{You} XH}, \bibinfo{author}{{You}
  ZY}, \bibinfo{author}{{Yu} YH}, \bibinfo{author}{{Yuan} Q},
  \bibinfo{author}{{Yue} H}, \bibinfo{author}{{Zeng} HD},
  \bibinfo{author}{{Zeng} TX}, \bibinfo{author}{{Zeng} W},
  \bibinfo{author}{{Zha} M}, \bibinfo{author}{{Zhang} BB},
  \bibinfo{author}{{Zhang} F}, \bibinfo{author}{{Zhang} H},
  \bibinfo{author}{{Zhang} HM}, \bibinfo{author}{{Zhang} HY},
  \bibinfo{author}{{Zhang} JL}, \bibinfo{author}{{Zhang} L},
  \bibinfo{author}{{Zhang} PF}, \bibinfo{author}{{Zhang} PP},
  \bibinfo{author}{{Zhang} R}, \bibinfo{author}{{Zhang} SB},
  \bibinfo{author}{{Zhang} SR}, \bibinfo{author}{{Zhang} SS},
  \bibinfo{author}{{Zhang} X}, \bibinfo{author}{{Zhang} XP},
  \bibinfo{author}{{Zhang} YF}, \bibinfo{author}{{Zhang} Y},
  \bibinfo{author}{{Zhang} Y}, \bibinfo{author}{{Zhao} B},
  \bibinfo{author}{{Zhao} J}, \bibinfo{author}{{Zhao} L},
  \bibinfo{author}{{Zhao} LZ}, \bibinfo{author}{{Zhao} SP},
  \bibinfo{author}{{Zhao} XH}, \bibinfo{author}{{Zheng} F},
  \bibinfo{author}{{Zhong} WJ}, \bibinfo{author}{{Zhou} B},
  \bibinfo{author}{{Zhou} H}, \bibinfo{author}{{Zhou} JN},
  \bibinfo{author}{{Zhou} M}, \bibinfo{author}{{Zhou} P},
  \bibinfo{author}{{Zhou} R}, \bibinfo{author}{{Zhou} XX},
  \bibinfo{author}{{Zhou} XX}, \bibinfo{author}{{Zhu} BY},
  \bibinfo{author}{{Zhu} CG}, \bibinfo{author}{{Zhu} FR},
  \bibinfo{author}{{Zhu} H}, \bibinfo{author}{{Zhu} KJ}, \bibinfo{author}{{Zou}
  YC}, \bibinfo{author}{{Zuo} X} and  \bibinfo{author}{{Lhaaso Collaboration}}
  (\bibinfo{year}{2024}), \bibinfo{month}{Aug.}
\bibinfo{title}{{Stringent Tests of Lorentz Invariance Violation from LHAASO
  Observations of GRB 221009A}}.
\bibinfo{journal}{{\em \prl}} \bibinfo{volume}{133} (\bibinfo{number}{7}),
  \bibinfo{eid}{071501}. \bibinfo{doi}{\doi{10.1103/PhysRevLett.133.071501}}.
\eprint{2402.06009}.

\bibtype{Article}%
\bibitem[{Cavallo} and {Rees}(1978)]{cav78}
\bibinfo{author}{{Cavallo} G} and  \bibinfo{author}{{Rees} MJ}
  (\bibinfo{year}{1978}), \bibinfo{month}{May}.
\bibinfo{title}{{A qualitative study of cosmic fireballs and gamma -ray
  bursts.}}
\bibinfo{journal}{{\em MNRAS}} \bibinfo{volume}{183}:
  \bibinfo{pages}{359--365}. \bibinfo{doi}{\doi{10.1093/mnras/183.3.359}}.

\bibtype{Article}%
\bibitem[{Cenko} et al.(2013)]{cenko13}
\bibinfo{author}{{Cenko} SB}, \bibinfo{author}{{Kulkarni} SR},
  \bibinfo{author}{{Horesh} A}, \bibinfo{author}{{Corsi} A},
  \bibinfo{author}{{Fox} DB}, \bibinfo{author}{{Carpenter} J},
  \bibinfo{author}{{Frail} DA}, \bibinfo{author}{{Nugent} PE},
  \bibinfo{author}{{Perley} DA}, \bibinfo{author}{{Gruber} D},
  \bibinfo{author}{{Gal-Yam} A}, \bibinfo{author}{{Groot} PJ},
  \bibinfo{author}{{Hallinan} G}, \bibinfo{author}{{Ofek} EO},
  \bibinfo{author}{{Rau} A}, \bibinfo{author}{{MacLeod} CL},
  \bibinfo{author}{{Miller} AA}, \bibinfo{author}{{Bloom} JS},
  \bibinfo{author}{{Filippenko} AV}, \bibinfo{author}{{Kasliwal} MM},
  \bibinfo{author}{{Law} NM}, \bibinfo{author}{{Morgan} AN},
  \bibinfo{author}{{Polishook} D}, \bibinfo{author}{{Poznanski} D},
  \bibinfo{author}{{Quimby} RM}, \bibinfo{author}{{Sesar} B},
  \bibinfo{author}{{Shen} KJ}, \bibinfo{author}{{Silverman} JM} and
  \bibinfo{author}{{Sternberg} A} (\bibinfo{year}{2013}), \bibinfo{month}{Jun.}
\bibinfo{title}{{Discovery of a Cosmological, Relativistic Outburst via its
  Rapidly Fading Optical Emission}}.
\bibinfo{journal}{{\em ApJ}} \bibinfo{volume}{769} (\bibinfo{number}{2}),
  \bibinfo{eid}{130}. \bibinfo{doi}{\doi{10.1088/0004-637X/769/2/130}}.
\eprint{1304.4236}.

\bibtype{Article}%
\bibitem[{Chornock} et al.(2017)]{Chornock+17}
\bibinfo{author}{{Chornock} R}, \bibinfo{author}{{Berger} E},
  \bibinfo{author}{{Kasen} D}, \bibinfo{author}{{Cowperthwaite} PS},
  \bibinfo{author}{{Nicholl} M}, \bibinfo{author}{{Villar} VA},
  \bibinfo{author}{{Alexander} KD}, \bibinfo{author}{{Blanchard} PK},
  \bibinfo{author}{{Eftekhari} T}, \bibinfo{author}{{Fong} W},
  \bibinfo{author}{{Margutti} R}, \bibinfo{author}{{Williams} PKG},
  \bibinfo{author}{{Annis} J}, \bibinfo{author}{{Brout} D},
  \bibinfo{author}{{Brown} DA}, \bibinfo{author}{{Chen} HY},
  \bibinfo{author}{{Drout} MR}, \bibinfo{author}{{Farr} B},
  \bibinfo{author}{{Foley} RJ}, \bibinfo{author}{{Frieman} JA},
  \bibinfo{author}{{Fryer} CL}, \bibinfo{author}{{Herner} K},
  \bibinfo{author}{{Holz} DE}, \bibinfo{author}{{Kessler} R},
  \bibinfo{author}{{Matheson} T}, \bibinfo{author}{{Metzger} BD},
  \bibinfo{author}{{Quataert} E}, \bibinfo{author}{{Rest} A},
  \bibinfo{author}{{Sako} M}, \bibinfo{author}{{Scolnic} DM},
  \bibinfo{author}{{Smith} N} and  \bibinfo{author}{{Soares-Santos} M}
  (\bibinfo{year}{2017}), \bibinfo{month}{Oct.}
\bibinfo{title}{{The Electromagnetic Counterpart of the Binary Neutron Star
  Merger LIGO/Virgo GW170817. IV. Detection of Near-infrared Signatures of
  r-process Nucleosynthesis with Gemini-South}}.
\bibinfo{journal}{{\em ApJl}} \bibinfo{volume}{848}, \bibinfo{eid}{L19}.
  \bibinfo{doi}{\doi{10.3847/2041-8213/aa905c}}.
\eprint{1710.05454}.

\bibtype{Article}%
\bibitem[{Cline} et al.(1980)]{cline80}
\bibinfo{author}{{Cline} TL}, \bibinfo{author}{{Desai} UD},
  \bibinfo{author}{{Pizzichini} G}, \bibinfo{author}{{Teegarden} BJ},
  \bibinfo{author}{{Evans} WD}, \bibinfo{author}{{Klebesadel} RW},
  \bibinfo{author}{{Laros} JG}, \bibinfo{author}{{Hurley} K},
  \bibinfo{author}{{Niel} M} and  \bibinfo{author}{{Vedrenne} G}
  (\bibinfo{year}{1980}), \bibinfo{month}{Apr.}
\bibinfo{title}{{Detection of a fast, intense and unusual gamma-ray
  transient.}}
\bibinfo{journal}{{\em ApJl}} \bibinfo{volume}{237}: \bibinfo{pages}{L1--L5}.
  \bibinfo{doi}{\doi{10.1086/183221}}.

\bibtype{Article}%
\bibitem[{Coppin} et al.(2020)]{coppin20}
\bibinfo{author}{{Coppin} P}, \bibinfo{author}{{de Vries} KD} and
  \bibinfo{author}{{van Eijndhoven} N} (\bibinfo{year}{2020}),
  \bibinfo{month}{Nov.}
\bibinfo{title}{{Identification of gamma-ray burst precursors in Fermi-GBM
  bursts}}.
\bibinfo{journal}{{\em \prd}} \bibinfo{volume}{102} (\bibinfo{number}{10}),
  \bibinfo{eid}{103014}. \bibinfo{doi}{\doi{10.1103/PhysRevD.102.103014}}.
\eprint{2004.03246}.

\bibtype{Article}%
\bibitem[{Costa} et al.(1997)]{costa97}
\bibinfo{author}{{Costa} E}, \bibinfo{author}{{Frontera} F},
  \bibinfo{author}{{Heise} J}, \bibinfo{author}{{Feroci} M},
  \bibinfo{author}{{in't Zand} J}, \bibinfo{author}{{Fiore} F},
  \bibinfo{author}{{Cinti} MN}, \bibinfo{author}{{Dal Fiume} D},
  \bibinfo{author}{{Nicastro} L}, \bibinfo{author}{{Orlandini} M},
  \bibinfo{author}{{Palazzi} E}, \bibinfo{author}{{Rapisarda\#} M},
  \bibinfo{author}{{Zavattini} G}, \bibinfo{author}{{Jager} R},
  \bibinfo{author}{{Parmar} A}, \bibinfo{author}{{Owens} A},
  \bibinfo{author}{{Molendi} S}, \bibinfo{author}{{Cusumano} G},
  \bibinfo{author}{{Maccarone} MC}, \bibinfo{author}{{Giarrusso} S},
  \bibinfo{author}{{Coletta} A}, \bibinfo{author}{{Antonelli} LA},
  \bibinfo{author}{{Giommi} P}, \bibinfo{author}{{Muller} JM},
  \bibinfo{author}{{Piro} L} and  \bibinfo{author}{{Butler} RC}
  (\bibinfo{year}{1997}), \bibinfo{month}{Jun.}
\bibinfo{title}{{Discovery of an X-ray afterglow associated with the
  {$\gamma$}-ray burst of 28 February 1997}}.
\bibinfo{journal}{{\em \nat}} \bibinfo{volume}{387}: \bibinfo{pages}{783--785}.
  \bibinfo{doi}{\doi{10.1038/42885}}.
\eprint{astro-ph/9706065}.

\bibtype{Article}%
\bibitem[{Coulter} et al.(2017{\natexlab{a}})]{Coulter+17}
\bibinfo{author}{{Coulter} DA}, \bibinfo{author}{{Foley} RJ},
  \bibinfo{author}{{Kilpatrick} CD}, \bibinfo{author}{{Drout} MR},
  \bibinfo{author}{{Piro} AL}, \bibinfo{author}{{Shappee} BJ},
  \bibinfo{author}{{Siebert} MR}, \bibinfo{author}{{Simon} JD},
  \bibinfo{author}{{Ulloa} N}, \bibinfo{author}{{Kasen} D},
  \bibinfo{author}{{Madore} BF}, \bibinfo{author}{{Murguia-Berthier} A},
  \bibinfo{author}{{Pan} YC}, \bibinfo{author}{{Prochaska} JX},
  \bibinfo{author}{{Ramirez-Ruiz} E}, \bibinfo{author}{{Rest} A} and
  \bibinfo{author}{{Rojas-Bravo} C} (\bibinfo{year}{2017}{\natexlab{a}}),
  \bibinfo{month}{Dec.}
\bibinfo{title}{{Swope Supernova Survey 2017a (SSS17a), the optical counterpart
  to a gravitational wave source}}.
\bibinfo{journal}{{\em Science}} \bibinfo{volume}{358}
  (\bibinfo{number}{6370}): \bibinfo{pages}{1556--1558}.
  \bibinfo{doi}{\doi{10.1126/science.aap9811}}.
\eprint{1710.05452}.

\bibtype{Article}%
\bibitem[{Coulter} et al.(2017{\natexlab{b}})]{coulter17}
\bibinfo{author}{{Coulter} DA}, \bibinfo{author}{{Foley} RJ},
  \bibinfo{author}{{Kilpatrick} CD}, \bibinfo{author}{{Drout} MR},
  \bibinfo{author}{{Piro} AL}, \bibinfo{author}{{Shappee} BJ},
  \bibinfo{author}{{Siebert} MR}, \bibinfo{author}{{Simon} JD},
  \bibinfo{author}{{Ulloa} N}, \bibinfo{author}{{Kasen} D},
  \bibinfo{author}{{Madore} BF}, \bibinfo{author}{{Murguia-Berthier} A},
  \bibinfo{author}{{Pan} YC}, \bibinfo{author}{{Prochaska} JX},
  \bibinfo{author}{{Ramirez-Ruiz} E}, \bibinfo{author}{{Rest} A} and
  \bibinfo{author}{{Rojas-Bravo} C} (\bibinfo{year}{2017}{\natexlab{b}}),
  \bibinfo{month}{Oct.}
\bibinfo{title}{{Swope Supernova Survey 2017a (SSS17a), the Optical Counterpart
  to a Gravitational Wave Source}}.
\bibinfo{journal}{{\em ArXiv e-prints}} \eprint{1710.05452}.

\bibtype{Article}%
\bibitem[{Daigne} and {Mochkovitch}(1998)]{daigne}
\bibinfo{author}{{Daigne} F} and  \bibinfo{author}{{Mochkovitch} R}
  (\bibinfo{year}{1998}), \bibinfo{month}{May}.
\bibinfo{title}{{Gamma-ray bursts from internal shocks in a relativistic wind:
  temporal and spectral properties}}.
\bibinfo{journal}{{\em MNRAS}} \bibinfo{volume}{296} (\bibinfo{number}{2}):
  \bibinfo{pages}{275--286}.
  \bibinfo{doi}{\doi{10.1046/j.1365-8711.1998.01305.x}}.
\eprint{astro-ph/9801245}.

\bibtype{Article}%
\bibitem[{Dainotti} et al.(2010)]{dainotti10}
\bibinfo{author}{{Dainotti} MG}, \bibinfo{author}{{Willingale} R},
  \bibinfo{author}{{Capozziello} S}, \bibinfo{author}{{Fabrizio Cardone} V} and
   \bibinfo{author}{{Ostrowski} M} (\bibinfo{year}{2010}),
  \bibinfo{month}{Oct.}
\bibinfo{title}{{Discovery of a Tight Correlation for Gamma-ray Burst
  Afterglows with ``Canonical'' Light Curves}}.
\bibinfo{journal}{{\em ApJl}} \bibinfo{volume}{722} (\bibinfo{number}{2}):
  \bibinfo{pages}{L215--L219}.
  \bibinfo{doi}{\doi{10.1088/2041-8205/722/2/L215}}.
\eprint{1009.1663}.

\bibtype{Article}%
\bibitem[{Drout} et al.(2017)]{Drout+17}
\bibinfo{author}{{Drout} MR}, \bibinfo{author}{{Piro} AL},
  \bibinfo{author}{{Shappee} BJ}, \bibinfo{author}{{Kilpatrick} CD},
  \bibinfo{author}{{Simon} JD}, \bibinfo{author}{{Contreras} C},
  \bibinfo{author}{{Coulter} DA}, \bibinfo{author}{{Foley} RJ},
  \bibinfo{author}{{Siebert} MR}, \bibinfo{author}{{Morrell} N},
  \bibinfo{author}{{Boutsia} K}, \bibinfo{author}{{Di Mille} F},
  \bibinfo{author}{{Holoien} TWS}, \bibinfo{author}{{Kasen} D},
  \bibinfo{author}{{Kollmeier} JA}, \bibinfo{author}{{Madore} BF},
  \bibinfo{author}{{Monson} AJ}, \bibinfo{author}{{Murguia-Berthier} A},
  \bibinfo{author}{{Pan} YC}, \bibinfo{author}{{Prochaska} JX},
  \bibinfo{author}{{Ramirez-Ruiz} E}, \bibinfo{author}{{Rest} A},
  \bibinfo{author}{{Adams} C}, \bibinfo{author}{{Alatalo} K},
  \bibinfo{author}{{Ba{\~n}ados} E}, \bibinfo{author}{{Baughman} J},
  \bibinfo{author}{{Beers} TC}, \bibinfo{author}{{Bernstein} RA},
  \bibinfo{author}{{Bitsakis} T}, \bibinfo{author}{{Campillay} A},
  \bibinfo{author}{{Hansen} TT}, \bibinfo{author}{{Higgs} CR},
  \bibinfo{author}{{Ji} AP}, \bibinfo{author}{{Maravelias} G},
  \bibinfo{author}{{Marshall} JL}, \bibinfo{author}{{Moni Bidin} C},
  \bibinfo{author}{{Prieto} JL}, \bibinfo{author}{{Rasmussen} KC},
  \bibinfo{author}{{Rojas-Bravo} C}, \bibinfo{author}{{Strom} AL},
  \bibinfo{author}{{Ulloa} N}, \bibinfo{author}{{Vargas-Gonz{\'a}lez} J},
  \bibinfo{author}{{Wan} Z} and  \bibinfo{author}{{Whitten} DD}
  (\bibinfo{year}{2017}), \bibinfo{month}{Dec.}
\bibinfo{title}{{Light curves of the neutron star merger GW170817/SSS17a:
  Implications for r-process nucleosynthesis}}.
\bibinfo{journal}{{\em Science}} \bibinfo{volume}{358}
  (\bibinfo{number}{6370}): \bibinfo{pages}{1570--1574}.
  \bibinfo{doi}{\doi{10.1126/science.aaq0049}}.
\eprint{1710.05443}.

\bibtype{Article}%
\bibitem[{Evans} et al.(1980)]{evans80}
\bibinfo{author}{{Evans} WD}, \bibinfo{author}{{Klebesadel} RW},
  \bibinfo{author}{{Laros} JG}, \bibinfo{author}{{Cline} TL},
  \bibinfo{author}{{Desai} UD}, \bibinfo{author}{{Teegarden} BJ},
  \bibinfo{author}{{Pizzichini} G}, \bibinfo{author}{{Hurley} K},
  \bibinfo{author}{{Niel} M} and  \bibinfo{author}{{Vedrenne} G}
  (\bibinfo{year}{1980}), \bibinfo{month}{Apr.}
\bibinfo{title}{{Location of the gamma-ray transient event of 1979 March 5}}.
\bibinfo{journal}{{\em ApJl}} \bibinfo{volume}{237}: \bibinfo{pages}{L7--L9}.
  \bibinfo{doi}{\doi{10.1086/183222}}.

\bibtype{Article}%
\bibitem[{Fishman} et al.(1994)]{fishman94}
\bibinfo{author}{{Fishman} GJ}, \bibinfo{author}{{Meegan} CA},
  \bibinfo{author}{{Wilson} RB}, \bibinfo{author}{{Brock} MN},
  \bibinfo{author}{{Horack} JM}, \bibinfo{author}{{Kouveliotou} C},
  \bibinfo{author}{{Howard} S}, \bibinfo{author}{{Paciesas} WS},
  \bibinfo{author}{{Briggs} MS}, \bibinfo{author}{{Pendleton} GN},
  \bibinfo{author}{{Koshut} TM}, \bibinfo{author}{{Mallozzi} RS},
  \bibinfo{author}{{Stollberg} M} and  \bibinfo{author}{{Lestrade} JP}
  (\bibinfo{year}{1994}), \bibinfo{month}{May}.
\bibinfo{title}{{The First BATSE Gamma-Ray Burst Catalog}}.
\bibinfo{journal}{{\em ApJS}} \bibinfo{volume}{92}: \bibinfo{pages}{229}.
  \bibinfo{doi}{\doi{10.1086/191968}}.

\bibtype{Article}%
\bibitem[{Fong} and {Berger}(2013)]{FongBerger13}
\bibinfo{author}{{Fong} W} and  \bibinfo{author}{{Berger} E}
  (\bibinfo{year}{2013}), \bibinfo{month}{Oct.}
\bibinfo{title}{{The Locations of Short Gamma-Ray Bursts as Evidence for
  Compact Object Binary Progenitors}}.
\bibinfo{journal}{{\em ApJ}} \bibinfo{volume}{776} (\bibinfo{number}{1}),
  \bibinfo{eid}{18}. \bibinfo{doi}{\doi{10.1088/0004-637X/776/1/18}}.
\eprint{1307.0819}.

\bibtype{Article}%
\bibitem[{Fong} et al.(2015)]{fong+15}
\bibinfo{author}{{Fong} W}, \bibinfo{author}{{Berger} E},
  \bibinfo{author}{{Margutti} R} and  \bibinfo{author}{{Zauderer} BA}
  (\bibinfo{year}{2015}), \bibinfo{month}{Dec}.
\bibinfo{title}{{A Decade of Short-duration Gamma-Ray Burst Broadband
  Afterglows: Energetics, Circumburst Densities, and Jet Opening Angles}}.
\bibinfo{journal}{{\em ApJ}} \bibinfo{volume}{815} (\bibinfo{number}{2}),
  \bibinfo{eid}{102}. \bibinfo{doi}{\doi{10.1088/0004-637X/815/2/102}}.
\eprint{1509.02922}.

\bibtype{Article}%
\bibitem[{Frail} et al.(2001)]{frail01}
\bibinfo{author}{{Frail} DA}, \bibinfo{author}{{Kulkarni} SR},
  \bibinfo{author}{{Sari} R}, \bibinfo{author}{{Djorgovski} SG},
  \bibinfo{author}{{Bloom} JS}, \bibinfo{author}{{Galama} TJ},
  \bibinfo{author}{{Reichart} DE}, \bibinfo{author}{{Berger} E},
  \bibinfo{author}{{Harrison} FA}, \bibinfo{author}{{Price} PA},
  \bibinfo{author}{{Yost} SA}, \bibinfo{author}{{Diercks} A},
  \bibinfo{author}{{Goodrich} RW} and  \bibinfo{author}{{Chaffee} F}
  (\bibinfo{year}{2001}), \bibinfo{month}{Nov.}
\bibinfo{title}{{Beaming in Gamma-Ray Bursts: Evidence for a Standard Energy
  Reservoir}}.
\bibinfo{journal}{{\em ApJl}} \bibinfo{volume}{562} (\bibinfo{number}{1}):
  \bibinfo{pages}{L55--L58}. \bibinfo{doi}{\doi{10.1086/338119}}.
\eprint{astro-ph/0102282}.

\bibtype{Article}%
\bibitem[{Frohmaier} et al.(2021)]{frohmaier21}
\bibinfo{author}{{Frohmaier} C}, \bibinfo{author}{{Angus} CR},
  \bibinfo{author}{{Vincenzi} M}, \bibinfo{author}{{Sullivan} M},
  \bibinfo{author}{{Smith} M}, \bibinfo{author}{{Nugent} PE},
  \bibinfo{author}{{Cenko} SB}, \bibinfo{author}{{Gal-Yam} A},
  \bibinfo{author}{{Kulkarni} SR}, \bibinfo{author}{{Law} NM} and
  \bibinfo{author}{{Quimby} RM} (\bibinfo{year}{2021}), \bibinfo{month}{Jan.}
\bibinfo{title}{{From core collapse to superluminous: the rates of massive
  stellar explosions from the Palomar Transient Factory}}.
\bibinfo{journal}{{\em MNRAS}} \bibinfo{volume}{500} (\bibinfo{number}{4}):
  \bibinfo{pages}{5142--5158}. \bibinfo{doi}{\doi{10.1093/mnras/staa3607}}.
\eprint{2010.15270}.

\bibtype{Article}%
\bibitem[{Fruchter} et al.(2006)]{fruchter06}
\bibinfo{author}{{Fruchter} AS}, \bibinfo{author}{{Levan} AJ},
  \bibinfo{author}{{Strolger} L}, \bibinfo{author}{{Vreeswijk} PM},
  \bibinfo{author}{{Thorsett} SE}, \bibinfo{author}{{Bersier} D},
  \bibinfo{author}{{Burud} I}, \bibinfo{author}{{Castro Cer{\'o}n} JM},
  \bibinfo{author}{{Castro-Tirado} AJ}, \bibinfo{author}{{Conselice} C},
  \bibinfo{author}{{Dahlen} T}, \bibinfo{author}{{Ferguson} HC},
  \bibinfo{author}{{Fynbo} JPU}, \bibinfo{author}{{Garnavich} PM},
  \bibinfo{author}{{Gibbons} RA}, \bibinfo{author}{{Gorosabel} J},
  \bibinfo{author}{{Gull} TR}, \bibinfo{author}{{Hjorth} J},
  \bibinfo{author}{{Holland} ST}, \bibinfo{author}{{Kouveliotou} C},
  \bibinfo{author}{{Levay} Z}, \bibinfo{author}{{Livio} M},
  \bibinfo{author}{{Metzger} MR}, \bibinfo{author}{{Nugent} PE},
  \bibinfo{author}{{Petro} L}, \bibinfo{author}{{Pian} E},
  \bibinfo{author}{{Rhoads} JE}, \bibinfo{author}{{Riess} AG},
  \bibinfo{author}{{Sahu} KC}, \bibinfo{author}{{Smette} A},
  \bibinfo{author}{{Tanvir} NR}, \bibinfo{author}{{Wijers} RAMJ} and
  \bibinfo{author}{{Woosley} SE} (\bibinfo{year}{2006}), \bibinfo{month}{May}.
\bibinfo{title}{{Long {\ensuremath{\gamma}}-ray bursts and core-collapse
  supernovae have different environments}}.
\bibinfo{journal}{{\em \nat}} \bibinfo{volume}{441} (\bibinfo{number}{7092}):
  \bibinfo{pages}{463--468}. \bibinfo{doi}{\doi{10.1038/nature04787}}.
\eprint{astro-ph/0603537}.

\bibtype{Article}%
\bibitem[{Fryer} et al.(1999)]{fryer}
\bibinfo{author}{{Fryer} CL}, \bibinfo{author}{{Woosley} SE} and
  \bibinfo{author}{{Hartmann} DH} (\bibinfo{year}{1999}), \bibinfo{month}{Nov.}
\bibinfo{title}{{Formation Rates of Black Hole Accretion Disk Gamma-Ray
  Bursts}}.
\bibinfo{journal}{{\em ApJ}} \bibinfo{volume}{526} (\bibinfo{number}{1}):
  \bibinfo{pages}{152--177}. \bibinfo{doi}{\doi{10.1086/307992}}.
\eprint{astro-ph/9904122}.

\bibtype{Article}%
\bibitem[{Galama} et al.(1998)]{Galama+98}
\bibinfo{author}{{Galama} TJ}, \bibinfo{author}{{Vreeswijk} PM},
  \bibinfo{author}{{van Paradijs} J}, \bibinfo{author}{{Kouveliotou} C},
  \bibinfo{author}{{Augusteijn} T}, \bibinfo{author}{{B{\"o}hnhardt} H},
  \bibinfo{author}{{Brewer} JP}, \bibinfo{author}{{Doublier} V},
  \bibinfo{author}{{Gonzalez} JF}, \bibinfo{author}{{Leibundgut} B},
  \bibinfo{author}{{Lidman} C}, \bibinfo{author}{{Hainaut} OR},
  \bibinfo{author}{{Patat} F}, \bibinfo{author}{{Heise} J},
  \bibinfo{author}{{in't Zand} J}, \bibinfo{author}{{Hurley} K},
  \bibinfo{author}{{Groot} PJ}, \bibinfo{author}{{Strom} RG},
  \bibinfo{author}{{Mazzali} PA}, \bibinfo{author}{{Iwamoto} K},
  \bibinfo{author}{{Nomoto} K}, \bibinfo{author}{{Umeda} H},
  \bibinfo{author}{{Nakamura} T}, \bibinfo{author}{{Young} TR},
  \bibinfo{author}{{Suzuki} T}, \bibinfo{author}{{Shigeyama} T},
  \bibinfo{author}{{Koshut} T}, \bibinfo{author}{{Kippen} M},
  \bibinfo{author}{{Robinson} C}, \bibinfo{author}{{de Wildt} P},
  \bibinfo{author}{{Wijers} RAMJ}, \bibinfo{author}{{Tanvir} N},
  \bibinfo{author}{{Greiner} J}, \bibinfo{author}{{Pian} E},
  \bibinfo{author}{{Palazzi} E}, \bibinfo{author}{{Frontera} F},
  \bibinfo{author}{{Masetti} N}, \bibinfo{author}{{Nicastro} L},
  \bibinfo{author}{{Feroci} M}, \bibinfo{author}{{Costa} E},
  \bibinfo{author}{{Piro} L}, \bibinfo{author}{{Peterson} BA},
  \bibinfo{author}{{Tinney} C}, \bibinfo{author}{{Boyle} B},
  \bibinfo{author}{{Cannon} R}, \bibinfo{author}{{Stathakis} R},
  \bibinfo{author}{{Sadler} E}, \bibinfo{author}{{Begam} MC} and
  \bibinfo{author}{{Ianna} P} (\bibinfo{year}{1998}), \bibinfo{month}{Oct.}
\bibinfo{title}{{An unusual supernova in the error box of the
  {\ensuremath{\gamma}}-ray burst of 25 April 1998}}.
\bibinfo{journal}{{\em \nat}} \bibinfo{volume}{395} (\bibinfo{number}{6703}):
  \bibinfo{pages}{670--672}. \bibinfo{doi}{\doi{10.1038/27150}}.
\eprint{astro-ph/9806175}.

\bibtype{Article}%
\bibitem[{Gehrels} et al.(2005)]{gehrels05}
\bibinfo{author}{{Gehrels} N}, \bibinfo{author}{{Sarazin} CL},
  \bibinfo{author}{{O'Brien} PT}, \bibinfo{author}{{Zhang} B},
  \bibinfo{author}{{Barbier} L}, \bibinfo{author}{{Barthelmy} SD},
  \bibinfo{author}{{Blustin} A}, \bibinfo{author}{{Burrows} DN},
  \bibinfo{author}{{Cannizzo} J}, \bibinfo{author}{{Cummings} JR},
  \bibinfo{author}{{Goad} M}, \bibinfo{author}{{Holland} ST},
  \bibinfo{author}{{Hurkett} CP}, \bibinfo{author}{{Kennea} JA},
  \bibinfo{author}{{Levan} A}, \bibinfo{author}{{Markwardt} CB},
  \bibinfo{author}{{Mason} KO}, \bibinfo{author}{{Meszaros} P},
  \bibinfo{author}{{Page} M}, \bibinfo{author}{{Palmer} DM},
  \bibinfo{author}{{Rol} E}, \bibinfo{author}{{Sakamoto} T},
  \bibinfo{author}{{Willingale} R}, \bibinfo{author}{{Angelini} L},
  \bibinfo{author}{{Beardmore} A}, \bibinfo{author}{{Boyd} PT},
  \bibinfo{author}{{Breeveld} A}, \bibinfo{author}{{Campana} S},
  \bibinfo{author}{{Chester} MM}, \bibinfo{author}{{Chincarini} G},
  \bibinfo{author}{{Cominsky} LR}, \bibinfo{author}{{Cusumano} G},
  \bibinfo{author}{{de Pasquale} M}, \bibinfo{author}{{Fenimore} EE},
  \bibinfo{author}{{Giommi} P}, \bibinfo{author}{{Gronwall} C},
  \bibinfo{author}{{Grupe} D}, \bibinfo{author}{{Hill} JE},
  \bibinfo{author}{{Hinshaw} D}, \bibinfo{author}{{Hjorth} J},
  \bibinfo{author}{{Hullinger} D}, \bibinfo{author}{{Hurley} KC},
  \bibinfo{author}{{Klose} S}, \bibinfo{author}{{Kobayashi} S},
  \bibinfo{author}{{Kouveliotou} C}, \bibinfo{author}{{Krimm} HA},
  \bibinfo{author}{{Mangano} V}, \bibinfo{author}{{Marshall} FE},
  \bibinfo{author}{{McGowan} K}, \bibinfo{author}{{Moretti} A},
  \bibinfo{author}{{Mushotzky} RF}, \bibinfo{author}{{Nakazawa} K},
  \bibinfo{author}{{Norris} JP}, \bibinfo{author}{{Nousek} JA},
  \bibinfo{author}{{Osborne} JP}, \bibinfo{author}{{Page} K},
  \bibinfo{author}{{Parsons} AM}, \bibinfo{author}{{Patel} S},
  \bibinfo{author}{{Perri} M}, \bibinfo{author}{{Poole} T},
  \bibinfo{author}{{Romano} P}, \bibinfo{author}{{Roming} PWA},
  \bibinfo{author}{{Rosen} S}, \bibinfo{author}{{Sato} G},
  \bibinfo{author}{{Schady} P}, \bibinfo{author}{{Smale} AP},
  \bibinfo{author}{{Sollerman} J}, \bibinfo{author}{{Starling} R},
  \bibinfo{author}{{Still} M}, \bibinfo{author}{{Suzuki} M},
  \bibinfo{author}{{Tagliaferri} G}, \bibinfo{author}{{Takahashi} T},
  \bibinfo{author}{{Tashiro} M}, \bibinfo{author}{{Tueller} J},
  \bibinfo{author}{{Wells} AA}, \bibinfo{author}{{White} NE} and
  \bibinfo{author}{{Wijers} RAMJ} (\bibinfo{year}{2005}), \bibinfo{month}{Oct.}
\bibinfo{title}{{A short {$\gamma$}-ray burst apparently associated with an
  elliptical galaxy at redshift z = 0.225}}.
\bibinfo{journal}{{\em \nat}} \bibinfo{volume}{437}: \bibinfo{pages}{851--854}.
  \bibinfo{doi}{\doi{10.1038/nature04142}}.
\eprint{astro-ph/0505630}.

\bibtype{Article}%
\bibitem[{Ghirlanda} et al.(2004)]{ghirlanda04}
\bibinfo{author}{{Ghirlanda} G}, \bibinfo{author}{{Ghisellini} G} and
  \bibinfo{author}{{Lazzati} D} (\bibinfo{year}{2004}), \bibinfo{month}{Nov.}
\bibinfo{title}{{The Collimation-corrected Gamma-Ray Burst Energies Correlate
  with the Peak Energy of Their {\ensuremath{\nu}}F$_{{\ensuremath{\nu}}}$
  Spectrum}}.
\bibinfo{journal}{{\em ApJ}} \bibinfo{volume}{616} (\bibinfo{number}{1}):
  \bibinfo{pages}{331--338}. \bibinfo{doi}{\doi{10.1086/424913}}.
\eprint{astro-ph/0405602}.

\bibtype{Article}%
\bibitem[{Ghirlanda} et al.(2019)]{Ghirlanda+19}
\bibinfo{author}{{Ghirlanda} G}, \bibinfo{author}{{Salafia} OS},
  \bibinfo{author}{{Paragi} Z}, \bibinfo{author}{{Giroletti} M},
  \bibinfo{author}{{Yang} J}, \bibinfo{author}{{Marcote} B},
  \bibinfo{author}{{Blanchard} J}, \bibinfo{author}{{Agudo} I},
  \bibinfo{author}{{An} T}, \bibinfo{author}{{Bernardini} MG},
  \bibinfo{author}{{Beswick} R}, \bibinfo{author}{{Branchesi} M},
  \bibinfo{author}{{Campana} S}, \bibinfo{author}{{Casadio} C},
  \bibinfo{author}{{Chassande-Mottin} E}, \bibinfo{author}{{Colpi} M},
  \bibinfo{author}{{Covino} S}, \bibinfo{author}{{D'Avanzo} P},
  \bibinfo{author}{{D'Elia} V}, \bibinfo{author}{{Frey} S},
  \bibinfo{author}{{Gawronski} M}, \bibinfo{author}{{Ghisellini} G},
  \bibinfo{author}{{Gurvits} LI}, \bibinfo{author}{{Jonker} PG},
  \bibinfo{author}{{van Langevelde} HJ}, \bibinfo{author}{{Melandri} A},
  \bibinfo{author}{{Moldon} J}, \bibinfo{author}{{Nava} L},
  \bibinfo{author}{{Perego} A}, \bibinfo{author}{{Perez-Torres} MA},
  \bibinfo{author}{{Reynolds} C}, \bibinfo{author}{{Salvaterra} R},
  \bibinfo{author}{{Tagliaferri} G}, \bibinfo{author}{{Venturi} T},
  \bibinfo{author}{{Vergani} SD} and  \bibinfo{author}{{Zhang} M}
  (\bibinfo{year}{2019}), \bibinfo{month}{Mar.}
\bibinfo{title}{{Compact radio emission indicates a structured jet was produced
  by a binary neutron star merger}}.
\bibinfo{journal}{{\em Science}} \bibinfo{volume}{363}
  (\bibinfo{number}{6430}): \bibinfo{pages}{968--971}.
  \bibinfo{doi}{\doi{10.1126/science.aau8815}}.
\eprint{1808.00469}.

\bibtype{Article}%
\bibitem[Goldstein et al.(2017)]{Goldstein+2017}
\bibinfo{author}{Goldstein A}, \bibinfo{author}{Veres P},
  \bibinfo{author}{Burns E}, \bibinfo{author}{Briggs MS},
  \bibinfo{author}{Hamburg R}, \bibinfo{author}{Kocevski D},
  \bibinfo{author}{Wilson-Hodge CA}, \bibinfo{author}{Preece RD},
  \bibinfo{author}{Poolakkil S}, \bibinfo{author}{Roberts OJ} and
  \bibinfo{author}{et~al.} (\bibinfo{year}{2017}), \bibinfo{month}{Oct}.
\bibinfo{title}{An ordinary short gamma-ray burst with extraordinary
  implications: Fermi -gbm detection of grb 170817a}.
\bibinfo{journal}{{\em The Astrophysical Journal}} \bibinfo{volume}{848}
  (\bibinfo{number}{2}): \bibinfo{pages}{L14}.
ISSN \bibinfo{issn}{2041-8213}. \bibinfo{doi}{\doi{10.3847/2041-8213/aa8f41}}.
\bibinfo{url}{\url{http://dx.doi.org/10.3847/2041-8213/aa8f41}}.

\bibtype{Article}%
\bibitem[{Gompertz} et al.(2018)]{Gompertz+18}
\bibinfo{author}{{Gompertz} BP}, \bibinfo{author}{{Levan} AJ},
  \bibinfo{author}{{Tanvir} NR}, \bibinfo{author}{{Hjorth} J},
  \bibinfo{author}{{Covino} S}, \bibinfo{author}{{Evans} PA},
  \bibinfo{author}{{Fruchter} AS},
  \bibinfo{author}{{Gonz{\'a}lez-Fern{\'a}ndez} C}, \bibinfo{author}{{Jin} ZP},
  \bibinfo{author}{{Lyman} JD}, \bibinfo{author}{{Oates} SR},
  \bibinfo{author}{{O'Brien} PT} and  \bibinfo{author}{{Wiersema} K}
  (\bibinfo{year}{2018}), \bibinfo{month}{Jun.}
\bibinfo{title}{{The Diversity of Kilonova Emission in Short Gamma-Ray
  Bursts}}.
\bibinfo{journal}{{\em ApJ}} \bibinfo{volume}{860} (\bibinfo{number}{1}),
  \bibinfo{eid}{62}. \bibinfo{doi}{\doi{10.3847/1538-4357/aac206}}.
\eprint{1710.05442}.

\bibtype{Article}%
\bibitem[{Graham} and {Fruchter}(2013)]{graham13}
\bibinfo{author}{{Graham} JF} and  \bibinfo{author}{{Fruchter} AS}
  (\bibinfo{year}{2013}), \bibinfo{month}{Sep.}
\bibinfo{title}{{The Metal Aversion of Long-duration Gamma-Ray Bursts}}.
\bibinfo{journal}{{\em ApJ}} \bibinfo{volume}{774}, \bibinfo{eid}{119}.
  \bibinfo{doi}{\doi{10.1088/0004-637X/774/2/119}}.
\eprint{1211.7068}.

\bibtype{Article}%
\bibitem[Hallinan et al.(2017)]{Hallinan+2017}
\bibinfo{author}{Hallinan G}, \bibinfo{author}{Corsi A},
  \bibinfo{author}{Mooley KP}, \bibinfo{author}{Hotokezaka K},
  \bibinfo{author}{Nakar E}, \bibinfo{author}{Kasliwal MM},
  \bibinfo{author}{Kaplan DL}, \bibinfo{author}{Frail DA},
  \bibinfo{author}{Myers ST}, \bibinfo{author}{Murphy T} and
  \bibinfo{author}{et~al.} (\bibinfo{year}{2017}), \bibinfo{month}{Oct}.
\bibinfo{title}{A radio counterpart to a neutron star merger}.
\bibinfo{journal}{{\em Science}} \bibinfo{volume}{358}
  (\bibinfo{number}{6370}): \bibinfo{pages}{1579?1583}.
ISSN \bibinfo{issn}{1095-9203}. \bibinfo{doi}{\doi{10.1126/science.aap9855}}.
\bibinfo{url}{\url{http://dx.doi.org/10.1126/science.aap9855}}.

\bibtype{Article}%
\bibitem[{Hjorth} et al.(2003)]{hjorth03}
\bibinfo{author}{{Hjorth} J}, \bibinfo{author}{{Sollerman} J},
  \bibinfo{author}{{M{\o}ller} P}, \bibinfo{author}{{Fynbo} JPU},
  \bibinfo{author}{{Woosley} SE}, \bibinfo{author}{{Kouveliotou} C},
  \bibinfo{author}{{Tanvir} NR}, \bibinfo{author}{{Greiner} J},
  \bibinfo{author}{{Andersen} MI}, \bibinfo{author}{{Castro-Tirado} AJ},
  \bibinfo{author}{{Castro Cer{\'o}n} JM}, \bibinfo{author}{{Fruchter} AS},
  \bibinfo{author}{{Gorosabel} J}, \bibinfo{author}{{Jakobsson} P},
  \bibinfo{author}{{Kaper} L}, \bibinfo{author}{{Klose} S},
  \bibinfo{author}{{Masetti} N}, \bibinfo{author}{{Pedersen} H},
  \bibinfo{author}{{Pedersen} K}, \bibinfo{author}{{Pian} E},
  \bibinfo{author}{{Palazzi} E}, \bibinfo{author}{{Rhoads} JE},
  \bibinfo{author}{{Rol} E}, \bibinfo{author}{{van den Heuvel} EPJ},
  \bibinfo{author}{{Vreeswijk} PM}, \bibinfo{author}{{Watson} D} and
  \bibinfo{author}{{Wijers} RAMJ} (\bibinfo{year}{2003}), \bibinfo{month}{Jun.}
\bibinfo{title}{{A very energetic supernova associated with the {$\gamma$}-ray
  burst of 29 March 2003}}.
\bibinfo{journal}{{\em Nature}} \bibinfo{volume}{423}:
  \bibinfo{pages}{847--850}. \bibinfo{doi}{\doi{10.1038/nature01750}}.
\eprint{astro-ph/0306347}.

\bibtype{Article}%
\bibitem[{IceCube Collaboration} et al.(2018)]{icecube18}
\bibinfo{author}{{IceCube Collaboration}}, \bibinfo{author}{{Aartsen} MG},
  \bibinfo{author}{{Ackermann} M}, \bibinfo{author}{{Adams} J},
  \bibinfo{author}{{Aguilar} JA}, \bibinfo{author}{{Ahlers} M},
  \bibinfo{author}{{Ahrens} M}, \bibinfo{author}{{Al Samarai} I},
  \bibinfo{author}{{Altmann} D}, \bibinfo{author}{{Andeen} K},
  \bibinfo{author}{{Anderson} T}, \bibinfo{author}{{Ansseau} I},
  \bibinfo{author}{{Anton} G}, \bibinfo{author}{{Arg{\"u}elles} C},
  \bibinfo{author}{{Auffenberg} J}, \bibinfo{author}{{Axani} S},
  \bibinfo{author}{{Bagherpour} H}, \bibinfo{author}{{Bai} X},
  \bibinfo{author}{{Barron} JP}, \bibinfo{author}{{Barwick} SW},
  \bibinfo{author}{{Baum} V}, \bibinfo{author}{{Bay} R},
  \bibinfo{author}{{Beatty} JJ}, \bibinfo{author}{{Becker Tjus} J},
  \bibinfo{author}{{Becker} KH}, \bibinfo{author}{{BenZvi} S},
  \bibinfo{author}{{Berley} D}, \bibinfo{author}{{Bernardini} E},
  \bibinfo{author}{{Besson} DZ}, \bibinfo{author}{{Binder} G},
  \bibinfo{author}{{Bindig} D}, \bibinfo{author}{{Blaufuss} E},
  \bibinfo{author}{{Blot} S}, \bibinfo{author}{{Bohm} C},
  \bibinfo{author}{{B{\"o}rner} M}, \bibinfo{author}{{Bos} F},
  \bibinfo{author}{{B{\"o}ser} S}, \bibinfo{author}{{Botner} O},
  \bibinfo{author}{{Bourbeau} E}, \bibinfo{author}{{Bourbeau} J},
  \bibinfo{author}{{Bradascio} F}, \bibinfo{author}{{Braun} J},
  \bibinfo{author}{{Brenzke} M}, \bibinfo{author}{{Bretz} HP},
  \bibinfo{author}{{Bron} S}, \bibinfo{author}{{Brostean-Kaiser} J},
  \bibinfo{author}{{Burgman} A}, \bibinfo{author}{{Busse} RS},
  \bibinfo{author}{{Carver} T}, \bibinfo{author}{{Cheung} E},
  \bibinfo{author}{{Chirkin} D}, \bibinfo{author}{{Christov} A},
  \bibinfo{author}{{Clark} K}, \bibinfo{author}{{Classen} L},
  \bibinfo{author}{{Coenders} S}, \bibinfo{author}{{Collin} GH},
  \bibinfo{author}{{Conrad} JM}, \bibinfo{author}{{Coppin} P},
  \bibinfo{author}{{Correa} P}, \bibinfo{author}{{Cowen} DF},
  \bibinfo{author}{{Cross} R}, \bibinfo{author}{{Dave} P},
  \bibinfo{author}{{Day} M}, \bibinfo{author}{{de Andr{\'e}} JPAM},
  \bibinfo{author}{{De Clercq} C}, \bibinfo{author}{{DeLaunay} JJ},
  \bibinfo{author}{{Dembinski} H}, \bibinfo{author}{{De Ridder} S},
  \bibinfo{author}{{Desiati} P}, \bibinfo{author}{{de Vries} KD},
  \bibinfo{author}{{de Wasseige} G}, \bibinfo{author}{{de With} M},
  \bibinfo{author}{{DeYoung} T}, \bibinfo{author}{{D{\'\i}az-V{\'e}lez} JC},
  \bibinfo{author}{{di Lorenzo} V}, \bibinfo{author}{{Dujmovic} H},
  \bibinfo{author}{{Dumm} JP}, \bibinfo{author}{{Dunkman} M},
  \bibinfo{author}{{Dvorak} E}, \bibinfo{author}{{Eberhardt} B},
  \bibinfo{author}{{Ehrhardt} T}, \bibinfo{author}{{Eichmann} B},
  \bibinfo{author}{{Eller} P}, \bibinfo{author}{{Evenson} PA},
  \bibinfo{author}{{Fahey} S}, \bibinfo{author}{{Fazely} AR},
  \bibinfo{author}{{Felde} J}, \bibinfo{author}{{Filimonov} K},
  \bibinfo{author}{{Finley} C}, \bibinfo{author}{{Flis} S},
  \bibinfo{author}{{Franckowiak} A}, \bibinfo{author}{{Friedman} E},
  \bibinfo{author}{{Fritz} A}, \bibinfo{author}{{Gaisser} TK},
  \bibinfo{author}{{Gallagher} J}, \bibinfo{author}{{Gerhardt} L},
  \bibinfo{author}{{Ghorbani} K}, \bibinfo{author}{{Glauch} T},
  \bibinfo{author}{{Gl{\"u}senkamp} T}, \bibinfo{author}{{Goldschmidt} A},
  \bibinfo{author}{{Gonzalez} JG}, \bibinfo{author}{{Grant} D},
  \bibinfo{author}{{Griffith} Z}, \bibinfo{author}{{Haack} C},
  \bibinfo{author}{{Hallgren} A}, \bibinfo{author}{{Halzen} F},
  \bibinfo{author}{{Hanson} K}, \bibinfo{author}{{Hebecker} D},
  \bibinfo{author}{{Heereman} D}, \bibinfo{author}{{Helbing} K},
  \bibinfo{author}{{Hellauer} R}, \bibinfo{author}{{Hickford} S},
  \bibinfo{author}{{Hignight} J}, \bibinfo{author}{{Hill} GC},
  \bibinfo{author}{{Hoffman} KD}, \bibinfo{author}{{Hoffmann} R},
  \bibinfo{author}{{Hoinka} T}, \bibinfo{author}{{Hokanson-Fasig} B},
  \bibinfo{author}{{Hoshina} K}, \bibinfo{author}{{Huang} F},
  \bibinfo{author}{{Huber} M}, \bibinfo{author}{{Hultqvist} K},
  \bibinfo{author}{{H{\"u}nnefeld} M}, \bibinfo{author}{{Hussain} R},
  \bibinfo{author}{{In} S}, \bibinfo{author}{{Iovine} N},
  \bibinfo{author}{{Ishihara} A}, \bibinfo{author}{{Jacobi} E},
  \bibinfo{author}{{Japaridze} GS}, \bibinfo{author}{{Jeong} M},
  \bibinfo{author}{{Jero} K}, \bibinfo{author}{{Jones} BJP},
  \bibinfo{author}{{Kalaczynski} P}, \bibinfo{author}{{Kang} W},
  \bibinfo{author}{{Kappes} A}, \bibinfo{author}{{Kappesser} D},
  \bibinfo{author}{{Karg} T}, \bibinfo{author}{{Karle} A},
  \bibinfo{author}{{Katz} U}, \bibinfo{author}{{Kauer} M},
  \bibinfo{author}{{Keivani} A}, \bibinfo{author}{{Kelley} JL},
  \bibinfo{author}{{Kheirandish} A}, \bibinfo{author}{{Kim} J},
  \bibinfo{author}{{Kim} M}, \bibinfo{author}{{Kintscher} T},
  \bibinfo{author}{{Kiryluk} J}, \bibinfo{author}{{Kittler} T},
  \bibinfo{author}{{Klein} SR}, \bibinfo{author}{{Koirala} R},
  \bibinfo{author}{{Kolanoski} H}, \bibinfo{author}{{K{\"o}pke} L},
  \bibinfo{author}{{Kopper} C}, \bibinfo{author}{{Kopper} S},
  \bibinfo{author}{{Koschinsky} JP}, \bibinfo{author}{{Koskinen} DJ},
  \bibinfo{author}{{Kowalski} M}, \bibinfo{author}{{Krings} K},
  \bibinfo{author}{{Kroll} M}, \bibinfo{author}{{Kr{\"u}ckl} G},
  \bibinfo{author}{{Kunwar} S}, \bibinfo{author}{{Kurahashi} N},
  \bibinfo{author}{{Kuwabara} T}, \bibinfo{author}{{Kyriacou} A},
  \bibinfo{author}{{Labare} M}, \bibinfo{author}{{Lanfranchi} JL},
  \bibinfo{author}{{Larson} MJ}, \bibinfo{author}{{Lauber} F},
  \bibinfo{author}{{Leonard} K}, \bibinfo{author}{{Lesiak-Bzdak} M},
  \bibinfo{author}{{Leuermann} M}, \bibinfo{author}{{Liu} QR},
  \bibinfo{author}{{Lozano Mariscal} CJ}, \bibinfo{author}{{Lu} L},
  \bibinfo{author}{{L{\"u}nemann} J}, \bibinfo{author}{{Luszczak} W},
  \bibinfo{author}{{Madsen} J}, \bibinfo{author}{{Maggi} G},
  \bibinfo{author}{{Mahn} KBM}, \bibinfo{author}{{Mancina} S},
  \bibinfo{author}{{Maruyama} R}, \bibinfo{author}{{Mase} K},
  \bibinfo{author}{{Maunu} R}, \bibinfo{author}{{Meagher} K},
  \bibinfo{author}{{Medici} M}, \bibinfo{author}{{Meier} M},
  \bibinfo{author}{{Menne} T}, \bibinfo{author}{{Merino} G},
  \bibinfo{author}{{Meures} T}, \bibinfo{author}{{Miarecki} S},
  \bibinfo{author}{{Micallef} J}, \bibinfo{author}{{Moment{\'e}} G},
  \bibinfo{author}{{Montaruli} T}, \bibinfo{author}{{Moore} RW},
  \bibinfo{author}{{Morse} R}, \bibinfo{author}{{Moulai} M},
  \bibinfo{author}{{Nahnhauer} R}, \bibinfo{author}{{Nakarmi} P},
  \bibinfo{author}{{Naumann} U}, \bibinfo{author}{{Neer} G},
  \bibinfo{author}{{Niederhausen} H}, \bibinfo{author}{{Nowicki} SC},
  \bibinfo{author}{{Nygren} DR}, \bibinfo{author}{{Obertacke Pollmann} A},
  \bibinfo{author}{{Olivas} A}, \bibinfo{author}{{O'Murchadha} A},
  \bibinfo{author}{{O'Sullivan} E}, \bibinfo{author}{{Palczewski} T},
  \bibinfo{author}{{Pandya} H}, \bibinfo{author}{{Pankova} DV},
  \bibinfo{author}{{Peiffer} P}, \bibinfo{author}{{Pepper} JA},
  \bibinfo{author}{{P{\'e}rez de los Heros} C}, \bibinfo{author}{{Pieloth} D},
  \bibinfo{author}{{Pinat} E}, \bibinfo{author}{{Plum} M},
  \bibinfo{author}{{Price} PB}, \bibinfo{author}{{Przybylski} GT},
  \bibinfo{author}{{Raab} C}, \bibinfo{author}{{R{\"a}del} L},
  \bibinfo{author}{{Rameez} M}, \bibinfo{author}{{Rauch} L},
  \bibinfo{author}{{Rawlins} K}, \bibinfo{author}{{Rea} IC},
  \bibinfo{author}{{Reimann} R}, \bibinfo{author}{{Relethford} B},
  \bibinfo{author}{{Relich} M}, \bibinfo{author}{{Resconi} E},
  \bibinfo{author}{{Rhode} W}, \bibinfo{author}{{Richman} M},
  \bibinfo{author}{{Robertson} S}, \bibinfo{author}{{Rongen} M},
  \bibinfo{author}{{Rott} C}, \bibinfo{author}{{Ruhe} T},
  \bibinfo{author}{{Ryckbosch} D}, \bibinfo{author}{{Rysewyk} D},
  \bibinfo{author}{{Safa} I}, \bibinfo{author}{{S{\"a}lzer} T},
  \bibinfo{author}{{Sanchez Herrera} SE}, \bibinfo{author}{{Sandrock} A},
  \bibinfo{author}{{Sandroos} J}, \bibinfo{author}{{Santander} M},
  \bibinfo{author}{{Sarkar} S}, \bibinfo{author}{{Sarkar} S},
  \bibinfo{author}{{Satalecka} K}, \bibinfo{author}{{Schlunder} P},
  \bibinfo{author}{{Schmidt} T}, \bibinfo{author}{{Schneider} A},
  \bibinfo{author}{{Schoenen} S}, \bibinfo{author}{{Sch{\"o}neberg} S},
  \bibinfo{author}{{Schumacher} L}, \bibinfo{author}{{Sclafani} S},
  \bibinfo{author}{{Seckel} D}, \bibinfo{author}{{Seunarine} S},
  \bibinfo{author}{{Soedingrekso} J}, \bibinfo{author}{{Soldin} D},
  \bibinfo{author}{{Song} M}, \bibinfo{author}{{Spiczak} GM},
  \bibinfo{author}{{Spiering} C}, \bibinfo{author}{{Stachurska} J},
  \bibinfo{author}{{Stamatikos} M}, \bibinfo{author}{{Stanev} T},
  \bibinfo{author}{{Stasik} A}, \bibinfo{author}{{Stein} R},
  \bibinfo{author}{{Stettner} J}, \bibinfo{author}{{Steuer} A},
  \bibinfo{author}{{Stezelberger} T}, \bibinfo{author}{{Stokstad} RG},
  \bibinfo{author}{{St{\"o}{\ss}l} A}, \bibinfo{author}{{Strotjohann} NL},
  \bibinfo{author}{{Stuttard} T}, \bibinfo{author}{{Sullivan} GW},
  \bibinfo{author}{{Sutherland} M}, \bibinfo{author}{{Taboada} I},
  \bibinfo{author}{{Tatar} J}, \bibinfo{author}{{Tenholt} F},
  \bibinfo{author}{{Ter-Antonyan} S}, \bibinfo{author}{{Terliuk} A},
  \bibinfo{author}{{Tilav} S}, \bibinfo{author}{{Toale} PA},
  \bibinfo{author}{{Tobin} MN}, \bibinfo{author}{{Toennis} C},
  \bibinfo{author}{{Toscano} S}, \bibinfo{author}{{Tosi} D},
  \bibinfo{author}{{Tselengidou} M}, \bibinfo{author}{{Tung} CF},
  \bibinfo{author}{{Turcati} A}, \bibinfo{author}{{Turley} CF},
  \bibinfo{author}{{Ty} B}, \bibinfo{author}{{Unger} E},
  \bibinfo{author}{{Usner} M}, \bibinfo{author}{{Vandenbroucke} J},
  \bibinfo{author}{{Van Driessche} W}, \bibinfo{author}{{van Eijk} D},
  \bibinfo{author}{{van Eijndhoven} N}, \bibinfo{author}{{Vanheule} S},
  \bibinfo{author}{{van Santen} J}, \bibinfo{author}{{Vogel} E},
  \bibinfo{author}{{Vraeghe} M}, \bibinfo{author}{{Walck} C},
  \bibinfo{author}{{Wallace} A}, \bibinfo{author}{{Wallraff} M},
  \bibinfo{author}{{Wandler} FD}, \bibinfo{author}{{Wandkowsky} N},
  \bibinfo{author}{{Waza} A}, \bibinfo{author}{{Weaver} C},
  \bibinfo{author}{{Weiss} MJ}, \bibinfo{author}{{Wendt} C},
  \bibinfo{author}{{Werthebach} J}, \bibinfo{author}{{Westerhoff} S},
  \bibinfo{author}{{Whelan} BJ}, \bibinfo{author}{{Whitehorn} N},
  \bibinfo{author}{{Wiebe} K}, \bibinfo{author}{{Wiebusch} CH},
  \bibinfo{author}{{Wille} L}, \bibinfo{author}{{Williams} DR},
  \bibinfo{author}{{Wills} L}, \bibinfo{author}{{Wolf} M},
  \bibinfo{author}{{Wood} J}, \bibinfo{author}{{Wood} TR},
  \bibinfo{author}{{Woschnagg} K}, \bibinfo{author}{{Xu} DL},
  \bibinfo{author}{{Xu} XW}, \bibinfo{author}{{Xu} Y}, \bibinfo{author}{{Yanez}
  JP}, \bibinfo{author}{{Yodh} G}, \bibinfo{author}{{Yoshida} S},
  \bibinfo{author}{{Yuan} T}, \bibinfo{author}{{Fermi-LAT Collaboration}},
  \bibinfo{author}{{Abdollahi} S}, \bibinfo{author}{{Ajello} M},
  \bibinfo{author}{{Angioni} R}, \bibinfo{author}{{Baldini} L},
  \bibinfo{author}{{Ballet} J}, \bibinfo{author}{{Barbiellini} G},
  \bibinfo{author}{{Bastieri} D}, \bibinfo{author}{{Bechtol} K},
  \bibinfo{author}{{Bellazzini} R}, \bibinfo{author}{{Berenji} B},
  \bibinfo{author}{{Bissaldi} E}, \bibinfo{author}{{Blandford} RD},
  \bibinfo{author}{{Bonino} R}, \bibinfo{author}{{Bottacini} E},
  \bibinfo{author}{{Bregeon} J}, \bibinfo{author}{{Bruel} P},
  \bibinfo{author}{{Buehler} R}, \bibinfo{author}{{Burnett} TH},
  \bibinfo{author}{{Burns} E}, \bibinfo{author}{{Buson} S},
  \bibinfo{author}{{Cameron} RA}, \bibinfo{author}{{Caputo} R},
  \bibinfo{author}{{Caraveo} PA}, \bibinfo{author}{{Cavazzuti} E},
  \bibinfo{author}{{Charles} E}, \bibinfo{author}{{Chen} S},
  \bibinfo{author}{{Cheung} CC}, \bibinfo{author}{{Chiang} J},
  \bibinfo{author}{{Chiaro} G}, \bibinfo{author}{{Ciprini} S},
  \bibinfo{author}{{Cohen-Tanugi} J}, \bibinfo{author}{{Conrad} J},
  \bibinfo{author}{{Costantin} D}, \bibinfo{author}{{Cutini} S},
  \bibinfo{author}{{D'Ammando} F}, \bibinfo{author}{{de Palma} F},
  \bibinfo{author}{{Digel} SW}, \bibinfo{author}{{Di Lalla} N},
  \bibinfo{author}{{Di Mauro} M}, \bibinfo{author}{{Di Venere} L},
  \bibinfo{author}{{Dom{\'\i}nguez} A}, \bibinfo{author}{{Favuzzi} C},
  \bibinfo{author}{{Franckowiak} A}, \bibinfo{author}{{Fukazawa} Y},
  \bibinfo{author}{{Funk} S}, \bibinfo{author}{{Fusco} P},
  \bibinfo{author}{{Gargano} F}, \bibinfo{author}{{Gasparrini} D},
  \bibinfo{author}{{Giglietto} N}, \bibinfo{author}{{Giomi} M},
  \bibinfo{author}{{Giommi} P}, \bibinfo{author}{{Giordano} F},
  \bibinfo{author}{{Giroletti} M}, \bibinfo{author}{{Glanzman} T},
  \bibinfo{author}{{Green} D}, \bibinfo{author}{{Grenier} IA},
  \bibinfo{author}{{Grondin} MH}, \bibinfo{author}{{Guiriec} S},
  \bibinfo{author}{{Harding} AK}, \bibinfo{author}{{Hayashida} M},
  \bibinfo{author}{{Hays} E}, \bibinfo{author}{{Hewitt} JW},
  \bibinfo{author}{{Horan} D}, \bibinfo{author}{{J{\'o}hannesson} G},
  \bibinfo{author}{{Kadler} M}, \bibinfo{author}{{Kensei} S},
  \bibinfo{author}{{Kocevski} D}, \bibinfo{author}{{Krauss} F},
  \bibinfo{author}{{Kreter} M}, \bibinfo{author}{{Kuss} M},
  \bibinfo{author}{{La Mura} G}, \bibinfo{author}{{Larsson} S},
  \bibinfo{author}{{Latronico} L}, \bibinfo{author}{{Lemoine-Goumard} M},
  \bibinfo{author}{{Li} J}, \bibinfo{author}{{Longo} F},
  \bibinfo{author}{{Loparco} F}, \bibinfo{author}{{Lovellette} MN},
  \bibinfo{author}{{Lubrano} P}, \bibinfo{author}{{Magill} JD},
  \bibinfo{author}{{Maldera} S}, \bibinfo{author}{{Malyshev} D},
  \bibinfo{author}{{Manfreda} A}, \bibinfo{author}{{Mazziotta} MN},
  \bibinfo{author}{{McEnery} JE}, \bibinfo{author}{{Meyer} M},
  \bibinfo{author}{{Michelson} PF}, \bibinfo{author}{{Mizuno} T},
  \bibinfo{author}{{Monzani} ME}, \bibinfo{author}{{Morselli} A},
  \bibinfo{author}{{Moskalenko} IV}, \bibinfo{author}{{Negro} M},
  \bibinfo{author}{{Nuss} E}, \bibinfo{author}{{Ojha} R},
  \bibinfo{author}{{Omodei} N}, \bibinfo{author}{{Orienti} M},
  \bibinfo{author}{{Orlando} E}, \bibinfo{author}{{Palatiello} M},
  \bibinfo{author}{{Paliya} VS}, \bibinfo{author}{{Perkins} JS},
  \bibinfo{author}{{Persic} M}, \bibinfo{author}{{Pesce-Rollins} M},
  \bibinfo{author}{{Piron} F}, \bibinfo{author}{{Porter} TA},
  \bibinfo{author}{{Principe} G}, \bibinfo{author}{{Rain{\`o}} S},
  \bibinfo{author}{{Rando} R}, \bibinfo{author}{{Rani} B},
  \bibinfo{author}{{Razzano} M}, \bibinfo{author}{{Razzaque} S},
  \bibinfo{author}{{Reimer} A}, \bibinfo{author}{{Reimer} O},
  \bibinfo{author}{{Renault-Tinacci} N}, \bibinfo{author}{{Ritz} S},
  \bibinfo{author}{{Rochester} LS}, \bibinfo{author}{{Saz Parkinson} PM},
  \bibinfo{author}{{Sgr{\`o}} C}, \bibinfo{author}{{Siskind} EJ},
  \bibinfo{author}{{Spandre} G}, \bibinfo{author}{{Spinelli} P},
  \bibinfo{author}{{Suson} DJ}, \bibinfo{author}{{Tajima} H},
  \bibinfo{author}{{Takahashi} M}, \bibinfo{author}{{Tanaka} Y},
  \bibinfo{author}{{Thayer} JB}, \bibinfo{author}{{Thompson} DJ},
  \bibinfo{author}{{Tibaldo} L}, \bibinfo{author}{{Torres} DF},
  \bibinfo{author}{{Torresi} E}, \bibinfo{author}{{Tosti} G},
  \bibinfo{author}{{Troja} E}, \bibinfo{author}{{Valverde} J},
  \bibinfo{author}{{Vianello} G}, \bibinfo{author}{{Vogel} M},
  \bibinfo{author}{{Wood} K}, \bibinfo{author}{{Wood} M},
  \bibinfo{author}{{Zaharijas} G}, \bibinfo{author}{{MAGIC Collaboration}},
  \bibinfo{author}{{Ahnen} ML}, \bibinfo{author}{{Ansoldi} S},
  \bibinfo{author}{{Antonelli} LA}, \bibinfo{author}{{Arcaro} C},
  \bibinfo{author}{{Baack} D}, \bibinfo{author}{{Babi{\'c}} A},
  \bibinfo{author}{{Banerjee} B}, \bibinfo{author}{{Bangale} P},
  \bibinfo{author}{{Barres de Almeida} U}, \bibinfo{author}{{Barrio} JA},
  \bibinfo{author}{{Becerra Gonz{\'a}lez} J}, \bibinfo{author}{{Bednarek} W},
  \bibinfo{author}{{Bernardini} E}, \bibinfo{author}{{Berti} A},
  \bibinfo{author}{{Bhattacharyya} W}, \bibinfo{author}{{Biland} A},
  \bibinfo{author}{{Blanch} O}, \bibinfo{author}{{Bonnoli} G},
  \bibinfo{author}{{Carosi} A}, \bibinfo{author}{{Carosi} R},
  \bibinfo{author}{{Ceribella} G}, \bibinfo{author}{{Chatterjee} A},
  \bibinfo{author}{{Colak} SM}, \bibinfo{author}{{Colin} P},
  \bibinfo{author}{{Colombo} E}, \bibinfo{author}{{Contreras} JL},
  \bibinfo{author}{{Cortina} J}, \bibinfo{author}{{Covino} S},
  \bibinfo{author}{{Cumani} P}, \bibinfo{author}{{Da Vela} P},
  \bibinfo{author}{{Dazzi} F}, \bibinfo{author}{{De Angelis} A},
  \bibinfo{author}{{De Lotto} B}, \bibinfo{author}{{Delfino} M},
  \bibinfo{author}{{Delgado} J}, \bibinfo{author}{{Di Pierro} F},
  \bibinfo{author}{{Dom{\'\i}nguez} A}, \bibinfo{author}{{Dominis Prester} D},
  \bibinfo{author}{{Dorner} D}, \bibinfo{author}{{Doro} M},
  \bibinfo{author}{{Einecke} S}, \bibinfo{author}{{Elsaesser} D},
  \bibinfo{author}{{Fallah Ramazani} V},
  \bibinfo{author}{{Fern{\'a}ndez-Barral} A}, \bibinfo{author}{{Fidalgo} D},
  \bibinfo{author}{{Foffano} L}, \bibinfo{author}{{Pfrang} K},
  \bibinfo{author}{{Fonseca} MV}, \bibinfo{author}{{Font} L},
  \bibinfo{author}{{Franceschini} A}, \bibinfo{author}{{Fruck} C},
  \bibinfo{author}{{Galindo} D}, \bibinfo{author}{{Gallozzi} S},
  \bibinfo{author}{{Garc{\'\i}a L{\'o}pez} RJ}, \bibinfo{author}{{Garczarczyk}
  M}, \bibinfo{author}{{Gaug} M}, \bibinfo{author}{{Giammaria} P},
  \bibinfo{author}{{Godinovi{\'c}} N}, \bibinfo{author}{{Gora} D},
  \bibinfo{author}{{Guberman} D}, \bibinfo{author}{{Hadasch} D},
  \bibinfo{author}{{Hahn} A}, \bibinfo{author}{{Hassan} T},
  \bibinfo{author}{{Hayashida} M}, \bibinfo{author}{{Herrera} J},
  \bibinfo{author}{{Hose} J}, \bibinfo{author}{{Hrupec} D},
  \bibinfo{author}{{Inoue} S}, \bibinfo{author}{{Ishio} K},
  \bibinfo{author}{{Konno} Y}, \bibinfo{author}{{Kubo} H},
  \bibinfo{author}{{Kushida} J}, \bibinfo{author}{{Lelas} D},
  \bibinfo{author}{{Lindfors} E}, \bibinfo{author}{{Lombardi} S},
  \bibinfo{author}{{Longo} F}, \bibinfo{author}{{L{\'o}pez} M},
  \bibinfo{author}{{Maggio} C}, \bibinfo{author}{{Majumdar} P},
  \bibinfo{author}{{Makariev} M}, \bibinfo{author}{{Maneva} G},
  \bibinfo{author}{{Manganaro} M}, \bibinfo{author}{{Mannheim} K},
  \bibinfo{author}{{Maraschi} L}, \bibinfo{author}{{Mariotti} M},
  \bibinfo{author}{{Mart{\'\i}nez} M}, \bibinfo{author}{{Masuda} S},
  \bibinfo{author}{{Mazin} D}, \bibinfo{author}{{Minev} M},
  \bibinfo{author}{{M} JM}, \bibinfo{author}{{Mirzoyan} R},
  \bibinfo{author}{{Moralejo} A}, \bibinfo{author}{{Moreno} V},
  \bibinfo{author}{{Moretti} E}, \bibinfo{author}{{Nagayoshi} T},
  \bibinfo{author}{{Neustroev} V}, \bibinfo{author}{{Niedzwiecki} A},
  \bibinfo{author}{{Nievas Rosillo} M}, \bibinfo{author}{{Nigro} C},
  \bibinfo{author}{{Nilsson} K}, \bibinfo{author}{{Ninci} D},
  \bibinfo{author}{{Nishijima} K}, \bibinfo{author}{{Noda} K},
  \bibinfo{author}{{Nogu{\'e}s} L}, \bibinfo{author}{{Paiano} S},
  \bibinfo{author}{{Palacio} J}, \bibinfo{author}{{Paneque} D},
  \bibinfo{author}{{Paoletti} R}, \bibinfo{author}{{Paredes} JM},
  \bibinfo{author}{{Pedaletti} G}, \bibinfo{author}{{Peresano} M},
  \bibinfo{author}{{Persic} M}, \bibinfo{author}{{Prada Moroni} PG},
  \bibinfo{author}{{Prandini} E}, \bibinfo{author}{{Puljak} I},
  \bibinfo{author}{{Rodriguez Garcia} J}, \bibinfo{author}{{Reichardt} I},
  \bibinfo{author}{{Rhode} W}, \bibinfo{author}{{Rib{\'o}} M},
  \bibinfo{author}{{Rico} J}, \bibinfo{author}{{Righi} C},
  \bibinfo{author}{{Rugliancich} A}, \bibinfo{author}{{Saito} T},
  \bibinfo{author}{{Satalecka} K}, \bibinfo{author}{{Schweizer} T},
  \bibinfo{author}{{Sitarek} J}, \bibinfo{author}{{{\v{S}}nidaric} I}, \bibinfo{author}{{Sobczynska} D},
  \bibinfo{author}{{Stamerra} A}, \bibinfo{author}{{Strzys} M},
  \bibinfo{author}{{Suri{\'c}} T}, \bibinfo{author}{{Takahashi} M},
  \bibinfo{author}{{Tavecchio} F}, \bibinfo{author}{{Temnikov} P},
  \bibinfo{author}{{Terzi{\'c}} T}, \bibinfo{author}{{Teshima} M},
  \bibinfo{author}{{Torres-Alb{\`a}} N}, \bibinfo{author}{{Treves} A},
  \bibinfo{author}{{Tsujimoto} S}, \bibinfo{author}{{Vanzo} G},
  \bibinfo{author}{{Vazquez Acosta} M}, \bibinfo{author}{{Vovk} I},
  \bibinfo{author}{{Ward} JE}, \bibinfo{author}{{Will} M},
  \bibinfo{author}{{S}}, \bibinfo{author}{{Zaric} D},
  \bibinfo{author}{{AGILE Team}}, \bibinfo{author}{{Lucarelli} F},
  \bibinfo{author}{{Tavani} M}, \bibinfo{author}{{Piano} G},
  \bibinfo{author}{{Donnarumma} I}, \bibinfo{author}{{Pittori} C},
  \bibinfo{author}{{Verrecchia} F}, \bibinfo{author}{{Barbiellini} G},
  \bibinfo{author}{{Bulgarelli} A}, \bibinfo{author}{{Caraveo} P},
  \bibinfo{author}{{Cattaneo} PW}, \bibinfo{author}{{Colafrancesco} S},
  \bibinfo{author}{{Costa} E}, \bibinfo{author}{{Di Cocco} G},
  \bibinfo{author}{{Ferrari} A}, \bibinfo{author}{{Gianotti} F},
  \bibinfo{author}{{Giuliani} A}, \bibinfo{author}{{Lipari} P},
  \bibinfo{author}{{Mereghetti} S}, \b (\bibinfo{year}{2018}),
  \bibinfo{month}{Jul.}
\bibinfo{title}{{Multimessenger observations of a flaring blazar coincident
  with high-energy neutrino IceCube-170922A}}.
\bibinfo{journal}{{\em Science}} \bibinfo{volume}{361}
  (\bibinfo{number}{6398}), \bibinfo{eid}{eaat1378}.
  \bibinfo{doi}{\doi{10.1126/science.aat1378}}.
\eprint{1807.08816}.

\bibtype{Article}%
\bibitem[{Jakobsson} et al.(2006)]{jakobsson06}
\bibinfo{author}{{Jakobsson} P}, \bibinfo{author}{{Levan} A},
  \bibinfo{author}{{Fynbo} JPU}, \bibinfo{author}{{Priddey} R},
  \bibinfo{author}{{Hjorth} J}, \bibinfo{author}{{Tanvir} N},
  \bibinfo{author}{{Watson} D}, \bibinfo{author}{{Jensen} BL},
  \bibinfo{author}{{Sollerman} J}, \bibinfo{author}{{Natarajan} P},
  \bibinfo{author}{{Gorosabel} J}, \bibinfo{author}{{Castro Cer{\'o}n} JM},
  \bibinfo{author}{{Pedersen} K}, \bibinfo{author}{{Pursimo} T},
  \bibinfo{author}{{{\'A}rnad{\'o}ttir} AS}, \bibinfo{author}{{Castro-Tirado}
  AJ}, \bibinfo{author}{{Davis} CJ}, \bibinfo{author}{{Deeg} HJ},
  \bibinfo{author}{{Fiuza} DA}, \bibinfo{author}{{Mikolaitis} S} and
  \bibinfo{author}{{Sousa} SG} (\bibinfo{year}{2006}), \bibinfo{month}{Mar.}
\bibinfo{title}{{A mean redshift of 2.8 for Swift gamma-ray bursts}}.
\bibinfo{journal}{{\em A\&A}} \bibinfo{volume}{447} (\bibinfo{number}{3}):
  \bibinfo{pages}{897--903}. \bibinfo{doi}{\doi{10.1051/0004-6361:20054287}}.
\eprint{astro-ph/0509888}.

\bibtype{Article}%
\bibitem[{Klebesadel} et al.(1973)]{kleb73}
\bibinfo{author}{{Klebesadel} RW}, \bibinfo{author}{{Strong} IB} and
  \bibinfo{author}{{Olson} RA} (\bibinfo{year}{1973}), \bibinfo{month}{Jun.}
\bibinfo{title}{{Observations of Gamma-Ray Bursts of Cosmic Origin}}.
\bibinfo{journal}{{\em ApJl}} \bibinfo{volume}{182}: \bibinfo{pages}{L85}.
  \bibinfo{doi}{\doi{10.1086/181225}}.

\bibtype{Article}%
\bibitem[{Komissarov} and {Barkov}(2009)]{kom09}
\bibinfo{author}{{Komissarov} SS} and  \bibinfo{author}{{Barkov} MV}
  (\bibinfo{year}{2009}), \bibinfo{month}{Aug.}
\bibinfo{title}{{Activation of the Blandford-Znajek mechanism in collapsing
  stars}}.
\bibinfo{journal}{{\em MNRAS}} \bibinfo{volume}{397} (\bibinfo{number}{3}):
  \bibinfo{pages}{1153--1168}.
  \bibinfo{doi}{\doi{10.1111/j.1365-2966.2009.14831.x}}.
\eprint{0902.2881}.

\bibtype{Article}%
\bibitem[{Kouveliotou} et al.(1993)]{Kouveliotou+93}
\bibinfo{author}{{Kouveliotou} C}, \bibinfo{author}{{Meegan} CA},
  \bibinfo{author}{{Fishman} GJ}, \bibinfo{author}{{Bhat} NP},
  \bibinfo{author}{{Briggs} MS}, \bibinfo{author}{{Koshut} TM},
  \bibinfo{author}{{Paciesas} WS} and  \bibinfo{author}{{Pendleton} GN}
  (\bibinfo{year}{1993}), \bibinfo{month}{Aug.}
\bibinfo{title}{{Identification of Two Classes of Gamma-Ray Bursts}}.
\bibinfo{journal}{{\em ApJl}} \bibinfo{volume}{413}: \bibinfo{pages}{L101}.
  \bibinfo{doi}{\doi{10.1086/186969}}.

\bibtype{Article}%
\bibitem[{Kouveliotou} et al.(1998)]{kouveliotou98}
\bibinfo{author}{{Kouveliotou} C}, \bibinfo{author}{{Dieters} S},
  \bibinfo{author}{{Strohmayer} T}, \bibinfo{author}{{van Paradijs} J},
  \bibinfo{author}{{Fishman} GJ}, \bibinfo{author}{{Meegan} CA},
  \bibinfo{author}{{Hurley} K}, \bibinfo{author}{{Kommers} J},
  \bibinfo{author}{{Smith} I}, \bibinfo{author}{{Frail} D} and
  \bibinfo{author}{{Murakami} T} (\bibinfo{year}{1998}), \bibinfo{month}{May}.
\bibinfo{title}{{An X-ray pulsar with a superstrong magnetic field in the soft
  {\ensuremath{\gamma}}-ray repeater SGR1806 - 20}}.
\bibinfo{journal}{{\em \nat}} \bibinfo{volume}{393} (\bibinfo{number}{6682}):
  \bibinfo{pages}{235--237}. \bibinfo{doi}{\doi{10.1038/30410}}.

\bibtype{Article}%
\bibitem[{Kumar} and {Crumley}(2015)]{kumar15}
\bibinfo{author}{{Kumar} P} and  \bibinfo{author}{{Crumley} P}
  (\bibinfo{year}{2015}), \bibinfo{month}{Oct.}
\bibinfo{title}{{Radiation from a relativistic Poynting jet: some general
  considerations}}.
\bibinfo{journal}{{\em MNRAS}} \bibinfo{volume}{453} (\bibinfo{number}{2}):
  \bibinfo{pages}{1820--1828}. \bibinfo{doi}{\doi{10.1093/mnras/stv1696}}.
\eprint{1509.00479}.

\bibtype{Article}%
\bibitem[{Kumar} and {Granot}(2003)]{kumar03}
\bibinfo{author}{{Kumar} P} and  \bibinfo{author}{{Granot} J}
  (\bibinfo{year}{2003}), \bibinfo{month}{Jul.}
\bibinfo{title}{{The Evolution of a Structured Relativistic Jet and Gamma-Ray
  Burst Afterglow Light Curves}}.
\bibinfo{journal}{{\em ApJ}} \bibinfo{volume}{591} (\bibinfo{number}{2}):
  \bibinfo{pages}{1075--1085}. \bibinfo{doi}{\doi{10.1086/375186}}.
\eprint{astro-ph/0303174}.

\bibtype{Article}%
\bibitem[{Laskar} et al.(2013)]{laskar13}
\bibinfo{author}{{Laskar} T}, \bibinfo{author}{{Berger} E},
  \bibinfo{author}{{Zauderer} BA}, \bibinfo{author}{{Margutti} R},
  \bibinfo{author}{{Soderberg} AM}, \bibinfo{author}{{Chakraborti} S},
  \bibinfo{author}{{Lunnan} R}, \bibinfo{author}{{Chornock} R},
  \bibinfo{author}{{Chandra} P} and  \bibinfo{author}{{Ray} A}
  (\bibinfo{year}{2013}), \bibinfo{month}{Oct.}
\bibinfo{title}{{A Reverse Shock in GRB 130427A}}.
\bibinfo{journal}{{\em Ap}} \bibinfo{volume}{776} (\bibinfo{number}{2}),
  \bibinfo{eid}{119}. \bibinfo{doi}{\doi{10.1088/0004-637X/776/2/119}}.
\eprint{1305.2453}.

\bibtype{Article}%
\bibitem[{Lesage} et al.(2023)]{lesage23}
\bibinfo{author}{{Lesage} S}, \bibinfo{author}{{Veres} P},
  \bibinfo{author}{{Briggs} MS}, \bibinfo{author}{{Goldstein} A},
  \bibinfo{author}{{Kocevski} D}, \bibinfo{author}{{Burns} E},
  \bibinfo{author}{{Wilson-Hodge} CA}, \bibinfo{author}{{Bhat} PN},
  \bibinfo{author}{{Huppenkothen} D}, \bibinfo{author}{{Fryer} CL},
  \bibinfo{author}{{Hamburg} R}, \bibinfo{author}{{Racusin} J},
  \bibinfo{author}{{Bissaldi} E}, \bibinfo{author}{{Cleveland} WH},
  \bibinfo{author}{{Dalessi} S}, \bibinfo{author}{{Fletcher} C},
  \bibinfo{author}{{Giles} MM}, \bibinfo{author}{{Hristov} BA},
  \bibinfo{author}{{Hui} CM}, \bibinfo{author}{{Mailyan} B},
  \bibinfo{author}{{Malacaria} C}, \bibinfo{author}{{Poolakkil} S},
  \bibinfo{author}{{Roberts} OJ}, \bibinfo{author}{{von Kienlin} A},
  \bibinfo{author}{{Wood} J}, \bibinfo{author}{{Ajello} M},
  \bibinfo{author}{{Arimoto} M}, \bibinfo{author}{{Baldini} L},
  \bibinfo{author}{{Ballet} J}, \bibinfo{author}{{Baring} MG},
  \bibinfo{author}{{Bastieri} D}, \bibinfo{author}{{Gonzalez} JB},
  \bibinfo{author}{{Bellazzini} R}, \bibinfo{author}{{Bissaldi} E},
  \bibinfo{author}{{Blandford} RD}, \bibinfo{author}{{Bonino} R},
  \bibinfo{author}{{Bruel} P}, \bibinfo{author}{{Buson} S},
  \bibinfo{author}{{Cameron} RA}, \bibinfo{author}{{Caputo} R},
  \bibinfo{author}{{Caraveo} PA}, \bibinfo{author}{{Cavazzuti} E},
  \bibinfo{author}{{Chiaro} G}, \bibinfo{author}{{Cibrario} N},
  \bibinfo{author}{{Ciprini} S}, \bibinfo{author}{{Orestano} PC},
  \bibinfo{author}{{Crnogorcevic} M}, \bibinfo{author}{{Cuoco} A},
  \bibinfo{author}{{Cutini} S}, \bibinfo{author}{{D'Ammando} F},
  \bibinfo{author}{{De Gaetano} S}, \bibinfo{author}{{Di Lalla} N},
  \bibinfo{author}{{Di Venere} L}, \bibinfo{author}{{Dom{\'\i}nguez} A},
  \bibinfo{author}{{Fegan} SJ}, \bibinfo{author}{{Ferrara} EC},
  \bibinfo{author}{{Fleischhack} H}, \bibinfo{author}{{Fukazawa} Y},
  \bibinfo{author}{{Funk} S}, \bibinfo{author}{{Fusco} P},
  \bibinfo{author}{{Galanti} G}, \bibinfo{author}{{Gammaldi} V},
  \bibinfo{author}{{Gargano} F}, \bibinfo{author}{{Gasbarra} C},
  \bibinfo{author}{{Gasparrini} D}, \bibinfo{author}{{Germani} S},
  \bibinfo{author}{{Giacchino} F}, \bibinfo{author}{{Giglietto} N},
  \bibinfo{author}{{Gill} R}, \bibinfo{author}{{Giroletti} M},
  \bibinfo{author}{{Granot} J}, \bibinfo{author}{{Green} D},
  \bibinfo{author}{{Grenier} IA}, \bibinfo{author}{{Guiriec} S},
  \bibinfo{author}{{Gustafsson} M}, \bibinfo{author}{{Hays} E},
  \bibinfo{author}{{Hewitt} JW}, \bibinfo{author}{{Horan} D},
  \bibinfo{author}{{Hou} X}, \bibinfo{author}{{Kuss} M},
  \bibinfo{author}{{Latronico} L}, \bibinfo{author}{{Laviron} A},
  \bibinfo{author}{{Lemoine-Goumard} M}, \bibinfo{author}{{Li} J},
  \bibinfo{author}{{Liodakis} I}, \bibinfo{author}{{Longo} F},
  \bibinfo{author}{{Loparco} F}, \bibinfo{author}{{Lorusso} L},
  \bibinfo{author}{{Lovellette} MN}, \bibinfo{author}{{Lubrano} P},
  \bibinfo{author}{{Maldera} S}, \bibinfo{author}{{Manfreda} A},
  \bibinfo{author}{{Mart{\'\i}-Devesa} G}, \bibinfo{author}{{Mazziotta} MN},
  \bibinfo{author}{{McEnery} JE}, \bibinfo{author}{{Mereu} I},
  \bibinfo{author}{{Meyer} M}, \bibinfo{author}{{Michelson} PF},
  \bibinfo{author}{{Mizuno} T}, \bibinfo{author}{{Monzani} ME},
  \bibinfo{author}{{Morselli} A}, \bibinfo{author}{{Moskalenko} IV},
  \bibinfo{author}{{Negro} M}, \bibinfo{author}{{Nuss} E},
  \bibinfo{author}{{Omodei} N}, \bibinfo{author}{{Orlando} E},
  \bibinfo{author}{{Ormes} JF}, \bibinfo{author}{{Paneque} D},
  \bibinfo{author}{{Panzarini} G}, \bibinfo{author}{{Persic} M},
  \bibinfo{author}{{Pesce-Rollins} M}, \bibinfo{author}{{Pillera} R},
  \bibinfo{author}{{Piron} F}, \bibinfo{author}{{Poon} H},
  \bibinfo{author}{{Porter} TA}, \bibinfo{author}{{Principe} G},
  \bibinfo{author}{{Rain{\`o}} S}, \bibinfo{author}{{Rando} R},
  \bibinfo{author}{{Rani} B}, \bibinfo{author}{{Razzano} M},
  \bibinfo{author}{{Razzaque} S}, \bibinfo{author}{{Reimer} A},
  \bibinfo{author}{{Reimer} O}, \bibinfo{author}{{Ryde} F},
  \bibinfo{author}{{S{\'a}nchez-Conde} M}, \bibinfo{author}{{Parkinson} PMS},
  \bibinfo{author}{{Scotton} L}, \bibinfo{author}{{Serini} D},
  \bibinfo{author}{{Sgr{\`o}} C}, \bibinfo{author}{{Sharma} V},
  \bibinfo{author}{{Siskind} EJ}, \bibinfo{author}{{Spandre} G},
  \bibinfo{author}{{Spinelli} P}, \bibinfo{author}{{Tajima} H},
  \bibinfo{author}{{Torres} DF}, \bibinfo{author}{{Valverde} J},
  \bibinfo{author}{{Venters} T}, \bibinfo{author}{{Wadiasingh} Z},
  \bibinfo{author}{{Wood} K} and  \bibinfo{author}{{Zaharijas} G}
  (\bibinfo{year}{2023}), \bibinfo{month}{Aug.}
\bibinfo{title}{{Fermi-GBM Discovery of GRB 221009A: An Extraordinarily Bright
  GRB from Onset to Afterglow}}.
\bibinfo{journal}{{\em ApJl}} \bibinfo{volume}{952} (\bibinfo{number}{2}),
  \bibinfo{eid}{L42}. \bibinfo{doi}{\doi{10.3847/2041-8213/ace5b4}}.
\eprint{2303.14172}.

\bibtype{Article}%
\bibitem[{Levan} et al.(2006)]{levan06}
\bibinfo{author}{{Levan} AJ}, \bibinfo{author}{{Wynn} GA},
  \bibinfo{author}{{Chapman} R}, \bibinfo{author}{{Davies} MB},
  \bibinfo{author}{{King} AR}, \bibinfo{author}{{Priddey} RS} and
  \bibinfo{author}{{Tanvir} NR} (\bibinfo{year}{2006}), \bibinfo{month}{May}.
\bibinfo{title}{{Short gamma-ray bursts in old populations: magnetars from
  white dwarf-white dwarf mergers}}.
\bibinfo{journal}{{\em MNRAS}} \bibinfo{volume}{368} (\bibinfo{number}{1}):
  \bibinfo{pages}{L1--L5}.
  \bibinfo{doi}{\doi{10.1111/j.1745-3933.2006.00144.x}}.
\eprint{astro-ph/0601332}.

\bibtype{Article}%
\bibitem[{Levan} et al.(2024{\natexlab{a}})]{levan24a}
\bibinfo{author}{{Levan} AJ}, \bibinfo{author}{{Gompertz} BP},
  \bibinfo{author}{{Salafia} OS}, \bibinfo{author}{{Bulla} M},
  \bibinfo{author}{{Burns} E}, \bibinfo{author}{{Hotokezaka} K},
  \bibinfo{author}{{Izzo} L}, \bibinfo{author}{{Lamb} GP},
  \bibinfo{author}{{Malesani} DB}, \bibinfo{author}{{Oates} SR},
  \bibinfo{author}{{Ravasio} ME}, \bibinfo{author}{{Rouco Escorial} A},
  \bibinfo{author}{{Schneider} B}, \bibinfo{author}{{Sarin} N},
  \bibinfo{author}{{Schulze} S}, \bibinfo{author}{{Tanvir} NR},
  \bibinfo{author}{{Ackley} K}, \bibinfo{author}{{Anderson} G},
  \bibinfo{author}{{Brammer} GB}, \bibinfo{author}{{Christensen} L},
  \bibinfo{author}{{Dhillon} VS}, \bibinfo{author}{{Evans} PA},
  \bibinfo{author}{{Fausnaugh} M}, \bibinfo{author}{{Fong} Wf},
  \bibinfo{author}{{Fruchter} AS}, \bibinfo{author}{{Fryer} C},
  \bibinfo{author}{{Fynbo} JPU}, \bibinfo{author}{{Gaspari} N},
  \bibinfo{author}{{Heintz} KE}, \bibinfo{author}{{Hjorth} J},
  \bibinfo{author}{{Kennea} JA}, \bibinfo{author}{{Kennedy} MR},
  \bibinfo{author}{{Laskar} T}, \bibinfo{author}{{Leloudas} G},
  \bibinfo{author}{{Mandel} I}, \bibinfo{author}{{Martin-Carrillo} A},
  \bibinfo{author}{{Metzger} BD}, \bibinfo{author}{{Nicholl} M},
  \bibinfo{author}{{Nugent} A}, \bibinfo{author}{{Palmerio} JT},
  \bibinfo{author}{{Pugliese} G}, \bibinfo{author}{{Rastinejad} J},
  \bibinfo{author}{{Rhodes} L}, \bibinfo{author}{{Rossi} A},
  \bibinfo{author}{{Saccardi} A}, \bibinfo{author}{{Smartt} SJ},
  \bibinfo{author}{{Stevance} HF}, \bibinfo{author}{{Tohuvavohu} A},
  \bibinfo{author}{{van der Horst} A}, \bibinfo{author}{{Vergani} SD},
  \bibinfo{author}{{Watson} D}, \bibinfo{author}{{Barclay} T},
  \bibinfo{author}{{Bhirombhakdi} K}, \bibinfo{author}{{Breedt} E},
  \bibinfo{author}{{Breeveld} AA}, \bibinfo{author}{{Brown} AJ},
  \bibinfo{author}{{Campana} S}, \bibinfo{author}{{Chrimes} AA},
  \bibinfo{author}{{D'Avanzo} P}, \bibinfo{author}{{D'Elia} V},
  \bibinfo{author}{{De Pasquale} M}, \bibinfo{author}{{Dyer} MJ},
  \bibinfo{author}{{Galloway} DK}, \bibinfo{author}{{Garbutt} JA},
  \bibinfo{author}{{Green} MJ}, \bibinfo{author}{{Hartmann} DH},
  \bibinfo{author}{{Jakobsson} P}, \bibinfo{author}{{Kerry} P},
  \bibinfo{author}{{Kouveliotou} C}, \bibinfo{author}{{Langeroodi} D},
  \bibinfo{author}{{Le Floc'h} E}, \bibinfo{author}{{Leung} JK},
  \bibinfo{author}{{Littlefair} SP}, \bibinfo{author}{{Munday} J},
  \bibinfo{author}{{O'Brien} P}, \bibinfo{author}{{Parsons} SG},
  \bibinfo{author}{{Pelisoli} I}, \bibinfo{author}{{Sahman} DI},
  \bibinfo{author}{{Salvaterra} R}, \bibinfo{author}{{Sbarufatti} B},
  \bibinfo{author}{{Steeghs} D}, \bibinfo{author}{{Tagliaferri} G},
  \bibinfo{author}{{Th{\"o}ne} CC}, \bibinfo{author}{{de Ugarte Postigo} A} and
   \bibinfo{author}{{Kann} DA} (\bibinfo{year}{2024}{\natexlab{a}}),
  \bibinfo{month}{Feb.}
\bibinfo{title}{{Heavy-element production in a compact object merger observed
  by JWST}}.
\bibinfo{journal}{{\em \nat}} \bibinfo{volume}{626} (\bibinfo{number}{8000}):
  \bibinfo{pages}{737--741}. \bibinfo{doi}{\doi{10.1038/s41586-023-06759-1}}.
\eprint{2307.02098}.

\bibtype{Article}%
\bibitem[{Levan} et al.(2024{\natexlab{b}})]{levan24b}
\bibinfo{author}{{Levan} AJ}, \bibinfo{author}{{Jonker} PG},
  \bibinfo{author}{{Saccardi} A}, \bibinfo{author}{{Bj{\o}rn Malesani} D},
  \bibinfo{author}{{Tanvir} NR}, \bibinfo{author}{{Izzo} L},
  \bibinfo{author}{{Heintz} KE}, \bibinfo{author}{{Mata S{\'a}nchez} D},
  \bibinfo{author}{{Quirola-V{\'a}squez} J}, \bibinfo{author}{{Torres} MAP},
  \bibinfo{author}{{Vergani} SD}, \bibinfo{author}{{Schulze} S},
  \bibinfo{author}{{Rossi} A}, \bibinfo{author}{{D'Avanzo} P},
  \bibinfo{author}{{Gompertz} B}, \bibinfo{author}{{Martin-Carrillo} A},
  \bibinfo{author}{{de Ugarte Postigo} A}, \bibinfo{author}{{Schneider} B},
  \bibinfo{author}{{Yuan} W}, \bibinfo{author}{{Ling} Z},
  \bibinfo{author}{{Zhang} W}, \bibinfo{author}{{Mao} X},
  \bibinfo{author}{{Liu} Y}, \bibinfo{author}{{Sun} H}, \bibinfo{author}{{Xu}
  D}, \bibinfo{author}{{Zhu} Z}, \bibinfo{author}{{Ag{\"u}{\'\i} Fern{\'a}ndez}
  JF}, \bibinfo{author}{{Amati} L}, \bibinfo{author}{{Bauer} FE},
  \bibinfo{author}{{Campana} S}, \bibinfo{author}{{Carotenuto} F},
  \bibinfo{author}{{Chrimes} A}, \bibinfo{author}{{van Dalen} JND},
  \bibinfo{author}{{D'Elia} V}, \bibinfo{author}{{Della Valle} M},
  \bibinfo{author}{{De Pasquale} M}, \bibinfo{author}{{Dhillon} VS},
  \bibinfo{author}{{Galbany} L}, \bibinfo{author}{{Gaspari} N},
  \bibinfo{author}{{Gianfagna} G}, \bibinfo{author}{{Gomboc} A},
  \bibinfo{author}{{Habeeb} N}, \bibinfo{author}{{van Hoof} APC},
  \bibinfo{author}{{Hu} Y}, \bibinfo{author}{{Jakobsson} P},
  \bibinfo{author}{{Julakanti} Y}, \bibinfo{author}{{Korth} J},
  \bibinfo{author}{{Kouveliotou} C}, \bibinfo{author}{{Laskar} T},
  \bibinfo{author}{{Littlefair} SP}, \bibinfo{author}{{Maiorano} E},
  \bibinfo{author}{{Mao} J}, \bibinfo{author}{{Melandri} A},
  \bibinfo{author}{{Miller} MC}, \bibinfo{author}{{Mukherjee} T},
  \bibinfo{author}{{Oates} SR}, \bibinfo{author}{{O'Brien} P},
  \bibinfo{author}{{Palmerio} JT}, \bibinfo{author}{{Parviainen} H},
  \bibinfo{author}{{Pieterse} DLA}, \bibinfo{author}{{Piranomonte} S},
  \bibinfo{author}{{Piro} L}, \bibinfo{author}{{Pugliese} G},
  \bibinfo{author}{{Ravasio} ME}, \bibinfo{author}{{Rayson} B},
  \bibinfo{author}{{Salvaterra} R}, \bibinfo{author}{{S{\'a}nchez-Ram{\'\i}rez}
  R}, \bibinfo{author}{{Sarin} N}, \bibinfo{author}{{Shilling} SPR},
  \bibinfo{author}{{Starling} RLC}, \bibinfo{author}{{Tagliaferri} G},
  \bibinfo{author}{{Linesh Thakur} A}, \bibinfo{author}{{Th{\"o}ne} CC},
  \bibinfo{author}{{Wiersema} K}, \bibinfo{author}{{Worssam} I} and
  \bibinfo{author}{{Zafar} T} (\bibinfo{year}{2024}{\natexlab{b}}),
  \bibinfo{month}{Apr.}
\bibinfo{title}{{The fast X-ray transient EP240315a: a z
  \raisebox{-0.5ex}\textasciitilde 5 gamma-ray burst in a Lyman continuum
  leaking galaxy}}.
\bibinfo{journal}{{\em arXiv e-prints}} ,
  \bibinfo{eid}{arXiv:2404.16350}\bibinfo{doi}{\doi{10.48550/arXiv.2404.16350}}.
\eprint{2404.16350}.

\bibtype{Article}%
\bibitem[{LHAASO Collaboration} et al.(2023)]{lhaaso23}
\bibinfo{author}{{LHAASO Collaboration}}, \bibinfo{author}{{Cao} Z},
  \bibinfo{author}{{Aharonian} F}, \bibinfo{author}{{An} Q},
  \bibinfo{author}{{Axikegu} A}, \bibinfo{author}{{Bai} LX},
  \bibinfo{author}{{Bai} YX}, \bibinfo{author}{{Bao} YW},
  \bibinfo{author}{{Bastieri} D}, \bibinfo{author}{{Bi} XJ},
  \bibinfo{author}{{Bi} YJ}, \bibinfo{author}{{Cai} JT}, \bibinfo{author}{{Cao}
  Q}, \bibinfo{author}{{Cao} WY}, \bibinfo{author}{{Cao} Z},
  \bibinfo{author}{{Chang} J}, \bibinfo{author}{{Chang} JF},
  \bibinfo{author}{{Chen} ES}, \bibinfo{author}{{Chen} L},
  \bibinfo{author}{{Chen} L}, \bibinfo{author}{{Chen} L},
  \bibinfo{author}{{Chen} MJ}, \bibinfo{author}{{Chen} ML},
  \bibinfo{author}{{Chen} QH}, \bibinfo{author}{{Chen} SH},
  \bibinfo{author}{{Chen} SZ}, \bibinfo{author}{{Chen} TL},
  \bibinfo{author}{{Chen} Y}, \bibinfo{author}{{Cheng} HL},
  \bibinfo{author}{{Cheng} N}, \bibinfo{author}{{Cheng} YD},
  \bibinfo{author}{{Cui} SW}, \bibinfo{author}{{Cui} XH},
  \bibinfo{author}{{Cui} YD}, \bibinfo{author}{{Dai} BZ},
  \bibinfo{author}{{Dai} HL}, \bibinfo{author}{{Danzengluobu} D},
  \bibinfo{author}{{Della Volpe} D}, \bibinfo{author}{{Dong} XQ},
  \bibinfo{author}{{Duan} KK}, \bibinfo{author}{{Fan} JH},
  \bibinfo{author}{{Fan} YZ}, \bibinfo{author}{{Fang} J},
  \bibinfo{author}{{Fang} K}, \bibinfo{author}{{Feng} CF},
  \bibinfo{author}{{Feng} L}, \bibinfo{author}{{Feng} SH},
  \bibinfo{author}{{Feng} XT}, \bibinfo{author}{{Feng} YL},
  \bibinfo{author}{{Gao} B}, \bibinfo{author}{{Gao} CD}, \bibinfo{author}{{Gao}
  LQ}, \bibinfo{author}{{Gao} Q}, \bibinfo{author}{{Gao} W},
  \bibinfo{author}{{Gao} WK}, \bibinfo{author}{{Ge} MM},
  \bibinfo{author}{{Geng} LS}, \bibinfo{author}{{Gong} GH},
  \bibinfo{author}{{Gou} QB}, \bibinfo{author}{{Gu} MH}, \bibinfo{author}{{Guo}
  FL}, \bibinfo{author}{{Guo} XL}, \bibinfo{author}{{Guo} YQ},
  \bibinfo{author}{{Guo} YY}, \bibinfo{author}{{Han} YA}, \bibinfo{author}{{He}
  HH}, \bibinfo{author}{{He} HN}, \bibinfo{author}{{He} JY},
  \bibinfo{author}{{He} XB}, \bibinfo{author}{{He} Y},
  \bibinfo{author}{{Heller} M}, \bibinfo{author}{{Hor} YK},
  \bibinfo{author}{{Hou} BW}, \bibinfo{author}{{Hou} C}, \bibinfo{author}{{Hou}
  X}, \bibinfo{author}{{Hu} HB}, \bibinfo{author}{{Hu} Q},
  \bibinfo{author}{{Hu} SC}, \bibinfo{author}{{Huang} DH},
  \bibinfo{author}{{Huang} TQ}, \bibinfo{author}{{Huang} WJ},
  \bibinfo{author}{{Huang} XT}, \bibinfo{author}{{Huang} ZC},
  \bibinfo{author}{{Ji} XL}, \bibinfo{author}{{Jia} HY}, \bibinfo{author}{{Jia}
  K}, \bibinfo{author}{{Jiang} K}, \bibinfo{author}{{Jiang} XW},
  \bibinfo{author}{{Jiang} ZJ}, \bibinfo{author}{{Jin} M},
  \bibinfo{author}{{Kang} MM}, \bibinfo{author}{{Ke} T},
  \bibinfo{author}{{Kuleshov} D}, \bibinfo{author}{{Kurinov} K},
  \bibinfo{author}{{Li} BB}, \bibinfo{author}{{Li} C}, \bibinfo{author}{{Li}
  C}, \bibinfo{author}{{Li} D}, \bibinfo{author}{{Li} F}, \bibinfo{author}{{Li}
  HB}, \bibinfo{author}{{Li} HC}, \bibinfo{author}{{Li} HY},
  \bibinfo{author}{{Li} J}, \bibinfo{author}{{Li} J}, \bibinfo{author}{{Li} J},
  \bibinfo{author}{{Li} K}, \bibinfo{author}{{Li} WL}, \bibinfo{author}{{Li}
  WL}, \bibinfo{author}{{Li} XR}, \bibinfo{author}{{Li} X},
  \bibinfo{author}{{Li} YZ}, \bibinfo{author}{{Li} Z}, \bibinfo{author}{{Li}
  Z}, \bibinfo{author}{{Liang} EW}, \bibinfo{author}{{Liang} YF},
  \bibinfo{author}{{Lin} SJ}, \bibinfo{author}{{Liu} B}, \bibinfo{author}{{Liu}
  C}, \bibinfo{author}{{Liu} D}, \bibinfo{author}{{Liu} H},
  \bibinfo{author}{{Liu} HD}, \bibinfo{author}{{Liu} J}, \bibinfo{author}{{Liu}
  JL}, \bibinfo{author}{{Liu} JL}, \bibinfo{author}{{Liu} JS},
  \bibinfo{author}{{Liu} JY}, \bibinfo{author}{{Liu} MY},
  \bibinfo{author}{{Liu} RY}, \bibinfo{author}{{Liu} SM},
  \bibinfo{author}{{Liu} W}, \bibinfo{author}{{Liu} Y}, \bibinfo{author}{{Liu}
  YN}, \bibinfo{author}{{Long} WJ}, \bibinfo{author}{{Lu} R},
  \bibinfo{author}{{Luo} Q}, \bibinfo{author}{{Lv} HK}, \bibinfo{author}{{Ma}
  BQ}, \bibinfo{author}{{Ma} LL}, \bibinfo{author}{{Ma} XH},
  \bibinfo{author}{{Mao} JR}, \bibinfo{author}{{Min} Z},
  \bibinfo{author}{{Mitthumsiri} W}, \bibinfo{author}{{Nan} YC},
  \bibinfo{author}{{Ou} ZW}, \bibinfo{author}{{Pang} BY},
  \bibinfo{author}{{Pattarakijwanich} P}, \bibinfo{author}{{Pei} ZY},
  \bibinfo{author}{{Qi} MY}, \bibinfo{author}{{Qi} YQ}, \bibinfo{author}{{Qiao}
  BQ}, \bibinfo{author}{{Qin} JJ}, \bibinfo{author}{{Ruffolo} D},
  \bibinfo{author}{{Saiz} A}, \bibinfo{author}{{Shao} CY},
  \bibinfo{author}{{Shao} L}, \bibinfo{author}{{Shchegolev} O},
  \bibinfo{author}{{Sheng} XD}, \bibinfo{author}{{Song} HC},
  \bibinfo{author}{{Stenkin} YV}, \bibinfo{author}{{Stepanov} V},
  \bibinfo{author}{{Su} Y}, \bibinfo{author}{{Sun} QN}, \bibinfo{author}{{Sun}
  XN}, \bibinfo{author}{{Sun} ZB}, \bibinfo{author}{{Tam} PHT},
  \bibinfo{author}{{Tang} ZB}, \bibinfo{author}{{Tian} WW},
  \bibinfo{author}{{Wang} C}, \bibinfo{author}{{Wang} CB},
  \bibinfo{author}{{Wang} GW}, \bibinfo{author}{{Wang} HG},
  \bibinfo{author}{{Wang} HH}, \bibinfo{author}{{Wang} JC},
  \bibinfo{author}{{Wang} JS}, \bibinfo{author}{{Wang} K},
  \bibinfo{author}{{Wang} LP}, \bibinfo{author}{{Wang} LY},
  \bibinfo{author}{{Wang} PH}, \bibinfo{author}{{Wang} R},
  \bibinfo{author}{{Wang} W}, \bibinfo{author}{{Wang} XG},
  \bibinfo{author}{{Wang} YD}, \bibinfo{author}{{Wang} YJ},
  \bibinfo{author}{{Wang} ZH}, \bibinfo{author}{{Wang} ZX},
  \bibinfo{author}{{Wang} Z}, \bibinfo{author}{{Wei} DM},
  \bibinfo{author}{{Wei} JJ}, \bibinfo{author}{{Wei} YJ},
  \bibinfo{author}{{Wen} T}, \bibinfo{author}{{Wu} CY}, \bibinfo{author}{{Wu}
  HR}, \bibinfo{author}{{Wu} S}, \bibinfo{author}{{Wu} XF},
  \bibinfo{author}{{Wu} YS}, \bibinfo{author}{{Xi} SQ}, \bibinfo{author}{{Xia}
  J}, \bibinfo{author}{{Xia} JJ}, \bibinfo{author}{{Xiang} GM},
  \bibinfo{author}{{Xiao} DX}, \bibinfo{author}{{Xiao} G},
  \bibinfo{author}{{Xin} GG}, \bibinfo{author}{{Xin} YL},
  \bibinfo{author}{{Xing} Y}, \bibinfo{author}{{Xiong} Z},
  \bibinfo{author}{{Xu} DL}, \bibinfo{author}{{Xu} RF}, \bibinfo{author}{{Xu}
  RX}, \bibinfo{author}{{Xue} L}, \bibinfo{author}{{Yan} DH},
  \bibinfo{author}{{Yan} JZ}, \bibinfo{author}{{Yan} T},
  \bibinfo{author}{{Yang} CW}, \bibinfo{author}{{Yang} F},
  \bibinfo{author}{{Yang} FF}, \bibinfo{author}{{Yang} HW},
  \bibinfo{author}{{Yang} JY}, \bibinfo{author}{{Yang} LL},
  \bibinfo{author}{{Yang} MJ}, \bibinfo{author}{{Yang} RZ},
  \bibinfo{author}{{Yang} SB}, \bibinfo{author}{{Yao} YH},
  \bibinfo{author}{{Ye} YM}, \bibinfo{author}{{Yin} LQ}, \bibinfo{author}{{Yin}
  N}, \bibinfo{author}{{You} XH}, \bibinfo{author}{{You} ZY},
  \bibinfo{author}{{Yu} YH}, \bibinfo{author}{{Yuan} Q}, \bibinfo{author}{{Yue}
  H}, \bibinfo{author}{{Zeng} HD}, \bibinfo{author}{{Zeng} TX},
  \bibinfo{author}{{Zeng} W}, \bibinfo{author}{{Zeng} ZK},
  \bibinfo{author}{{Zhang} B}, \bibinfo{author}{{Zhang} BB},
  \bibinfo{author}{{Zhang} F}, \bibinfo{author}{{Zhang} HM},
  \bibinfo{author}{{Zhang} HY}, \bibinfo{author}{{Zhang} JL},
  \bibinfo{author}{{Zhang} LX}, \bibinfo{author}{{Zhang} L},
  \bibinfo{author}{{Zhang} PF}, \bibinfo{author}{{Zhang} PP},
  \bibinfo{author}{{Zhang} R}, \bibinfo{author}{{Zhang} SB},
  \bibinfo{author}{{Zhang} SR}, \bibinfo{author}{{Zhang} SS},
  \bibinfo{author}{{Zhang} X}, \bibinfo{author}{{Zhang} XP},
  \bibinfo{author}{{Zhang} YF}, \bibinfo{author}{{Zhang} Y},
  \bibinfo{author}{{Zhang} Y}, \bibinfo{author}{{Zhao} B},
  \bibinfo{author}{{Zhao} J}, \bibinfo{author}{{Zhao} L},
  \bibinfo{author}{{Zhao} LZ}, \bibinfo{author}{{Zhao} SP},
  \bibinfo{author}{{Zheng} F}, \bibinfo{author}{{Zhou} B},
  \bibinfo{author}{{Zhou} H}, \bibinfo{author}{{Zhou} JN},
  \bibinfo{author}{{Zhou} P}, \bibinfo{author}{{Zhou} R},
  \bibinfo{author}{{Zhou} XX}, \bibinfo{author}{{Zhu} CG},
  \bibinfo{author}{{Zhu} FR}, \bibinfo{author}{{Zhu} H}, \bibinfo{author}{{Zhu}
  KJ} and  \bibinfo{author}{{Zuo} X} (\bibinfo{year}{2023}),
  \bibinfo{month}{Jun.}
\bibinfo{title}{{A tera-electron volt afterglow from a narrow jet in an
  extremely bright gamma-ray burst.}}
\bibinfo{journal}{{\em Science}} \bibinfo{volume}{380}
  (\bibinfo{number}{6652}): \bibinfo{pages}{1390--1396}.
  \bibinfo{doi}{\doi{10.1126/science.adg9328}}.
\eprint{2306.06372}.

\bibtype{Article}%
\bibitem[{Lipunov} et al.(2017)]{Lipunov+17}
\bibinfo{author}{{Lipunov} VM}, \bibinfo{author}{{Gorbovskoy} E},
  \bibinfo{author}{{Kornilov} VG}, \bibinfo{author}{{.~Tyurina} N},
  \bibinfo{author}{{Balanutsa} P}, \bibinfo{author}{{Kuznetsov} A},
  \bibinfo{author}{{Vlasenko} D}, \bibinfo{author}{{Kuvshinov} D},
  \bibinfo{author}{{Gorbunov} I}, \bibinfo{author}{{Buckley} DAH},
  \bibinfo{author}{{Krylov} AV}, \bibinfo{author}{{Podesta} R},
  \bibinfo{author}{{Lopez} C}, \bibinfo{author}{{Podesta} F},
  \bibinfo{author}{{Levato} H}, \bibinfo{author}{{Saffe} C},
  \bibinfo{author}{{Mallamachi} C}, \bibinfo{author}{{Potter} S},
  \bibinfo{author}{{Budnev} NM}, \bibinfo{author}{{Gress} O},
  \bibinfo{author}{{Ishmuhametova} Y}, \bibinfo{author}{{Vladimirov} V},
  \bibinfo{author}{{Zimnukhov} D}, \bibinfo{author}{{Yurkov} V},
  \bibinfo{author}{{Sergienko} Y}, \bibinfo{author}{{Gabovich} A},
  \bibinfo{author}{{Rebolo} R}, \bibinfo{author}{{Serra-Ricart} M},
  \bibinfo{author}{{Israelyan} G}, \bibinfo{author}{{Chazov} V},
  \bibinfo{author}{{Wang} X}, \bibinfo{author}{{Tlatov} A} and
  \bibinfo{author}{{Panchenko} MI} (\bibinfo{year}{2017}),
  \bibinfo{month}{Nov.}
\bibinfo{title}{{MASTER Optical Detection of the First LIGO/Virgo Neutron Star
  Binary Merger GW170817}}.
\bibinfo{journal}{{\em ApJl}} \bibinfo{volume}{850}, \bibinfo{eid}{L1}.
  \bibinfo{doi}{\doi{10.3847/2041-8213/aa92c0}}.
\eprint{1710.05461}.

\bibtype{Article}%
\bibitem[{Littlejohns} et al.(2013)]{Littlejohns+13}
\bibinfo{author}{{Littlejohns} O}, \bibinfo{author}{{Butler} N},
  \bibinfo{author}{{Watson} AM}, \bibinfo{author}{{Kutyrev} A},
  \bibinfo{author}{{Lee} WH}, \bibinfo{author}{{Richer} MG},
  \bibinfo{author}{{Klein} C}, \bibinfo{author}{{Fox} O},
  \bibinfo{author}{{Prochaska} JX}, \bibinfo{author}{{Bloom} J},
  \bibinfo{author}{{Cucchiara} A}, \bibinfo{author}{{Troja} E},
  \bibinfo{author}{{Ramirez-Ruiz} E}, \bibinfo{author}{{de Diego} JA},
  \bibinfo{author}{{Georgiev} L}, \bibinfo{author}{{Gonzalez} J},
  \bibinfo{author}{{Roman-Zuniga} C}, \bibinfo{author}{{Gehrels} N} and
  \bibinfo{author}{{Moseley} H} (\bibinfo{year}{2013}), \bibinfo{month}{Jan.}
\bibinfo{title}{{GRB 131004A: RATIR optical and NIR upper limits.}}
\bibinfo{journal}{{\em GRB Coordinates Network}} \bibinfo{volume}{15312}:
  \bibinfo{pages}{1}.

\bibtype{Article}%
\bibitem[{Liu} et al.(2024)]{liu24}
\bibinfo{author}{{Liu} Y}, \bibinfo{author}{{Sun} H}, \bibinfo{author}{{Xu} D},
  \bibinfo{author}{{Svinkin} DS}, \bibinfo{author}{{Delaunay} J},
  \bibinfo{author}{{Tanvir} NR}, \bibinfo{author}{{Gao} H},
  \bibinfo{author}{{Zhang} C}, \bibinfo{author}{{Chen} Y},
  \bibinfo{author}{{Wu} XF}, \bibinfo{author}{{Zhang} B},
  \bibinfo{author}{{Yuan} W}, \bibinfo{author}{{An} J},
  \bibinfo{author}{{Bruni} G}, \bibinfo{author}{{Frederiks} DD},
  \bibinfo{author}{{Ghirlanda} G}, \bibinfo{author}{{Hu} JW},
  \bibinfo{author}{{Li} A}, \bibinfo{author}{{Li} CK}, \bibinfo{author}{{Li}
  JD}, \bibinfo{author}{{Malesani} DB}, \bibinfo{author}{{Piro} L},
  \bibinfo{author}{{Raman} G}, \bibinfo{author}{{Ricci} R},
  \bibinfo{author}{{Troja} E}, \bibinfo{author}{{Vergani} SD},
  \bibinfo{author}{{Wu} QY}, \bibinfo{author}{{Yang} J},
  \bibinfo{author}{{Zhang} BB}, \bibinfo{author}{{Zhu} ZP},
  \bibinfo{author}{{de Ugarte Postigo} A}, \bibinfo{author}{{Demin} AG},
  \bibinfo{author}{{Dobie} D}, \bibinfo{author}{{Fan} Z}, \bibinfo{author}{{Fu}
  SY}, \bibinfo{author}{{Fynbo} JPU}, \bibinfo{author}{{Geng} JJ},
  \bibinfo{author}{{Gianfagna} G}, \bibinfo{author}{{Hu} YD},
  \bibinfo{author}{{Huang} YF}, \bibinfo{author}{{Jiang} SQ},
  \bibinfo{author}{{Jonker} PG}, \bibinfo{author}{{Julakanti} Y},
  \bibinfo{author}{{Kennea} JA}, \bibinfo{author}{{Kokomov} AA},
  \bibinfo{author}{{Kuulkers} E}, \bibinfo{author}{{Lei} WH},
  \bibinfo{author}{{Leung} JK}, \bibinfo{author}{{Levan} AJ},
  \bibinfo{author}{{Li} DY}, \bibinfo{author}{{Li} Y},
  \bibinfo{author}{{Littlefair} SP}, \bibinfo{author}{{Liu} X},
  \bibinfo{author}{{Lysenko} AL}, \bibinfo{author}{{Ma} YN},
  \bibinfo{author}{{Martin-Carrillo} A}, \bibinfo{author}{{O'Brien} P},
  \bibinfo{author}{{Parsotan} T}, \bibinfo{author}{{Quirola-Vasquez} J},
  \bibinfo{author}{{Ridnaia} AV}, \bibinfo{author}{{Ronchini} S},
  \bibinfo{author}{{Rossi} A}, \bibinfo{author}{{Mata-Sanchez} D},
  \bibinfo{author}{{Schneider} B}, \bibinfo{author}{{Shen} RF},
  \bibinfo{author}{{Thakur} AL}, \bibinfo{author}{{Tohuvavohu} A},
  \bibinfo{author}{{Torres} MAP}, \bibinfo{author}{{Tsvetkova} AE},
  \bibinfo{author}{{Ulanov} MV}, \bibinfo{author}{{Wei} JJ},
  \bibinfo{author}{{Xiao} D}, \bibinfo{author}{{Yin} YHI},
  \bibinfo{author}{{Bai} M}, \bibinfo{author}{{Burwitz} V},
  \bibinfo{author}{{Cai} ZM}, \bibinfo{author}{{Chen} FS},
  \bibinfo{author}{{Chen} HL}, \bibinfo{author}{{Chen} TX},
  \bibinfo{author}{{Chen} W}, \bibinfo{author}{{Chen} YF},
  \bibinfo{author}{{Chen} YH}, \bibinfo{author}{{Cheng} HQ},
  \bibinfo{author}{{Cui} CZ}, \bibinfo{author}{{Cui} WW},
  \bibinfo{author}{{Dai} YF}, \bibinfo{author}{{Dai} ZG},
  \bibinfo{author}{{Eder} J}, \bibinfo{author}{{Fan} DW},
  \bibinfo{author}{{Feldman} C}, \bibinfo{author}{{Feng} H},
  \bibinfo{author}{{Feng} Z}, \bibinfo{author}{{Friedrich} P},
  \bibinfo{author}{{Gao} X}, \bibinfo{author}{{Guan} J}, \bibinfo{author}{{Han}
  DW}, \bibinfo{author}{{Han} J}, \bibinfo{author}{{Hou} DJ},
  \bibinfo{author}{{Hu} HB}, \bibinfo{author}{{Hu} T}, \bibinfo{author}{{Huang}
  MH}, \bibinfo{author}{{Huo} J}, \bibinfo{author}{{Hutchinson} I},
  \bibinfo{author}{{Ji} Z}, \bibinfo{author}{{Jia} SM}, \bibinfo{author}{{Jia}
  ZQ}, \bibinfo{author}{{Jiang} BW}, \bibinfo{author}{{Jin} CC},
  \bibinfo{author}{{Jin} G}, \bibinfo{author}{{Jin} JJ},
  \bibinfo{author}{{Keereman} A}, \bibinfo{author}{{Lerman} H},
  \bibinfo{author}{{Li} JF}, \bibinfo{author}{{Li} LH}, \bibinfo{author}{{Li}
  MS}, \bibinfo{author}{{Li} W}, \bibinfo{author}{{Li} ZD},
  \bibinfo{author}{{Lian} TY}, \bibinfo{author}{{Liang} EW},
  \bibinfo{author}{{Ling} ZX}, \bibinfo{author}{{Liu} CZ},
  \bibinfo{author}{{Liu} HY}, \bibinfo{author}{{Liu} HQ},
  \bibinfo{author}{{Liu} MJ}, \bibinfo{author}{{Liu} YR}, \bibinfo{author}{{Lu}
  FJ}, \bibinfo{author}{{LU} HJ}, \bibinfo{author}{{Luo} LD},
  \bibinfo{author}{{Ma} FL}, \bibinfo{author}{{Ma} J}, \bibinfo{author}{{Mao}
  JR}, \bibinfo{author}{{Mao} X}, \bibinfo{author}{{McHugh} M},
  \bibinfo{author}{{Meidinger} N}, \bibinfo{author}{{Nandra} K},
  \bibinfo{author}{{Osborne} JP}, \bibinfo{author}{{Pan} HW},
  \bibinfo{author}{{Pan} X}, \bibinfo{author}{{Ravasio} ME},
  \bibinfo{author}{{Rau} A}, \bibinfo{author}{{Rea} N},
  \bibinfo{author}{{Rehman} U}, \bibinfo{author}{{Sanders} J},
  \bibinfo{author}{{Santovincenzo} A}, \bibinfo{author}{{Song} LM},
  \bibinfo{author}{{Su} J}, \bibinfo{author}{{Sun} LJ}, \bibinfo{author}{{Sun}
  SL}, \bibinfo{author}{{Sun} XJ}, \bibinfo{author}{{Tan} YY},
  \bibinfo{author}{{Tang} QJ}, \bibinfo{author}{{Tao} YH},
  \bibinfo{author}{{Tong} JZ}, \bibinfo{author}{{Wang} H},
  \bibinfo{author}{{Wang} J}, \bibinfo{author}{{Wang} L},
  \bibinfo{author}{{Wang} WX}, \bibinfo{author}{{Wang} XF},
  \bibinfo{author}{{Wang} XY}, \bibinfo{author}{{Wang} YL},
  \bibinfo{author}{{Wang} YS}, \bibinfo{author}{{Wei} DM},
  \bibinfo{author}{{Willingale} R}, \bibinfo{author}{{Xiong} SL},
  \bibinfo{author}{{Xu} HT}, \bibinfo{author}{{Xu} JJ}, \bibinfo{author}{{Xu}
  XP}, \bibinfo{author}{{Xu} YF}, \bibinfo{author}{{Xu} Z},
  \bibinfo{author}{{Xue} CB}, \bibinfo{author}{{Xue} YL},
  \bibinfo{author}{{Yan} AL}, \bibinfo{author}{{Yang} F},
  \bibinfo{author}{{Yang} HN}, \bibinfo{author}{{Yang} XT},
  \bibinfo{author}{{Yang} YJ}, \bibinfo{author}{{Yu} YW},
  \bibinfo{author}{{Zhang} J}, \bibinfo{author}{{Zhang} M},
  \bibinfo{author}{{Zhang} SN}, \bibinfo{author}{{Zhang} WD},
  \bibinfo{author}{{Zhang} WJ}, \bibinfo{author}{{Zhang} YH},
  \bibinfo{author}{{Zhang} Z}, \bibinfo{author}{{Zhang} Z},
  \bibinfo{author}{{Zhang} ZL}, \bibinfo{author}{{Zhao} DH},
  \bibinfo{author}{{Zhao} HS}, \bibinfo{author}{{Zhao} XF},
  \bibinfo{author}{{Zhao} ZJ}, \bibinfo{author}{{Zhou} LX},
  \bibinfo{author}{{Zhou} YL}, \bibinfo{author}{{Zhu} YX},
  \bibinfo{author}{{Zhu} ZC} and  \bibinfo{author}{{Zuo} XX}
  (\bibinfo{year}{2024}), \bibinfo{month}{Apr.}
\bibinfo{title}{{Soft X-ray prompt emission from a high-redshift gamma-ray
  burst EP240315a}}.
\bibinfo{journal}{{\em arXiv e-prints}} ,
  \bibinfo{eid}{arXiv:2404.16425}\bibinfo{doi}{\doi{10.48550/arXiv.2404.16425}}.
\eprint{2404.16425}.

\bibtype{Article}%
\bibitem[{MAGIC Collaboration} et al.(2019)]{magic19}
\bibinfo{author}{{MAGIC Collaboration}}, \bibinfo{author}{{Acciari} VA},
  \bibinfo{author}{{Ansoldi} S}, \bibinfo{author}{{Antonelli} LA},
  \bibinfo{author}{{Engels} AA}, \bibinfo{author}{{Baack} D},
  \bibinfo{author}{{Babi{\'c}} A}, \bibinfo{author}{{Banerjee} B},
  \bibinfo{author}{{Barres de Almeida} U}, \bibinfo{author}{{Barrio} JA},
  \bibinfo{author}{{Becerra Gonz{\'a}lez} J}, \bibinfo{author}{{Bednarek} W},
  \bibinfo{author}{{Bellizzi} L}, \bibinfo{author}{{Bernardini} E},
  \bibinfo{author}{{Berti} A}, \bibinfo{author}{{Besenrieder} J},
  \bibinfo{author}{{Bhattacharyya} W}, \bibinfo{author}{{Bigongiari} C},
  \bibinfo{author}{{Biland} A}, \bibinfo{author}{{Blanch} O},
  \bibinfo{author}{{Bonnoli} G}, \bibinfo{author}{{Bo{\v{s}}njak} {\v{Z}}},
  \bibinfo{author}{{Busetto} G}, \bibinfo{author}{{Carosi} R},
  \bibinfo{author}{{Ceribella} G}, \bibinfo{author}{{Chai} Y},
  \bibinfo{author}{{Chilingaryan} A}, \bibinfo{author}{{Cikota} S},
  \bibinfo{author}{{Colak} SM}, \bibinfo{author}{{Colin} U},
  \bibinfo{author}{{Colombo} E}, \bibinfo{author}{{Contreras} JL},
  \bibinfo{author}{{Cortina} J}, \bibinfo{author}{{Covino} S},
  \bibinfo{author}{{D'Elia} V}, \bibinfo{author}{{da Vela} P},
  \bibinfo{author}{{Dazzi} F}, \bibinfo{author}{{de Angelis} A},
  \bibinfo{author}{{de Lotto} B}, \bibinfo{author}{{Delfino} M},
  \bibinfo{author}{{Delgado} J}, \bibinfo{author}{{Depaoli} D},
  \bibinfo{author}{{di Pierro} F}, \bibinfo{author}{{di Venere} L},
  \bibinfo{author}{{Do Souto Espi{\~n}eira} E}, \bibinfo{author}{{Dominis
  Prester} D}, \bibinfo{author}{{Donini} A}, \bibinfo{author}{{Dorner} D},
  \bibinfo{author}{{Doro} M}, \bibinfo{author}{{Elsaesser} D},
  \bibinfo{author}{{Fallah Ramazani} V}, \bibinfo{author}{{Fattorini} A},
  \bibinfo{author}{{Ferrara} G}, \bibinfo{author}{{Fidalgo} D},
  \bibinfo{author}{{Foffano} L}, \bibinfo{author}{{Fonseca} MV},
  \bibinfo{author}{{Font} L}, \bibinfo{author}{{Fruck} C},
  \bibinfo{author}{{Fukami} S}, \bibinfo{author}{{Garc{\'\i}a L{\'o}pez} RJ},
  \bibinfo{author}{{Garczarczyk} M}, \bibinfo{author}{{Gasparyan} S},
  \bibinfo{author}{{Gaug} M}, \bibinfo{author}{{Giglietto} N},
  \bibinfo{author}{{Giordano} F}, \bibinfo{author}{{Godinovi{\'c}} N},
  \bibinfo{author}{{Green} D}, \bibinfo{author}{{Guberman} D},
  \bibinfo{author}{{Hadasch} D}, \bibinfo{author}{{Hahn} A},
  \bibinfo{author}{{Herrera} J}, \bibinfo{author}{{Hoang} J},
  \bibinfo{author}{{Hrupec} D}, \bibinfo{author}{{H{\"u}tten} M},
  \bibinfo{author}{{Inada} T}, \bibinfo{author}{{Inoue} S},
  \bibinfo{author}{{Ishio} K}, \bibinfo{author}{{Iwamura} Y},
  \bibinfo{author}{{Jouvin} L}, \bibinfo{author}{{Kerszberg} D},
  \bibinfo{author}{{Kubo} H}, \bibinfo{author}{{Kushida} J},
  \bibinfo{author}{{Lamastra} A}, \bibinfo{author}{{Lelas} D},
  \bibinfo{author}{{Leone} F}, \bibinfo{author}{{Lindfors} E},
  \bibinfo{author}{{Lombardi} S}, \bibinfo{author}{{Longo} F},
  \bibinfo{author}{{L{\'o}pez} M}, \bibinfo{author}{{L{\'o}pez-Coto} R},
  \bibinfo{author}{{L{\'o}pez-Oramas} A}, \bibinfo{author}{{Loporchio} S},
  \bibinfo{author}{{Machado de Oliveira Fraga} B}, \bibinfo{author}{{Maggio}
  C}, \bibinfo{author}{{Majumdar} P}, \bibinfo{author}{{Makariev} M},
  \bibinfo{author}{{Mallamaci} M}, \bibinfo{author}{{Maneva} G},
  \bibinfo{author}{{Manganaro} M}, \bibinfo{author}{{Mannheim} K},
  \bibinfo{author}{{Maraschi} L}, \bibinfo{author}{{Mariotti} M},
  \bibinfo{author}{{Mart{\'\i}nez} M}, \bibinfo{author}{{Mazin} D},
  \bibinfo{author}{{Mi{\'c}anovi{\'c}} S}, \bibinfo{author}{{Miceli} D},
  \bibinfo{author}{{Minev} M}, \bibinfo{author}{{Miranda} JM},
  \bibinfo{author}{{Mirzoyan} R}, \bibinfo{author}{{Molina} E},
  \bibinfo{author}{{Moralejo} A}, \bibinfo{author}{{Morcuende} D},
  \bibinfo{author}{{Moreno} V}, \bibinfo{author}{{Moretti} E},
  \bibinfo{author}{{Munar-Adrover} P}, \bibinfo{author}{{Neustroev} V},
  \bibinfo{author}{{Nigro} C}, \bibinfo{author}{{Nilsson} K},
  \bibinfo{author}{{Ninci} D}, \bibinfo{author}{{Nishijima} K},
  \bibinfo{author}{{Noda} K}, \bibinfo{author}{{Nogu{\'e}s} L},
  \bibinfo{author}{{Nozaki} S}, \bibinfo{author}{{Paiano} S},
  \bibinfo{author}{{Palatiello} M}, \bibinfo{author}{{Paneque} D},
  \bibinfo{author}{{Paoletti} R}, \bibinfo{author}{{Paredes} JM},
  \bibinfo{author}{{Pe{\~n}il} P}, \bibinfo{author}{{Peresano} M},
  \bibinfo{author}{{Persic} M}, \bibinfo{author}{{Moroni} PGP},
  \bibinfo{author}{{Prandini} E}, \bibinfo{author}{{Puljak} I},
  \bibinfo{author}{{Rhode} W}, \bibinfo{author}{{Rib{\'o}} M},
  \bibinfo{author}{{Rico} J}, \bibinfo{author}{{Righi} C},
  \bibinfo{author}{{Rugliancich} A}, \bibinfo{author}{{Saha} L},
  \bibinfo{author}{{Sahakyan} N}, \bibinfo{author}{{Saito} T},
  \bibinfo{author}{{Sakurai} S}, \bibinfo{author}{{Satalecka} K},
  \bibinfo{author}{{Schmidt} K}, \bibinfo{author}{{Schweizer} T},
  \bibinfo{author}{{Sitarek} J}, \bibinfo{author}{{{\v{S}}nidari{\'c}} I},
  \bibinfo{author}{{Sobczynska} D}, \bibinfo{author}{{Somero} A},
  \bibinfo{author}{{Stamerra} A}, \bibinfo{author}{{Strom} D},
  \bibinfo{author}{{Strzys} M}, \bibinfo{author}{{Suda} Y},
  \bibinfo{author}{{Suri{\'c}} T}, \bibinfo{author}{{Takahashi} M},
  \bibinfo{author}{{Tavecchio} F}, \bibinfo{author}{{Temnikov} P},
  \bibinfo{author}{{Terzi{\'c}} T}, \bibinfo{author}{{Teshima} M},
  \bibinfo{author}{{Torres-Alb{\`a}} N}, \bibinfo{author}{{Tosti} L},
  \bibinfo{author}{{Vagelli} V}, \bibinfo{author}{{van Scherpenberg} J},
  \bibinfo{author}{{Vanzo} G}, \bibinfo{author}{{Vazquez Acosta} M},
  \bibinfo{author}{{Vigorito} CF}, \bibinfo{author}{{Vitale} V},
  \bibinfo{author}{{Vovk} I}, \bibinfo{author}{{Will} M},
  \bibinfo{author}{{Zari{\'c}} D}, \bibinfo{author}{{Nava} L},
  \bibinfo{author}{{Veres} P}, \bibinfo{author}{{Bhat} PN},
  \bibinfo{author}{{Briggs} MS}, \bibinfo{author}{{Cleveland} WH},
  \bibinfo{author}{{Hamburg} R}, \bibinfo{author}{{Hui} CM},
  \bibinfo{author}{{Mailyan} B}, \bibinfo{author}{{Preece} RD},
  \bibinfo{author}{{Roberts} OJ}, \bibinfo{author}{{von Kienlin} A},
  \bibinfo{author}{{Wilson-Hodge} CA}, \bibinfo{author}{{Kocevski} D},
  \bibinfo{author}{{Arimoto} M}, \bibinfo{author}{{Tak} D},
  \bibinfo{author}{{Asano} K}, \bibinfo{author}{{Axelsson} M},
  \bibinfo{author}{{Barbiellini} G}, \bibinfo{author}{{Bissaldi} E},
  \bibinfo{author}{{Dirirsa} FF}, \bibinfo{author}{{Gill} R},
  \bibinfo{author}{{Granot} J}, \bibinfo{author}{{McEnery} J},
  \bibinfo{author}{{Omodei} N}, \bibinfo{author}{{Razzaque} S},
  \bibinfo{author}{{Piron} F}, \bibinfo{author}{{Racusin} JL},
  \bibinfo{author}{{Thompson} DJ}, \bibinfo{author}{{Campana} S},
  \bibinfo{author}{{Bernardini} MG}, \bibinfo{author}{{Kuin} NPM},
  \bibinfo{author}{{Siegel} MH}, \bibinfo{author}{{Cenko} SB},
  \bibinfo{author}{{O'Brien} P}, \bibinfo{author}{{Capalbi} M},
  \bibinfo{author}{{Da{\i}} A}, \bibinfo{author}{{de Pasquale} M},
  \bibinfo{author}{{Gropp} J}, \bibinfo{author}{{Klingler} N},
  \bibinfo{author}{{Osborne} JP}, \bibinfo{author}{{Perri} M},
  \bibinfo{author}{{Starling} RLC}, \bibinfo{author}{{Tagliaferri} G},
  \bibinfo{author}{{Tohuvavohu} A}, \bibinfo{author}{{Ursi} A},
  \bibinfo{author}{{Tavani} M}, \bibinfo{author}{{Cardillo} M},
  \bibinfo{author}{{Casentini} C}, \bibinfo{author}{{Piano} G},
  \bibinfo{author}{{Evangelista} Y}, \bibinfo{author}{{Verrecchia} F},
  \bibinfo{author}{{Pittori} C}, \bibinfo{author}{{Lucarelli} F},
  \bibinfo{author}{{Bulgarelli} A}, \bibinfo{author}{{Parmiggiani} N},
  \bibinfo{author}{{Anderson} GE}, \bibinfo{author}{{Anderson} JP},
  \bibinfo{author}{{Bernardi} G}, \bibinfo{author}{{Bolmer} J},
  \bibinfo{author}{{Caballero-Garc{\'\i}a} MD}, \bibinfo{author}{{Carrasco}
  IM}, \bibinfo{author}{{Castell{\'o}n} A}, \bibinfo{author}{{Castro Segura}
  N}, \bibinfo{author}{{Castro-Tirado} AJ}, \bibinfo{author}{{Cherukuri} SV},
  \bibinfo{author}{{Cockeram} AM}, \bibinfo{author}{{D'Avanzo} P},
  \bibinfo{author}{{di Dato} A}, \bibinfo{author}{{Diretse} R},
  \bibinfo{author}{{Fender} RP}, \bibinfo{author}{{Fern{\'a}ndez-Garc{\'\i}a}
  E}, \bibinfo{author}{{Fynbo} JPU}, \bibinfo{author}{{Fruchter} AS},
  \bibinfo{author}{{Greiner} J}, \bibinfo{author}{{Gromadzki} M},
  \bibinfo{author}{{Heintz} KE}, \bibinfo{author}{{Heywood} I},
  \bibinfo{author}{{van der Horst} AJ}, \bibinfo{author}{{Hu} YD},
  \bibinfo{author}{{Inserra} C}, \bibinfo{author}{{Izzo} L},
  \bibinfo{author}{{Jaiswal} V}, \bibinfo{author}{{Jakobsson} P},
  \bibinfo{author}{{Japelj} J}, \bibinfo{author}{{Kankare} E},
  \bibinfo{author}{{Kann} DA}, \bibinfo{author}{{Kouveliotou} C},
  \bibinfo{author}{{Klose} S}, \bibinfo{author}{{Levan} AJ},
  \bibinfo{author}{{Li} XY}, \bibinfo{author}{{Lotti} S},
  \bibinfo{author}{{Maguire} K}, \bibinfo{author}{{Malesani} DB},
  \bibinfo{author}{{Manulis} I}, \bibinfo{author}{{Marongiu} M},
  \bibinfo{author}{{Martin} S}, \bibinfo{author}{{Melandri} A},
  \bibinfo{author}{{Micha{\l}owski} MJ}, \bibinfo{author}{{Miller-Jones} JCA},
  \bibinfo{author}{{Misra} K}, \bibinfo{author}{{Moin} A},
  \bibinfo{author}{{Mooley} KP}, \bibinfo{author}{{Nasri} S},
  \bibinfo{author}{{Nicholl} M}, \bibinfo{author}{{Noschese} A},
  \bibinfo{author}{{Novara} G}, \bibinfo{author}{{Pandey} SB},
  \bibinfo{author}{{Peretti} E}, \bibinfo{author}{{P{\'e}rez Del Pulgar} CJ},
  \bibinfo{author}{{P{\'e}rez-Torres} MA}, \bibinfo{author}{{Perley} DA},
  \bibinfo{author}{{Piro} L}, \bibinfo{author}{{Ragosta} F},
  \bibinfo{author}{{Resmi} L}, \bibinfo{author}{{Ricci} R},
  \bibinfo{author}{{Rossi} A}, \bibinfo{author}{{S{\'a}nchez-Ram{\'\i}rez} R},
  \bibinfo{author}{{Selsing} J}, \bibinfo{author}{{Schulze} S},
  \bibinfo{author}{{Smartt} SJ}, \bibinfo{author}{{Smith} IA},
  \bibinfo{author}{{Sokolov} VV}, \bibinfo{author}{{Stevens} J},
  \bibinfo{author}{{Tanvir} NR}, \bibinfo{author}{{Th{\"o}ne} CC},
  \bibinfo{author}{{Tiengo} A}, \bibinfo{author}{{Tremou} E},
  \bibinfo{author}{{Troja} E}, \bibinfo{author}{{de Ugarte Postigo} A},
  \bibinfo{author}{{Valeev} AF}, \bibinfo{author}{{Vergani} SD},
  \bibinfo{author}{{Wieringa} M}, \bibinfo{author}{{Woudt} PA},
  \bibinfo{author}{{Xu} D}, \bibinfo{author}{{Yaron} O} and
  \bibinfo{author}{{Young} DR} (\bibinfo{year}{2019}), \bibinfo{month}{Nov.}
\bibinfo{title}{{Observation of inverse Compton emission from a long
  {\ensuremath{\gamma}}-ray burst}}.
\bibinfo{journal}{{\em \nat}} \bibinfo{volume}{575} (\bibinfo{number}{7783}):
  \bibinfo{pages}{459--463}. \bibinfo{doi}{\doi{10.1038/s41586-019-1754-6}}.
\eprint{2006.07251}.

\bibtype{Article}%
\bibitem[{Malesani} et al.(2023)]{malesani23}
\bibinfo{author}{{Malesani} DB}, \bibinfo{author}{{Levan} AJ},
  \bibinfo{author}{{Izzo} L}, \bibinfo{author}{{de Ugarte Postigo} A},
  \bibinfo{author}{{Ghirlanda} G}, \bibinfo{author}{{Heintz} KE},
  \bibinfo{author}{{Kann} DA}, \bibinfo{author}{{Lamb} GP},
  \bibinfo{author}{{Palmerio} J}, \bibinfo{author}{{Salafia} OS},
  \bibinfo{author}{{Salvaterra} R}, \bibinfo{author}{{Tanvir} NR},
  \bibinfo{author}{{Ag{\"u}{\'\i} Fern{\'a}ndez} JF},
  \bibinfo{author}{{Campana} S}, \bibinfo{author}{{Chrimes} AA},
  \bibinfo{author}{{D'Avanzo} P}, \bibinfo{author}{{D'Elia} V},
  \bibinfo{author}{{Della Valle} M}, \bibinfo{author}{{De Pasquale} M},
  \bibinfo{author}{{Fynbo} JPU}, \bibinfo{author}{{Gaspari} N},
  \bibinfo{author}{{Gompertz} BP}, \bibinfo{author}{{Hartmann} DH},
  \bibinfo{author}{{Hjorth} J}, \bibinfo{author}{{Jakobsson} P},
  \bibinfo{author}{{Palazzi} E}, \bibinfo{author}{{Pian} E},
  \bibinfo{author}{{Pugliese} G}, \bibinfo{author}{{Ravasio} ME},
  \bibinfo{author}{{Rossi} A}, \bibinfo{author}{{Saccardi} A},
  \bibinfo{author}{{Schady} P}, \bibinfo{author}{{Schneider} B},
  \bibinfo{author}{{Sollerman} J}, \bibinfo{author}{{Starling} RLC},
  \bibinfo{author}{{Th{\"o}ne} CC}, \bibinfo{author}{{van der Horst} AJ},
  \bibinfo{author}{{Vergani} SD}, \bibinfo{author}{{Watson} D},
  \bibinfo{author}{{Wiersema} K}, \bibinfo{author}{{Xu} D} and
  \bibinfo{author}{{Zafar} T} (\bibinfo{year}{2023}), \bibinfo{month}{Feb.}
\bibinfo{title}{{The brightest GRB ever detected: GRB 221009A as a highly
  luminous event at z = 0.151}}.
\bibinfo{journal}{{\em arXiv e-prints}} ,
  \bibinfo{eid}{arXiv:2302.07891}\bibinfo{doi}{\doi{10.48550/arXiv.2302.07891}}.
\eprint{2302.07891}.

\bibtype{Article}%
\bibitem[{Mazets} et al.(1979)]{mazets79}
\bibinfo{author}{{Mazets} EP}, \bibinfo{author}{{Golentskii} SV},
  \bibinfo{author}{{Ilinskii} VN}, \bibinfo{author}{{Aptekar} RL} and
  \bibinfo{author}{{Guryan} IA} (\bibinfo{year}{1979}), \bibinfo{month}{Dec.}
\bibinfo{title}{{Observations of a flaring X-ray pulsar in Dorado}}.
\bibinfo{journal}{{\em \nat}} \bibinfo{volume}{282} (\bibinfo{number}{5739}):
  \bibinfo{pages}{587--589}. \bibinfo{doi}{\doi{10.1038/282587a0}}.

\bibtype{Article}%
\bibitem[{Meegan} et al.(1992)]{meegan92}
\bibinfo{author}{{Meegan} CA}, \bibinfo{author}{{Fishman} GJ},
  \bibinfo{author}{{Wilson} RB}, \bibinfo{author}{{Paciesas} WS},
  \bibinfo{author}{{Pendleton} GN}, \bibinfo{author}{{Horack} JM},
  \bibinfo{author}{{Brock} MN} and  \bibinfo{author}{{Kouveliotou} C}
  (\bibinfo{year}{1992}), \bibinfo{month}{Jan.}
\bibinfo{title}{{Spatial distribution of {\ensuremath{\gamma}}-ray bursts
  observed by BATSE}}.
\bibinfo{journal}{{\em \nat}} \bibinfo{volume}{355} (\bibinfo{number}{6356}):
  \bibinfo{pages}{143--145}. \bibinfo{doi}{\doi{10.1038/355143a0}}.

\bibtype{Article}%
\bibitem[{Meszaros} and {Rees}(1993)]{mez93}
\bibinfo{author}{{Meszaros} P} and  \bibinfo{author}{{Rees} MJ}
  (\bibinfo{year}{1993}), \bibinfo{month}{Mar.}
\bibinfo{title}{{Relativistic Fireballs and Their Impact on External Matter:
  Models for Cosmological Gamma-Ray Bursts}}.
\bibinfo{journal}{{\em ApJ}} \bibinfo{volume}{405}: \bibinfo{pages}{278}.
  \bibinfo{doi}{\doi{10.1086/172360}}.

\bibtype{Article}%
\bibitem[{Metzger} et al.(1997)]{metzger97}
\bibinfo{author}{{Metzger} MR}, \bibinfo{author}{{Djorgovski} SG},
  \bibinfo{author}{{Kulkarni} SR}, \bibinfo{author}{{Steidel} CC},
  \bibinfo{author}{{Adelberger} KL}, \bibinfo{author}{{Frail} DA},
  \bibinfo{author}{{Costa} E} and  \bibinfo{author}{{Frontera} F}
  (\bibinfo{year}{1997}), \bibinfo{month}{Jun.}
\bibinfo{title}{{Spectral constraints on the redshift of the optical
  counterpart to the {$\gamma$}-ray burst of 8 May 1997}}.
\bibinfo{journal}{{\em \nat}} \bibinfo{volume}{387}: \bibinfo{pages}{878--880}.
  \bibinfo{doi}{\doi{10.1038/43132}}.

\bibtype{Article}%
\bibitem[{Metzger} et al.(2008)]{Metzger+08_magnetarEE}
\bibinfo{author}{{Metzger} BD}, \bibinfo{author}{{Quataert} E} and
  \bibinfo{author}{{Thompson} TA} (\bibinfo{year}{2008}), \bibinfo{month}{Apr.}
\bibinfo{title}{{Short-duration gamma-ray bursts with extended emission from
  protomagnetar spin-down}}.
\bibinfo{journal}{{\em MNRAS}} \bibinfo{volume}{385} (\bibinfo{number}{3}):
  \bibinfo{pages}{1455--1460}.
  \bibinfo{doi}{\doi{10.1111/j.1365-2966.2008.12923.x}}.
\eprint{0712.1233}.

\bibtype{Article}%
\bibitem[{Metzger} et al.(2015)]{Metzger+15}
\bibinfo{author}{{Metzger} BD}, \bibinfo{author}{{Bauswein} A},
  \bibinfo{author}{{Goriely} S} and  \bibinfo{author}{{Kasen} D}
  (\bibinfo{year}{2015}), \bibinfo{month}{Jan.}
\bibinfo{title}{{Neutron-powered precursors of kilonovae}}.
\bibinfo{journal}{{\em MNRAS}} \bibinfo{volume}{446} (\bibinfo{number}{1}):
  \bibinfo{pages}{1115--1120}. \bibinfo{doi}{\doi{10.1093/mnras/stu2225}}.
\eprint{1409.0544}.

\bibtype{Article}%
\bibitem[{Modjaz} et al.(2016)]{modjaz18}
\bibinfo{author}{{Modjaz} M}, \bibinfo{author}{{Liu} YQ},
  \bibinfo{author}{{Bianco} FB} and  \bibinfo{author}{{Graur} O}
  (\bibinfo{year}{2016}), \bibinfo{month}{Dec.}
\bibinfo{title}{{The Spectral SN-GRB Connection: Systematic Spectral
  Comparisons between Type Ic Supernovae and Broad-lined Type Ic Supernovae
  with and without Gamma-Ray Bursts}}.
\bibinfo{journal}{{\em ApJ}} \bibinfo{volume}{832} (\bibinfo{number}{2}),
  \bibinfo{eid}{108}. \bibinfo{doi}{\doi{10.3847/0004-637X/832/2/108}}.
\eprint{1509.07124}.

\bibtype{Article}%
\bibitem[{Mooley} et al.(2018)]{Mooley+18}
\bibinfo{author}{{Mooley} KP}, \bibinfo{author}{{Deller} AT},
  \bibinfo{author}{{Gottlieb} O}, \bibinfo{author}{{Nakar} E},
  \bibinfo{author}{{Hallinan} G}, \bibinfo{author}{{Bourke} S},
  \bibinfo{author}{{Frail} DA}, \bibinfo{author}{{Horesh} A},
  \bibinfo{author}{{Corsi} A} and  \bibinfo{author}{{Hotokezaka} K}
  (\bibinfo{year}{2018}), \bibinfo{month}{Sep.}
\bibinfo{title}{{Superluminal motion of a relativistic jet in the neutron-star
  merger GW170817}}.
\bibinfo{journal}{{\em \nat}} \bibinfo{volume}{561} (\bibinfo{number}{7723}):
  \bibinfo{pages}{355--359}. \bibinfo{doi}{\doi{10.1038/s41586-018-0486-3}}.
\eprint{1806.09693}.

\bibtype{Article}%
\bibitem[{Osten} et al.(2010)]{osten10}
\bibinfo{author}{{Osten} RA}, \bibinfo{author}{{Godet} O},
  \bibinfo{author}{{Drake} S}, \bibinfo{author}{{Tueller} J},
  \bibinfo{author}{{Cummings} J}, \bibinfo{author}{{Krimm} H},
  \bibinfo{author}{{Pye} J}, \bibinfo{author}{{Pal'shin} V},
  \bibinfo{author}{{Golenetskii} S}, \bibinfo{author}{{Reale} F},
  \bibinfo{author}{{Oates} SR}, \bibinfo{author}{{Page} MJ} and
  \bibinfo{author}{{Melandri} A} (\bibinfo{year}{2010}), \bibinfo{month}{Sep.}
\bibinfo{title}{{The Mouse That Roared: A Superflare from the dMe Flare Star EV
  Lac Detected by Swift and Konus-Wind}}.
\bibinfo{journal}{{\em ApJ}} \bibinfo{volume}{721} (\bibinfo{number}{1}):
  \bibinfo{pages}{785--801}. \bibinfo{doi}{\doi{10.1088/0004-637X/721/1/785}}.
\eprint{1007.5300}.

\bibtype{Article}%
\bibitem[{Perley} et al.(2014)]{perley14}
\bibinfo{author}{{Perley} DA}, \bibinfo{author}{{Cenko} SB},
  \bibinfo{author}{{Corsi} A}, \bibinfo{author}{{Tanvir} NR},
  \bibinfo{author}{{Levan} AJ}, \bibinfo{author}{{Kann} DA},
  \bibinfo{author}{{Sonbas} E}, \bibinfo{author}{{Wiersema} K},
  \bibinfo{author}{{Zheng} W}, \bibinfo{author}{{Zhao} XH},
  \bibinfo{author}{{Bai} JM}, \bibinfo{author}{{Bremer} M},
  \bibinfo{author}{{Castro-Tirado} AJ}, \bibinfo{author}{{Chang} L},
  \bibinfo{author}{{Clubb} KI}, \bibinfo{author}{{Frail} D},
  \bibinfo{author}{{Fruchter} A}, \bibinfo{author}{{G{\"o}{\u{g}}{\"u}{\c{s}}}
  E}, \bibinfo{author}{{Greiner} J}, \bibinfo{author}{{G{\"u}ver} T},
  \bibinfo{author}{{Horesh} A}, \bibinfo{author}{{Filippenko} AV},
  \bibinfo{author}{{Klose} S}, \bibinfo{author}{{Mao} J},
  \bibinfo{author}{{Morgan} AN}, \bibinfo{author}{{Pozanenko} AS},
  \bibinfo{author}{{Schmidl} S}, \bibinfo{author}{{Stecklum} B},
  \bibinfo{author}{{Tanga} M}, \bibinfo{author}{{Volnova} AA},
  \bibinfo{author}{{Volvach} AE}, \bibinfo{author}{{Wang} JG},
  \bibinfo{author}{{Winters} JM} and  \bibinfo{author}{{Xin} YX}
  (\bibinfo{year}{2014}), \bibinfo{month}{Jan.}
\bibinfo{title}{{The Afterglow of GRB 130427A from 1 to {}10$^{16}$ GHz}}.
\bibinfo{journal}{{\em ApJ}} \bibinfo{volume}{781} (\bibinfo{number}{1}),
  \bibinfo{eid}{37}. \bibinfo{doi}{\doi{10.1088/0004-637X/781/1/37}}.
\eprint{1307.4401}.

\bibtype{Article}%
\bibitem[{Perley} et al.(2016)]{perley16}
\bibinfo{author}{{Perley} DA}, \bibinfo{author}{{Tanvir} NR},
  \bibinfo{author}{{Hjorth} J}, \bibinfo{author}{{Laskar} T},
  \bibinfo{author}{{Berger} E}, \bibinfo{author}{{Chary} R},
  \bibinfo{author}{{de Ugarte Postigo} A}, \bibinfo{author}{{Fynbo} JPU},
  \bibinfo{author}{{Kr{\"u}hler} T}, \bibinfo{author}{{Levan} AJ},
  \bibinfo{author}{{Micha{\l}owski} MJ} and  \bibinfo{author}{{Schulze} S}
  (\bibinfo{year}{2016}), \bibinfo{month}{Jan.}
\bibinfo{title}{{The Swift GRB Host Galaxy Legacy Survey. II. Rest-frame
  Near-IR Luminosity Distribution and Evidence for a Near-solar Metallicity
  Threshold}}.
\bibinfo{journal}{{\em ApJ}} \bibinfo{volume}{817}, \bibinfo{eid}{8}.
  \bibinfo{doi}{\doi{10.3847/0004-637X/817/1/8}}.
\eprint{1504.02479}.

\bibtype{Article}%
\bibitem[{Pian} et al.(2017)]{Pian+17}
\bibinfo{author}{{Pian} E}, \bibinfo{author}{{D'Avanzo} P},
  \bibinfo{author}{{Benetti} S}, \bibinfo{author}{{Branchesi} M},
  \bibinfo{author}{{Brocato} E}, \bibinfo{author}{{Campana} S},
  \bibinfo{author}{{Cappellaro} E}, \bibinfo{author}{{Covino} S},
  \bibinfo{author}{{D'Elia} V}, \bibinfo{author}{{Fynbo} JPU},
  \bibinfo{author}{{Getman} F}, \bibinfo{author}{{Ghirlanda} G},
  \bibinfo{author}{{Ghisellini} G}, \bibinfo{author}{{Grado} A},
  \bibinfo{author}{{Greco} G}, \bibinfo{author}{{Hjorth} J},
  \bibinfo{author}{{Kouveliotou} C}, \bibinfo{author}{{Levan} A},
  \bibinfo{author}{{Limatola} L}, \bibinfo{author}{{Malesani} D},
  \bibinfo{author}{{Mazzali} PA}, \bibinfo{author}{{Melandri} A},
  \bibinfo{author}{{M{\o}ller} P}, \bibinfo{author}{{Nicastro} L},
  \bibinfo{author}{{Palazzi} E}, \bibinfo{author}{{Piranomonte} S},
  \bibinfo{author}{{Rossi} A}, \bibinfo{author}{{Salafia} OS},
  \bibinfo{author}{{Selsing} J}, \bibinfo{author}{{Stratta} G},
  \bibinfo{author}{{Tanaka} M}, \bibinfo{author}{{Tanvir} NR},
  \bibinfo{author}{{Tomasella} L}, \bibinfo{author}{{Watson} D},
  \bibinfo{author}{{Yang} S}, \bibinfo{author}{{Amati} L},
  \bibinfo{author}{{Antonelli} LA}, \bibinfo{author}{{Ascenzi} S},
  \bibinfo{author}{{Bernardini} MG}, \bibinfo{author}{{Bo{\"e}r} M},
  \bibinfo{author}{{Bufano} F}, \bibinfo{author}{{Bulgarelli} A},
  \bibinfo{author}{{Capaccioli} M}, \bibinfo{author}{{Casella} P},
  \bibinfo{author}{{Castro-Tirado} AJ}, \bibinfo{author}{{Chassande-Mottin} E},
  \bibinfo{author}{{Ciolfi} R}, \bibinfo{author}{{Copperwheat} CM},
  \bibinfo{author}{{Dadina} M}, \bibinfo{author}{{De Cesare} G},
  \bibinfo{author}{{di Paola} A}, \bibinfo{author}{{Fan} YZ},
  \bibinfo{author}{{Gendre} B}, \bibinfo{author}{{Giuffrida} G},
  \bibinfo{author}{{Giunta} A}, \bibinfo{author}{{Hunt} LK},
  \bibinfo{author}{{Israel} GL}, \bibinfo{author}{{Jin} ZP},
  \bibinfo{author}{{Kasliwal} MM}, \bibinfo{author}{{Klose} S},
  \bibinfo{author}{{Lisi} M}, \bibinfo{author}{{Longo} F},
  \bibinfo{author}{{Maiorano} E}, \bibinfo{author}{{Mapelli} M},
  \bibinfo{author}{{Masetti} N}, \bibinfo{author}{{Nava} L},
  \bibinfo{author}{{Patricelli} B}, \bibinfo{author}{{Perley} D},
  \bibinfo{author}{{Pescalli} A}, \bibinfo{author}{{Piran} T},
  \bibinfo{author}{{Possenti} A}, \bibinfo{author}{{Pulone} L},
  \bibinfo{author}{{Razzano} M}, \bibinfo{author}{{Salvaterra} R},
  \bibinfo{author}{{Schipani} P}, \bibinfo{author}{{Spera} M},
  \bibinfo{author}{{Stamerra} A}, \bibinfo{author}{{Stella} L},
  \bibinfo{author}{{Tagliaferri} G}, \bibinfo{author}{{Testa} V},
  \bibinfo{author}{{Troja} E}, \bibinfo{author}{{Turatto} M},
  \bibinfo{author}{{Vergani} SD} and  \bibinfo{author}{{Vergani} D}
  (\bibinfo{year}{2017}), \bibinfo{month}{Nov.}
\bibinfo{title}{{Spectroscopic identification of r-process nucleosynthesis in a
  double neutron-star merger}}.
\bibinfo{journal}{{\em \nat}} \bibinfo{volume}{551}: \bibinfo{pages}{67--70}.
  \bibinfo{doi}{\doi{10.1038/nature24298}}.
\eprint{1710.05858}.

\bibtype{Article}%
\bibitem[{Quirola-V{\'a}squez} et al.(2023)]{qv23}
\bibinfo{author}{{Quirola-V{\'a}squez} J}, \bibinfo{author}{{Bauer} FE},
  \bibinfo{author}{{Jonker} PG}, \bibinfo{author}{{Brandt} WN},
  \bibinfo{author}{{Yang} G}, \bibinfo{author}{{Levan} AJ},
  \bibinfo{author}{{Xue} YQ}, \bibinfo{author}{{Eappachen} D},
  \bibinfo{author}{{Camacho} E}, \bibinfo{author}{{Ravasio} ME},
  \bibinfo{author}{{Zheng} XC} and  \bibinfo{author}{{Luo} B}
  (\bibinfo{year}{2023}), \bibinfo{month}{Jul.}
\bibinfo{title}{{Extragalactic fast X-ray transient candidates discovered by
  Chandra (2014-2022)}}.
\bibinfo{journal}{{\em A\&A}} \bibinfo{volume}{675}, \bibinfo{eid}{A44}.
  \bibinfo{doi}{\doi{10.1051/0004-6361/202345912}}.
\eprint{2304.13795}.

\bibtype{Article}%
\bibitem[{Racusin} et al.(2008)]{racusin08}
\bibinfo{author}{{Racusin} JL}, \bibinfo{author}{{Karpov} SV},
  \bibinfo{author}{{Sokolowski} M}, \bibinfo{author}{{Granot} J},
  \bibinfo{author}{{Wu} XF}, \bibinfo{author}{{Pal'Shin} V},
  \bibinfo{author}{{Covino} S}, \bibinfo{author}{{van der Horst} AJ},
  \bibinfo{author}{{Oates} SR}, \bibinfo{author}{{Schady} P},
  \bibinfo{author}{{Smith} RJ}, \bibinfo{author}{{Cummings} J},
  \bibinfo{author}{{Starling} RLC}, \bibinfo{author}{{Piotrowski} LW},
  \bibinfo{author}{{Zhang} B}, \bibinfo{author}{{Evans} PA},
  \bibinfo{author}{{Holland} ST}, \bibinfo{author}{{Malek} K},
  \bibinfo{author}{{Page} MT}, \bibinfo{author}{{Vetere} L},
  \bibinfo{author}{{Margutti} R}, \bibinfo{author}{{Guidorzi} C},
  \bibinfo{author}{{Kamble} AP}, \bibinfo{author}{{Curran} PA},
  \bibinfo{author}{{Beardmore} A}, \bibinfo{author}{{Kouveliotou} C},
  \bibinfo{author}{{Mankiewicz} L}, \bibinfo{author}{{Melandri} A},
  \bibinfo{author}{{O'Brien} PT}, \bibinfo{author}{{Page} KL},
  \bibinfo{author}{{Piran} T}, \bibinfo{author}{{Tanvir} NR},
  \bibinfo{author}{{Wrochna} G}, \bibinfo{author}{{Aptekar} RL},
  \bibinfo{author}{{Barthelmy} S}, \bibinfo{author}{{Bartolini} C},
  \bibinfo{author}{{Beskin} GM}, \bibinfo{author}{{Bondar} S},
  \bibinfo{author}{{Bremer} M}, \bibinfo{author}{{Campana} S},
  \bibinfo{author}{{Castro-Tirado} A}, \bibinfo{author}{{Cucchiara} A},
  \bibinfo{author}{{Cwiok} M}, \bibinfo{author}{{D'Avanzo} P},
  \bibinfo{author}{{D'Elia} V}, \bibinfo{author}{{Della Valle} M},
  \bibinfo{author}{{de Ugarte Postigo} A}, \bibinfo{author}{{Dominik} W},
  \bibinfo{author}{{Falcone} A}, \bibinfo{author}{{Fiore} F},
  \bibinfo{author}{{Fox} DB}, \bibinfo{author}{{Frederiks} DD},
  \bibinfo{author}{{Fruchter} AS}, \bibinfo{author}{{Fugazza} D},
  \bibinfo{author}{{Garrett} MA}, \bibinfo{author}{{Gehrels} N},
  \bibinfo{author}{{Golenetskii} S}, \bibinfo{author}{{Gomboc} A},
  \bibinfo{author}{{Gorosabel} J}, \bibinfo{author}{{Greco} G},
  \bibinfo{author}{{Guarnieri} A}, \bibinfo{author}{{Immler} S},
  \bibinfo{author}{{Jelinek} M}, \bibinfo{author}{{Kasprowicz} G},
  \bibinfo{author}{{La Parola} V}, \bibinfo{author}{{Levan} AJ},
  \bibinfo{author}{{Mangano} V}, \bibinfo{author}{{Mazets} EP},
  \bibinfo{author}{{Molinari} E}, \bibinfo{author}{{Moretti} A},
  \bibinfo{author}{{Nawrocki} K}, \bibinfo{author}{{Oleynik} PP},
  \bibinfo{author}{{Osborne} JP}, \bibinfo{author}{{Pagani} C},
  \bibinfo{author}{{Pandey} SB}, \bibinfo{author}{{Paragi} Z},
  \bibinfo{author}{{Perri} M}, \bibinfo{author}{{Piccioni} A},
  \bibinfo{author}{{Ramirez-Ruiz} E}, \bibinfo{author}{{Roming} PWA},
  \bibinfo{author}{{Steele} IA}, \bibinfo{author}{{Strom} RG},
  \bibinfo{author}{{Testa} V}, \bibinfo{author}{{Tosti} G},
  \bibinfo{author}{{Ulanov} MV}, \bibinfo{author}{{Wiersema} K},
  \bibinfo{author}{{Wijers} RAMJ}, \bibinfo{author}{{Winters} JM},
  \bibinfo{author}{{Zarnecki} AF}, \bibinfo{author}{{Zerbi} F},
  \bibinfo{author}{{M{\'e}sz{\'a}ros} P}, \bibinfo{author}{{Chincarini} G} and
  \bibinfo{author}{{Burrows} DN} (\bibinfo{year}{2008}), \bibinfo{month}{Sep.}
\bibinfo{title}{{Broadband observations of the naked-eye {$\gamma$}-ray burst
  GRB080319B}}.
\bibinfo{journal}{{\em \nat}} \bibinfo{volume}{455}: \bibinfo{pages}{183--188}.
  \bibinfo{doi}{\doi{10.1038/nature07270}}.
\eprint{0805.1557}.

\bibtype{Article}%
\bibitem[{Rastinejad} et al.(2022)]{rastinejad22}
\bibinfo{author}{{Rastinejad} JC}, \bibinfo{author}{{Gompertz} BP},
  \bibinfo{author}{{Levan} AJ}, \bibinfo{author}{{Fong} W},
  \bibinfo{author}{{Nicholl} M}, \bibinfo{author}{{Lamb} GP},
  \bibinfo{author}{{Malesani} DB}, \bibinfo{author}{{Nugent} AE},
  \bibinfo{author}{{Oates} SR}, \bibinfo{author}{{Tanvir} NR},
  \bibinfo{author}{{de Ugarte Postigo} A}, \bibinfo{author}{{Kilpatrick} CD},
  \bibinfo{author}{{Moore} CJ}, \bibinfo{author}{{Metzger} BD},
  \bibinfo{author}{{Ravasio} ME}, \bibinfo{author}{{Rossi} A},
  \bibinfo{author}{{Schroeder} G}, \bibinfo{author}{{Jencson} J},
  \bibinfo{author}{{Sand} DJ}, \bibinfo{author}{{Smith} N},
  \bibinfo{author}{{Ag{\"u}{\'\i} Fern{\'a}ndez} JF}, \bibinfo{author}{{Berger}
  E}, \bibinfo{author}{{Blanchard} PK}, \bibinfo{author}{{Chornock} R},
  \bibinfo{author}{{Cobb} BE}, \bibinfo{author}{{De Pasquale} M},
  \bibinfo{author}{{Fynbo} JPU}, \bibinfo{author}{{Izzo} L},
  \bibinfo{author}{{Kann} DA}, \bibinfo{author}{{Laskar} T},
  \bibinfo{author}{{Marini} E}, \bibinfo{author}{{Paterson} K},
  \bibinfo{author}{{Rouco Escorial} A}, \bibinfo{author}{{Sears} HM} and
  \bibinfo{author}{{Th{\"o}ne} CC} (\bibinfo{year}{2022}),
  \bibinfo{month}{Apr.}
\bibinfo{title}{{A Kilonova Following a Long-Duration Gamma-Ray Burst at 350
  Mpc}}.
\bibinfo{journal}{{\em arXiv e-prints}} ,
  \bibinfo{eid}{arXiv:2204.10864}\eprint{2204.10864}.

\bibtype{Article}%
\bibitem[{Ravasio} et al.(2019)]{ravasio19}
\bibinfo{author}{{Ravasio} ME}, \bibinfo{author}{{Ghirlanda} G},
  \bibinfo{author}{{Nava} L} and  \bibinfo{author}{{Ghisellini} G}
  (\bibinfo{year}{2019}), \bibinfo{month}{May}.
\bibinfo{title}{{Evidence of two spectral breaks in the prompt emission of
  gamma-ray bursts}}.
\bibinfo{journal}{{\em A\&A}} \bibinfo{volume}{625}, \bibinfo{eid}{A60}.
  \bibinfo{doi}{\doi{10.1051/0004-6361/201834987}}.
\eprint{1903.02555}.

\bibtype{Article}%
\bibitem[{Ravasio} et al.(2024)]{ravasio24}
\bibinfo{author}{{Ravasio} ME}, \bibinfo{author}{{Salafia} OS},
  \bibinfo{author}{{Oganesyan} G}, \bibinfo{author}{{Mei} A},
  \bibinfo{author}{{Ghirlanda} G}, \bibinfo{author}{{Ascenzi} S},
  \bibinfo{author}{{Banerjee} B}, \bibinfo{author}{{Macera} S},
  \bibinfo{author}{{Branchesi} M}, \bibinfo{author}{{Jonker} PG},
  \bibinfo{author}{{Levan} AJ}, \bibinfo{author}{{Malesani} DB},
  \bibinfo{author}{{Mulrey} KB}, \bibinfo{author}{{Giuliani} A},
  \bibinfo{author}{{Celotti} A} and  \bibinfo{author}{{Ghisellini} G}
  (\bibinfo{year}{2024}), \bibinfo{month}{Jul.}
\bibinfo{title}{{A mega{\textendash}electron volt emission line in the spectrum
  of a gamma-ray burst}}.
\bibinfo{journal}{{\em Science}} \bibinfo{volume}{385}
  (\bibinfo{number}{6707}): \bibinfo{pages}{452--455}.
  \bibinfo{doi}{\doi{10.1126/science.adj3638}}.
\eprint{2303.16223}.

\bibtype{Article}%
\bibitem[{Rossi} et al.(2022)]{rossi21}
\bibinfo{author}{{Rossi} A}, \bibinfo{author}{{Rothberg} B},
  \bibinfo{author}{{Palazzi} E}, \bibinfo{author}{{Kann} DA},
  \bibinfo{author}{{D'Avanzo} P}, \bibinfo{author}{{Amati} L},
  \bibinfo{author}{{Klose} S}, \bibinfo{author}{{Perego} A},
  \bibinfo{author}{{Pian} E}, \bibinfo{author}{{Guidorzi} C},
  \bibinfo{author}{{Pozanenko} AS}, \bibinfo{author}{{Savaglio} S},
  \bibinfo{author}{{Stratta} G}, \bibinfo{author}{{Agapito} G},
  \bibinfo{author}{{Covino} S}, \bibinfo{author}{{Cusano} F},
  \bibinfo{author}{{D'Elia} V}, \bibinfo{author}{{De Pasquale} M},
  \bibinfo{author}{{Della Valle} M}, \bibinfo{author}{{Kuhn} O},
  \bibinfo{author}{{Izzo} L}, \bibinfo{author}{{Loffredo} E},
  \bibinfo{author}{{Masetti} N}, \bibinfo{author}{{Melandri} A},
  \bibinfo{author}{{Minaev} PY}, \bibinfo{author}{{Guelbenzu} AN},
  \bibinfo{author}{{Paris} D}, \bibinfo{author}{{Paiano} S},
  \bibinfo{author}{{Plantet} C}, \bibinfo{author}{{Rossi} F},
  \bibinfo{author}{{Salvaterra} R}, \bibinfo{author}{{Schulze} S},
  \bibinfo{author}{{Veillet} C} and  \bibinfo{author}{{Volnova} AA}
  (\bibinfo{year}{2022}), \bibinfo{month}{Jun.}
\bibinfo{title}{{The Peculiar Short-duration GRB 200826A and Its Supernova}}.
\bibinfo{journal}{{\em ApJ}} \bibinfo{volume}{932} (\bibinfo{number}{1}),
  \bibinfo{eid}{1}. \bibinfo{doi}{\doi{10.3847/1538-4357/ac60a2}}.
\eprint{2105.03829}.

\bibtype{Article}%
\bibitem[{Ryde}(2005)]{ryde05}
\bibinfo{author}{{Ryde} F} (\bibinfo{year}{2005}), \bibinfo{month}{Jun.}
\bibinfo{title}{{Is Thermal Emission in Gamma-Ray Bursts Ubiquitous?}}
\bibinfo{journal}{{\em ApJl}} \bibinfo{volume}{625} (\bibinfo{number}{2}):
  \bibinfo{pages}{L95--L98}. \bibinfo{doi}{\doi{10.1086/431239}}.
\eprint{astro-ph/0504450}.

\bibtype{Article}%
\bibitem[{Saccardi} et al.(2023)]{saccardi23}
\bibinfo{author}{{Saccardi} A}, \bibinfo{author}{{Vergani} SD},
  \bibinfo{author}{{De Cia} A}, \bibinfo{author}{{D'Elia} V},
  \bibinfo{author}{{Heintz} KE}, \bibinfo{author}{{Izzo} L},
  \bibinfo{author}{{Palmerio} JT}, \bibinfo{author}{{Petitjean} P},
  \bibinfo{author}{{Rossi} A}, \bibinfo{author}{{de Ugarte Postigo} A},
  \bibinfo{author}{{Christensen} L}, \bibinfo{author}{{Konstantopoulou} C},
  \bibinfo{author}{{Levan} AJ}, \bibinfo{author}{{Malesani} DB},
  \bibinfo{author}{{M{\o}ller} P}, \bibinfo{author}{{Ramburuth-Hurt} T},
  \bibinfo{author}{{Salvaterra} R}, \bibinfo{author}{{Tanvir} NR},
  \bibinfo{author}{{Th{\"o}ne} CC}, \bibinfo{author}{{Vejlgaard} S},
  \bibinfo{author}{{Fynbo} JPU}, \bibinfo{author}{{Kann} DA},
  \bibinfo{author}{{Schady} P}, \bibinfo{author}{{Watson} DJ},
  \bibinfo{author}{{Wiersema} K}, \bibinfo{author}{{Campana} S},
  \bibinfo{author}{{Covino} S}, \bibinfo{author}{{De Pasquale} M},
  \bibinfo{author}{{Fausey} H}, \bibinfo{author}{{Hartmann} DH},
  \bibinfo{author}{{van der Horst} AJ}, \bibinfo{author}{{Jakobsson} P},
  \bibinfo{author}{{Palazzi} E}, \bibinfo{author}{{Pugliese} G},
  \bibinfo{author}{{Savaglio} S}, \bibinfo{author}{{Starling} RLC},
  \bibinfo{author}{{Stratta} G} and  \bibinfo{author}{{Zafar} T}
  (\bibinfo{year}{2023}), \bibinfo{month}{Mar.}
\bibinfo{title}{{Dissecting the interstellar medium of a z = 6.3 galaxy.
  X-shooter spectroscopy and HST imaging of the afterglow and environment of
  the Swift GRB 210905A}}.
\bibinfo{journal}{{\em A\&A}} \bibinfo{volume}{671}, \bibinfo{eid}{A84}.
  \bibinfo{doi}{\doi{10.1051/0004-6361/202244205}}.
\eprint{2211.16524}.

\bibtype{Article}%
\bibitem[{Sakamoto} et al.(2005)]{sakamoto05}
\bibinfo{author}{{Sakamoto} T}, \bibinfo{author}{{Lamb} DQ},
  \bibinfo{author}{{Kawai} N}, \bibinfo{author}{{Yoshida} A},
  \bibinfo{author}{{Graziani} C}, \bibinfo{author}{{Fenimore} EE},
  \bibinfo{author}{{Donaghy} TQ}, \bibinfo{author}{{Matsuoka} M},
  \bibinfo{author}{{Suzuki} M}, \bibinfo{author}{{Ricker} G},
  \bibinfo{author}{{Atteia} JL}, \bibinfo{author}{{Shirasaki} Y},
  \bibinfo{author}{{Tamagawa} T}, \bibinfo{author}{{Torii} K},
  \bibinfo{author}{{Galassi} M}, \bibinfo{author}{{Doty} J},
  \bibinfo{author}{{Vanderspek} R}, \bibinfo{author}{{Crew} GB},
  \bibinfo{author}{{Villasenor} J}, \bibinfo{author}{{Butler} N},
  \bibinfo{author}{{Prigozhin} G}, \bibinfo{author}{{Jernigan} JG},
  \bibinfo{author}{{Barraud} C}, \bibinfo{author}{{Boer} M},
  \bibinfo{author}{{Dezalay} JP}, \bibinfo{author}{{Olive} JF},
  \bibinfo{author}{{Hurley} K}, \bibinfo{author}{{Levine} A},
  \bibinfo{author}{{Monnelly} G}, \bibinfo{author}{{Martel} F},
  \bibinfo{author}{{Morgan} E}, \bibinfo{author}{{Woosley} SE},
  \bibinfo{author}{{Cline} T}, \bibinfo{author}{{Braga} J},
  \bibinfo{author}{{Manchanda} R}, \bibinfo{author}{{Pizzichini} G},
  \bibinfo{author}{{Takagishi} K} and  \bibinfo{author}{{Yamauchi} M}
  (\bibinfo{year}{2005}), \bibinfo{month}{Aug.}
\bibinfo{title}{{Global Characteristics of X-Ray Flashes and X-Ray-Rich
  Gamma-Ray Bursts Observed by HETE-2}}.
\bibinfo{journal}{{\em ApJ}} \bibinfo{volume}{629} (\bibinfo{number}{1}):
  \bibinfo{pages}{311--327}. \bibinfo{doi}{\doi{10.1086/431235}}.

\bibtype{Article}%
\bibitem[{Salvaterra} et al.(2009)]{salvaterra09}
\bibinfo{author}{{Salvaterra} R}, \bibinfo{author}{{Della Valle} M},
  \bibinfo{author}{{Campana} S}, \bibinfo{author}{{Chincarini} G},
  \bibinfo{author}{{Covino} S}, \bibinfo{author}{{D'Avanzo} P},
  \bibinfo{author}{{Fern{\'a}ndez-Soto} A}, \bibinfo{author}{{Guidorzi} C},
  \bibinfo{author}{{Mannucci} F}, \bibinfo{author}{{Margutti} R},
  \bibinfo{author}{{Th{\"o}ne} CC}, \bibinfo{author}{{Antonelli} LA},
  \bibinfo{author}{{Barthelmy} SD}, \bibinfo{author}{{de Pasquale} M},
  \bibinfo{author}{{D'Elia} V}, \bibinfo{author}{{Fiore} F},
  \bibinfo{author}{{Fugazza} D}, \bibinfo{author}{{Hunt} LK},
  \bibinfo{author}{{Maiorano} E}, \bibinfo{author}{{Marinoni} S},
  \bibinfo{author}{{Marshall} FE}, \bibinfo{author}{{Molinari} E},
  \bibinfo{author}{{Nousek} J}, \bibinfo{author}{{Pian} E},
  \bibinfo{author}{{Racusin} JL}, \bibinfo{author}{{Stella} L},
  \bibinfo{author}{{Amati} L}, \bibinfo{author}{{Andreuzzi} G},
  \bibinfo{author}{{Cusumano} G}, \bibinfo{author}{{Fenimore} EE},
  \bibinfo{author}{{Ferrero} P}, \bibinfo{author}{{Giommi} P},
  \bibinfo{author}{{Guetta} D}, \bibinfo{author}{{Holland} ST},
  \bibinfo{author}{{Hurley} K}, \bibinfo{author}{{Israel} GL},
  \bibinfo{author}{{Mao} J}, \bibinfo{author}{{Markwardt} CB},
  \bibinfo{author}{{Masetti} N}, \bibinfo{author}{{Pagani} C},
  \bibinfo{author}{{Palazzi} E}, \bibinfo{author}{{Palmer} DM},
  \bibinfo{author}{{Piranomonte} S}, \bibinfo{author}{{Tagliaferri} G} and
  \bibinfo{author}{{Testa} V} (\bibinfo{year}{2009}), \bibinfo{month}{Oct.}
\bibinfo{title}{{GRB090423 at a redshift of z\~{}8.1}}.
\bibinfo{journal}{{\em \nat}} \bibinfo{volume}{461}:
  \bibinfo{pages}{1258--1260}. \bibinfo{doi}{\doi{10.1038/nature08445}}.
\eprint{0906.1578}.

\bibtype{Article}%
\bibitem[{Sari} et al.(1998)]{Sari+98}
\bibinfo{author}{{Sari} R}, \bibinfo{author}{{Piran} T} and
  \bibinfo{author}{{Narayan} R} (\bibinfo{year}{1998}), \bibinfo{month}{Apr.}
\bibinfo{title}{{Spectra and Light Curves of Gamma-Ray Burst Afterglows}}.
\bibinfo{journal}{{\em ApJl}} \bibinfo{volume}{497} (\bibinfo{number}{1}):
  \bibinfo{pages}{L17--L20}. \bibinfo{doi}{\doi{10.1086/311269}}.
\eprint{astro-ph/9712005}.

\bibtype{Article}%
\bibitem[{Savchenko} et al.(2017)]{Savchenko+17}
\bibinfo{author}{{Savchenko} V}, \bibinfo{author}{{Ferrigno} C},
  \bibinfo{author}{{Kuulkers} E}, \bibinfo{author}{{Bazzano} A},
  \bibinfo{author}{{Bozzo} E}, \bibinfo{author}{{Brandt} S},
  \bibinfo{author}{{Chenevez} J}, \bibinfo{author}{{Courvoisier} TJL},
  \bibinfo{author}{{Diehl} R}, \bibinfo{author}{{Domingo} A},
  \bibinfo{author}{{Hanlon} L}, \bibinfo{author}{{Jourdain} E},
  \bibinfo{author}{{von Kienlin} A}, \bibinfo{author}{{Laurent} P},
  \bibinfo{author}{{Lebrun} F}, \bibinfo{author}{{Lutovinov} A},
  \bibinfo{author}{{Martin-Carrillo} A}, \bibinfo{author}{{Mereghetti} S},
  \bibinfo{author}{{Natalucci} L}, \bibinfo{author}{{Rodi} J},
  \bibinfo{author}{{Roques} JP}, \bibinfo{author}{{Sunyaev} R} and
  \bibinfo{author}{{Ubertini} P} (\bibinfo{year}{2017}), \bibinfo{month}{Oct.}
\bibinfo{title}{{INTEGRAL Detection of the First Prompt Gamma-Ray Signal
  Coincident with the Gravitational-wave Event GW170817}}.
\bibinfo{journal}{{\em ApJl}} \bibinfo{volume}{848} (\bibinfo{number}{2}),
  \bibinfo{eid}{L15}. \bibinfo{doi}{\doi{10.3847/2041-8213/aa8f94}}.
\eprint{1710.05449}.

\bibtype{Article}%
\bibitem[{Selsing} et al.(2019)]{Selsing+19}
\bibinfo{author}{{Selsing} J}, \bibinfo{author}{{Malesani} D},
  \bibinfo{author}{{Goldoni} P}, \bibinfo{author}{{Fynbo} JPU},
  \bibinfo{author}{{Kr{\"u}hler} T}, \bibinfo{author}{{Antonelli} LA},
  \bibinfo{author}{{Arabsalmani} M}, \bibinfo{author}{{Bolmer} J},
  \bibinfo{author}{{Cano} Z}, \bibinfo{author}{{Christensen} L},
  \bibinfo{author}{{Covino} S}, \bibinfo{author}{{D'Avanzo} P},
  \bibinfo{author}{{D'Elia} V}, \bibinfo{author}{{De Cia} A},
  \bibinfo{author}{{de Ugarte Postigo} A}, \bibinfo{author}{{Flores} H},
  \bibinfo{author}{{Friis} M}, \bibinfo{author}{{Gomboc} A},
  \bibinfo{author}{{Greiner} J}, \bibinfo{author}{{Groot} P},
  \bibinfo{author}{{Hammer} F}, \bibinfo{author}{{Hartoog} OE},
  \bibinfo{author}{{Heintz} KE}, \bibinfo{author}{{Hjorth} J},
  \bibinfo{author}{{Jakobsson} P}, \bibinfo{author}{{Japelj} J},
  \bibinfo{author}{{Kann} DA}, \bibinfo{author}{{Kaper} L},
  \bibinfo{author}{{Ledoux} C}, \bibinfo{author}{{Leloudas} G},
  \bibinfo{author}{{Levan} AJ}, \bibinfo{author}{{Maiorano} E},
  \bibinfo{author}{{Melandri} A}, \bibinfo{author}{{Milvang-Jensen} B},
  \bibinfo{author}{{Palazzi} E}, \bibinfo{author}{{Palmerio} JT},
  \bibinfo{author}{{Perley} DA}, \bibinfo{author}{{Pian} E},
  \bibinfo{author}{{Piranomonte} S}, \bibinfo{author}{{Pugliese} G},
  \bibinfo{author}{{S{\'a}nchez-Ram{\'\i}rez} R}, \bibinfo{author}{{Savaglio}
  S}, \bibinfo{author}{{Schady} P}, \bibinfo{author}{{Schulze} S},
  \bibinfo{author}{{Sollerman} J}, \bibinfo{author}{{Sparre} M},
  \bibinfo{author}{{Tagliaferri} G}, \bibinfo{author}{{Tanvir} NR},
  \bibinfo{author}{{Th{\"o}ne} CC}, \bibinfo{author}{{Vergani} SD},
  \bibinfo{author}{{Vreeswijk} P}, \bibinfo{author}{{Watson} D},
  \bibinfo{author}{{Wiersema} K}, \bibinfo{author}{{Wijers} R},
  \bibinfo{author}{{Xu} D} and  \bibinfo{author}{{Zafar} T}
  (\bibinfo{year}{2019}), \bibinfo{month}{Mar.}
\bibinfo{title}{{The X-shooter GRB afterglow legacy sample (XS-GRB)}}.
\bibinfo{journal}{{\em A\&A}} \bibinfo{volume}{623}, \bibinfo{eid}{A92}.
  \bibinfo{doi}{\doi{10.1051/0004-6361/201832835}}.
\eprint{1802.07727}.

\bibtype{Article}%
\bibitem[{Shappee} et al.(2017)]{Shappee+17}
\bibinfo{author}{{Shappee} BJ}, \bibinfo{author}{{Simon} JD},
  \bibinfo{author}{{Drout} MR}, \bibinfo{author}{{Piro} AL},
  \bibinfo{author}{{Morrell} N}, \bibinfo{author}{{Prieto} JL},
  \bibinfo{author}{{Kasen} D}, \bibinfo{author}{{Holoien} TWS},
  \bibinfo{author}{{Kollmeier} JA}, \bibinfo{author}{{Kelson} DD},
  \bibinfo{author}{{Coulter} DA}, \bibinfo{author}{{Foley} RJ},
  \bibinfo{author}{{Kilpatrick} CD}, \bibinfo{author}{{Siebert} MR},
  \bibinfo{author}{{Madore} BF}, \bibinfo{author}{{Murguia-Berthier} A},
  \bibinfo{author}{{Pan} YC}, \bibinfo{author}{{Prochaska} JX},
  \bibinfo{author}{{Ramirez-Ruiz} E}, \bibinfo{author}{{Rest} A},
  \bibinfo{author}{{Adams} C}, \bibinfo{author}{{Alatalo} K},
  \bibinfo{author}{{Ba{\~n}ados} E}, \bibinfo{author}{{Baughman} J},
  \bibinfo{author}{{Bernstein} RA}, \bibinfo{author}{{Bitsakis} T},
  \bibinfo{author}{{Boutsia} K}, \bibinfo{author}{{Bravo} JR},
  \bibinfo{author}{{Di Mille} F}, \bibinfo{author}{{Higgs} CR},
  \bibinfo{author}{{Ji} AP}, \bibinfo{author}{{Maravelias} G},
  \bibinfo{author}{{Marshall} JL}, \bibinfo{author}{{Placco} VM},
  \bibinfo{author}{{Prieto} G} and  \bibinfo{author}{{Wan} Z}
  (\bibinfo{year}{2017}), \bibinfo{month}{Dec.}
\bibinfo{title}{{Early spectra of the gravitational wave source GW170817:
  Evolution of a neutron star merger}}.
\bibinfo{journal}{{\em Science}} \bibinfo{volume}{358}
  (\bibinfo{number}{6370}): \bibinfo{pages}{1574--1578}.
  \bibinfo{doi}{\doi{10.1126/science.aaq0186}}.
\eprint{1710.05432}.

\bibtype{Article}%
\bibitem[{Smartt} et al.(2017)]{Smartt+17}
\bibinfo{author}{{Smartt} SJ}, \bibinfo{author}{{Chen} TW},
  \bibinfo{author}{{Jerkstrand} A}, \bibinfo{author}{{Coughlin} M},
  \bibinfo{author}{{Kankare} E}, \bibinfo{author}{{Sim} SA},
  \bibinfo{author}{{Fraser} M}, \bibinfo{author}{{Inserra} C},
  \bibinfo{author}{{Maguire} K}, \bibinfo{author}{{Chambers} KC},
  \bibinfo{author}{{Huber} ME}, \bibinfo{author}{{Kr{\"u}hler} T},
  \bibinfo{author}{{Leloudas} G}, \bibinfo{author}{{Magee} M},
  \bibinfo{author}{{Shingles} LJ}, \bibinfo{author}{{Smith} KW},
  \bibinfo{author}{{Young} DR}, \bibinfo{author}{{Tonry} J},
  \bibinfo{author}{{Kotak} R}, \bibinfo{author}{{Gal-Yam} A},
  \bibinfo{author}{{Lyman} JD}, \bibinfo{author}{{Homan} DS},
  \bibinfo{author}{{Agliozzo} C}, \bibinfo{author}{{Anderson} JP},
  \bibinfo{author}{{Angus} CR}, \bibinfo{author}{{Ashall} C},
  \bibinfo{author}{{Barbarino} C}, \bibinfo{author}{{Bauer} FE},
  \bibinfo{author}{{Berton} M}, \bibinfo{author}{{Botticella} MT},
  \bibinfo{author}{{Bulla} M}, \bibinfo{author}{{Bulger} J},
  \bibinfo{author}{{Cannizzaro} G}, \bibinfo{author}{{Cano} Z},
  \bibinfo{author}{{Cartier} R}, \bibinfo{author}{{Cikota} A},
  \bibinfo{author}{{Clark} P}, \bibinfo{author}{{De Cia} A},
  \bibinfo{author}{{Della Valle} M}, \bibinfo{author}{{Denneau} L},
  \bibinfo{author}{{Dennefeld} M}, \bibinfo{author}{{Dessart} L},
  \bibinfo{author}{{Dimitriadis} G}, \bibinfo{author}{{Elias-Rosa} N},
  \bibinfo{author}{{Firth} RE}, \bibinfo{author}{{Flewelling} H},
  \bibinfo{author}{{Fl{\"o}rs} A}, \bibinfo{author}{{Franckowiak} A},
  \bibinfo{author}{{Frohmaier} C}, \bibinfo{author}{{Galbany} L},
  \bibinfo{author}{{Gonz{\'a}lez-Gait{\'a}n} S}, \bibinfo{author}{{Greiner} J},
  \bibinfo{author}{{Gromadzki} M}, \bibinfo{author}{{Guelbenzu} AN},
  \bibinfo{author}{{Guti{\'e}rrez} CP}, \bibinfo{author}{{Hamanowicz} A},
  \bibinfo{author}{{Hanlon} L}, \bibinfo{author}{{Harmanen} J},
  \bibinfo{author}{{Heintz} KE}, \bibinfo{author}{{Heinze} A},
  \bibinfo{author}{{Hernandez} MS}, \bibinfo{author}{{Hodgkin} ST},
  \bibinfo{author}{{Hook} IM}, \bibinfo{author}{{Izzo} L},
  \bibinfo{author}{{James} PA}, \bibinfo{author}{{Jonker} PG},
  \bibinfo{author}{{Kerzendorf} WE}, \bibinfo{author}{{Klose} S},
  \bibinfo{author}{{Kostrzewa-Rutkowska} Z}, \bibinfo{author}{{Kowalski} M},
  \bibinfo{author}{{Kromer} M}, \bibinfo{author}{{Kuncarayakti} H},
  \bibinfo{author}{{Lawrence} A}, \bibinfo{author}{{Lowe} TB},
  \bibinfo{author}{{Magnier} EA}, \bibinfo{author}{{Manulis} I},
  \bibinfo{author}{{Martin-Carrillo} A}, \bibinfo{author}{{Mattila} S},
  \bibinfo{author}{{McBrien} O}, \bibinfo{author}{{M{\"u}ller} A},
  \bibinfo{author}{{Nordin} J}, \bibinfo{author}{{O'Neill} D},
  \bibinfo{author}{{Onori} F}, \bibinfo{author}{{Palmerio} JT},
  \bibinfo{author}{{Pastorello} A}, \bibinfo{author}{{Patat} F},
  \bibinfo{author}{{Pignata} G}, \bibinfo{author}{{Podsiadlowski} P},
  \bibinfo{author}{{Pumo} ML}, \bibinfo{author}{{Prentice} SJ},
  \bibinfo{author}{{Rau} A}, \bibinfo{author}{{Razza} A},
  \bibinfo{author}{{Rest} A}, \bibinfo{author}{{Reynolds} T},
  \bibinfo{author}{{Roy} R}, \bibinfo{author}{{Ruiter} AJ},
  \bibinfo{author}{{Rybicki} KA}, \bibinfo{author}{{Salmon} L},
  \bibinfo{author}{{Schady} P}, \bibinfo{author}{{Schultz} ASB},
  \bibinfo{author}{{Schweyer} T}, \bibinfo{author}{{Seitenzahl} IR},
  \bibinfo{author}{{Smith} M}, \bibinfo{author}{{Sollerman} J},
  \bibinfo{author}{{Stalder} B}, \bibinfo{author}{{Stubbs} CW},
  \bibinfo{author}{{Sullivan} M}, \bibinfo{author}{{Szegedi} H},
  \bibinfo{author}{{Taddia} F}, \bibinfo{author}{{Taubenberger} S},
  \bibinfo{author}{{Terreran} G}, \bibinfo{author}{{van Soelen} B},
  \bibinfo{author}{{Vos} J}, \bibinfo{author}{{Wainscoat} RJ},
  \bibinfo{author}{{Walton} NA}, \bibinfo{author}{{Waters} C},
  \bibinfo{author}{{Weiland} H}, \bibinfo{author}{{Willman} M},
  \bibinfo{author}{{Wiseman} P}, \bibinfo{author}{{Wright} DE},
  \bibinfo{author}{{Wyrzykowski} {\L}} and  \bibinfo{author}{{Yaron} O}
  (\bibinfo{year}{2017}), \bibinfo{month}{Nov}.
\bibinfo{title}{{A kilonova as the electromagnetic counterpart to a
  gravitational-wave source}}.
\bibinfo{journal}{{\em \nat}} \bibinfo{volume}{551} (\bibinfo{number}{7678}):
  \bibinfo{pages}{75--79}. \bibinfo{doi}{\doi{10.1038/nature24303}}.
\eprint{1710.05841}.

\bibtype{Article}%
\bibitem[{Stanek} et al.(2003)]{stanek03}
\bibinfo{author}{{Stanek} KZ}, \bibinfo{author}{{Matheson} T},
  \bibinfo{author}{{Garnavich} PM}, \bibinfo{author}{{Martini} P},
  \bibinfo{author}{{Berlind} P}, \bibinfo{author}{{Caldwell} N},
  \bibinfo{author}{{Challis} P}, \bibinfo{author}{{Brown} WR},
  \bibinfo{author}{{Schild} R}, \bibinfo{author}{{Krisciunas} K},
  \bibinfo{author}{{Calkins} ML}, \bibinfo{author}{{Lee} JC},
  \bibinfo{author}{{Hathi} N}, \bibinfo{author}{{Jansen} RA},
  \bibinfo{author}{{Windhorst} R}, \bibinfo{author}{{Echevarria} L},
  \bibinfo{author}{{Eisenstein} DJ}, \bibinfo{author}{{Pindor} B},
  \bibinfo{author}{{Olszewski} EW}, \bibinfo{author}{{Harding} P},
  \bibinfo{author}{{Holland} ST} and  \bibinfo{author}{{Bersier} D}
  (\bibinfo{year}{2003}), \bibinfo{month}{Jul.}
\bibinfo{title}{{Spectroscopic Discovery of the Supernova 2003dh Associated
  with GRB 030329}}.
\bibinfo{journal}{{\em ApJl}} \bibinfo{volume}{591}:
  \bibinfo{pages}{L17--L20}. \bibinfo{doi}{\doi{10.1086/376976}}.
\eprint{astro-ph/0304173}.

\bibtype{Article}%
\bibitem[{Sun} et al.(2015)]{sun15}
\bibinfo{author}{{Sun} H}, \bibinfo{author}{{Zhang} B} and
  \bibinfo{author}{{Li} Z} (\bibinfo{year}{2015}), \bibinfo{month}{Oct.}
\bibinfo{title}{{Extragalactic High-energy Transients: Event Rate Densities and
  Luminosity Functions}}.
\bibinfo{journal}{{\em ApJ}} \bibinfo{volume}{812} (\bibinfo{number}{1}),
  \bibinfo{eid}{33}. \bibinfo{doi}{\doi{10.1088/0004-637X/812/1/33}}.
\eprint{1509.01592}.

\bibtype{Article}%
\bibitem[{Tanvir} et al.(2009)]{tanvir09}
\bibinfo{author}{{Tanvir} NR}, \bibinfo{author}{{Fox} DB},
  \bibinfo{author}{{Levan} AJ}, \bibinfo{author}{{Berger} E},
  \bibinfo{author}{{Wiersema} K}, \bibinfo{author}{{Fynbo} JPU},
  \bibinfo{author}{{Cucchiara} A}, \bibinfo{author}{{Kr{\"u}hler} T},
  \bibinfo{author}{{Gehrels} N}, \bibinfo{author}{{Bloom} JS},
  \bibinfo{author}{{Greiner} J}, \bibinfo{author}{{Evans} PA},
  \bibinfo{author}{{Rol} E}, \bibinfo{author}{{Olivares} F},
  \bibinfo{author}{{Hjorth} J}, \bibinfo{author}{{Jakobsson} P},
  \bibinfo{author}{{Farihi} J}, \bibinfo{author}{{Willingale} R},
  \bibinfo{author}{{Starling} RLC}, \bibinfo{author}{{Cenko} SB},
  \bibinfo{author}{{Perley} D}, \bibinfo{author}{{Maund} JR},
  \bibinfo{author}{{Duke} J}, \bibinfo{author}{{Wijers} RAMJ},
  \bibinfo{author}{{Adamson} AJ}, \bibinfo{author}{{Allan} A},
  \bibinfo{author}{{Bremer} MN}, \bibinfo{author}{{Burrows} DN},
  \bibinfo{author}{{Castro-Tirado} AJ}, \bibinfo{author}{{Cavanagh} B},
  \bibinfo{author}{{de Ugarte Postigo} A}, \bibinfo{author}{{Dopita} MA},
  \bibinfo{author}{{Fatkhullin} TA}, \bibinfo{author}{{Fruchter} AS},
  \bibinfo{author}{{Foley} RJ}, \bibinfo{author}{{Gorosabel} J},
  \bibinfo{author}{{Kennea} J}, \bibinfo{author}{{Kerr} T},
  \bibinfo{author}{{Klose} S}, \bibinfo{author}{{Krimm} HA},
  \bibinfo{author}{{Komarova} VN}, \bibinfo{author}{{Kulkarni} SR},
  \bibinfo{author}{{Moskvitin} AS}, \bibinfo{author}{{Mundell} CG},
  \bibinfo{author}{{Naylor} T}, \bibinfo{author}{{Page} K},
  \bibinfo{author}{{Penprase} BE}, \bibinfo{author}{{Perri} M},
  \bibinfo{author}{{Podsiadlowski} P}, \bibinfo{author}{{Roth} K},
  \bibinfo{author}{{Rutledge} RE}, \bibinfo{author}{{Sakamoto} T},
  \bibinfo{author}{{Schady} P}, \bibinfo{author}{{Schmidt} BP},
  \bibinfo{author}{{Soderberg} AM}, \bibinfo{author}{{Sollerman} J},
  \bibinfo{author}{{Stephens} AW}, \bibinfo{author}{{Stratta} G},
  \bibinfo{author}{{Ukwatta} TN}, \bibinfo{author}{{Watson} D},
  \bibinfo{author}{{Westra} E}, \bibinfo{author}{{Wold} T} and
  \bibinfo{author}{{Wolf} C} (\bibinfo{year}{2009}), \bibinfo{month}{Oct.}
\bibinfo{title}{{A {$\gamma$}-ray burst at a redshift of z\~{}8.2}}.
\bibinfo{journal}{{\em Nature}} \bibinfo{volume}{461}:
  \bibinfo{pages}{1254--1257}. \bibinfo{doi}{\doi{10.1038/nature08459}}.
\eprint{0906.1577}.

\bibtype{Article}%
\bibitem[{Tanvir} et al.(2012)]{tanvir12}
\bibinfo{author}{{Tanvir} NR}, \bibinfo{author}{{Mackey} AD},
  \bibinfo{author}{{Ferguson} AMN}, \bibinfo{author}{{Huxor} A},
  \bibinfo{author}{{Read} JI}, \bibinfo{author}{{Lewis} GF},
  \bibinfo{author}{{Irwin} MJ}, \bibinfo{author}{{Chapman} S},
  \bibinfo{author}{{Ibata} R}, \bibinfo{author}{{Wilkinson} MI},
  \bibinfo{author}{{McConnachie} AW}, \bibinfo{author}{{Martin} NF},
  \bibinfo{author}{{Davies} MB} and  \bibinfo{author}{{Bridges} TJ}
  (\bibinfo{year}{2012}), \bibinfo{month}{May}.
\bibinfo{title}{{The structure of star clusters in the outer halo of M31}}.
\bibinfo{journal}{{\em MNRAS}} \bibinfo{volume}{422}:
  \bibinfo{pages}{162--184}.
  \bibinfo{doi}{\doi{10.1111/j.1365-2966.2012.20590.x}}.
\eprint{1202.2100}.

\bibtype{Article}%
\bibitem[{Tanvir} et al.(2013)]{tanvir13}
\bibinfo{author}{{Tanvir} NR}, \bibinfo{author}{{Levan} AJ},
  \bibinfo{author}{{Fruchter} AS}, \bibinfo{author}{{Hjorth} J},
  \bibinfo{author}{{Hounsell} RA}, \bibinfo{author}{{Wiersema} K} and
  \bibinfo{author}{{Tunnicliffe} RL} (\bibinfo{year}{2013}),
  \bibinfo{month}{Aug.}
\bibinfo{title}{{A `kilonova' associated with the short-duration {$\gamma$}-ray
  burst GRB 130603B}}.
\bibinfo{journal}{{\em \nat}} \bibinfo{volume}{500}: \bibinfo{pages}{547--549}.
  \bibinfo{doi}{\doi{10.1038/nature12505}}.
\eprint{1306.4971}.

\bibtype{Article}%
\bibitem[{Tanvir} et al.(2017)]{Tanvir+17}
\bibinfo{author}{{Tanvir} NR}, \bibinfo{author}{{Levan} AJ},
  \bibinfo{author}{{Gonz{\'a}lez-Fern{\'a}ndez} C}, \bibinfo{author}{{Korobkin}
  O}, \bibinfo{author}{{Mandel} I}, \bibinfo{author}{{Rosswog} S},
  \bibinfo{author}{{Hjorth} J}, \bibinfo{author}{{D'Avanzo} P},
  \bibinfo{author}{{Fruchter} AS}, \bibinfo{author}{{Fryer} CL},
  \bibinfo{author}{{Kangas} T}, \bibinfo{author}{{Milvang-Jensen} B},
  \bibinfo{author}{{Rosetti} S}, \bibinfo{author}{{Steeghs} D},
  \bibinfo{author}{{Wollaeger} RT}, \bibinfo{author}{{Cano} Z},
  \bibinfo{author}{{Copperwheat} CM}, \bibinfo{author}{{Covino} S},
  \bibinfo{author}{{D'Elia} V}, \bibinfo{author}{{de Ugarte Postigo} A},
  \bibinfo{author}{{Evans} PA}, \bibinfo{author}{{Even} WP},
  \bibinfo{author}{{Fairhurst} S}, \bibinfo{author}{{Figuera Jaimes} R},
  \bibinfo{author}{{Fontes} CJ}, \bibinfo{author}{{Fujii} YI},
  \bibinfo{author}{{Fynbo} JPU}, \bibinfo{author}{{Gompertz} BP},
  \bibinfo{author}{{Greiner} J}, \bibinfo{author}{{Hodosan} G},
  \bibinfo{author}{{Irwin} MJ}, \bibinfo{author}{{Jakobsson} P},
  \bibinfo{author}{{J{\o}rgensen} UG}, \bibinfo{author}{{Kann} DA},
  \bibinfo{author}{{Lyman} JD}, \bibinfo{author}{{Malesani} D},
  \bibinfo{author}{{McMahon} RG}, \bibinfo{author}{{Melandri} A},
  \bibinfo{author}{{O'Brien} PT}, \bibinfo{author}{{Osborne} JP},
  \bibinfo{author}{{Palazzi} E}, \bibinfo{author}{{Perley} DA},
  \bibinfo{author}{{Pian} E}, \bibinfo{author}{{Piranomonte} S},
  \bibinfo{author}{{Rabus} M}, \bibinfo{author}{{Rol} E},
  \bibinfo{author}{{Rowlinson} A}, \bibinfo{author}{{Schulze} S},
  \bibinfo{author}{{Sutton} P}, \bibinfo{author}{{Th{\"o}ne} CC},
  \bibinfo{author}{{Ulaczyk} K}, \bibinfo{author}{{Watson} D},
  \bibinfo{author}{{Wiersema} K} and  \bibinfo{author}{{Wijers} RAMJ}
  (\bibinfo{year}{2017}), \bibinfo{month}{Oct.}
\bibinfo{title}{{The Emergence of a Lanthanide-rich Kilonova Following the
  Merger of Two Neutron Stars}}.
\bibinfo{journal}{{\em ApJl}} \bibinfo{volume}{848} (\bibinfo{number}{2}),
  \bibinfo{eid}{L27}. \bibinfo{doi}{\doi{10.3847/2041-8213/aa90b6}}.
\eprint{1710.05455}.

\bibtype{Article}%
\bibitem[{Tanvir} et al.(2019)]{tanvir19}
\bibinfo{author}{{Tanvir} NR}, \bibinfo{author}{{Fynbo} JPU},
  \bibinfo{author}{{de Ugarte Postigo} A}, \bibinfo{author}{{Japelj} J},
  \bibinfo{author}{{Wiersema} K}, \bibinfo{author}{{Malesani} D},
  \bibinfo{author}{{Perley} DA}, \bibinfo{author}{{Levan} AJ},
  \bibinfo{author}{{Selsing} J}, \bibinfo{author}{{Cenko} SB},
  \bibinfo{author}{{Kann} DA}, \bibinfo{author}{{Milvang-Jensen} B},
  \bibinfo{author}{{Berger} E}, \bibinfo{author}{{Cano} Z},
  \bibinfo{author}{{Chornock} R}, \bibinfo{author}{{Covino} S},
  \bibinfo{author}{{Cucchiara} A}, \bibinfo{author}{{D'Elia} V},
  \bibinfo{author}{{Gargiulo} A}, \bibinfo{author}{{Goldoni} P},
  \bibinfo{author}{{Gomboc} A}, \bibinfo{author}{{Heintz} KE},
  \bibinfo{author}{{Hjorth} J}, \bibinfo{author}{{Izzo} L},
  \bibinfo{author}{{Jakobsson} P}, \bibinfo{author}{{Kaper} L},
  \bibinfo{author}{{Kr{\"u}hler} T}, \bibinfo{author}{{Laskar} T},
  \bibinfo{author}{{Myers} M}, \bibinfo{author}{{Piranomonte} S},
  \bibinfo{author}{{Pugliese} G}, \bibinfo{author}{{Rossi} A},
  \bibinfo{author}{{S{\'a}nchez-Ram{\'\i}rez} R}, \bibinfo{author}{{Schulze}
  S}, \bibinfo{author}{{Sparre} M}, \bibinfo{author}{{Stanway} ER},
  \bibinfo{author}{{Tagliaferri} G}, \bibinfo{author}{{Th{\"o}ne} CC},
  \bibinfo{author}{{Vergani} S}, \bibinfo{author}{{Vreeswijk} PM},
  \bibinfo{author}{{Wijers} RAMJ}, \bibinfo{author}{{Watson} D} and
  \bibinfo{author}{{Xu} D} (\bibinfo{year}{2019}), \bibinfo{month}{Mar.}
\bibinfo{title}{{The fraction of ionizing radiation from massive stars that
  escapes to the intergalactic medium}}.
\bibinfo{journal}{{\em MNRAS}} \bibinfo{volume}{483} (\bibinfo{number}{4}):
  \bibinfo{pages}{5380--5408}. \bibinfo{doi}{\doi{10.1093/mnras/sty3460}}.
\eprint{1805.07318}.

\bibtype{Article}%
\bibitem[{Toma} et al.(2016)]{toma16}
\bibinfo{author}{{Toma} K}, \bibinfo{author}{{Yoon} SC} and
  \bibinfo{author}{{Bromm} V} (\bibinfo{year}{2016}), \bibinfo{month}{Dec.}
\bibinfo{title}{{Gamma-Ray Bursts and Population III Stars}}.
\bibinfo{journal}{{\em Space Science Reviews}} \bibinfo{volume}{202} (\bibinfo{number}{1-4}):
  \bibinfo{pages}{159--180}. \bibinfo{doi}{\doi{10.1007/s11214-016-0250-7}}.
\eprint{1603.04640}.

\bibtype{Article}%
\bibitem[{Troja} et al.(2017)]{Troja+17}
\bibinfo{author}{{Troja} E}, \bibinfo{author}{{Piro} L}, \bibinfo{author}{{van
  Eerten} H}, \bibinfo{author}{{Wollaeger} RT}, \bibinfo{author}{{Im} M},
  \bibinfo{author}{{Fox} OD}, \bibinfo{author}{{Butler} NR},
  \bibinfo{author}{{Cenko} SB}, \bibinfo{author}{{Sakamoto} T},
  \bibinfo{author}{{Fryer} CL}, \bibinfo{author}{{Ricci} R},
  \bibinfo{author}{{Lien} A}, \bibinfo{author}{{Ryan} RE},
  \bibinfo{author}{{Korobkin} O}, \bibinfo{author}{{Lee} SK},
  \bibinfo{author}{{Burgess} JM}, \bibinfo{author}{{Lee} WH},
  \bibinfo{author}{{Watson} AM}, \bibinfo{author}{{Choi} C},
  \bibinfo{author}{{Covino} S}, \bibinfo{author}{{D'Avanzo} P},
  \bibinfo{author}{{Fontes} CJ}, \bibinfo{author}{{Gonz{\'a}lez} JB},
  \bibinfo{author}{{Khandrika} HG}, \bibinfo{author}{{Kim} J},
  \bibinfo{author}{{Kim} SL}, \bibinfo{author}{{Lee} CU},
  \bibinfo{author}{{Lee} HM}, \bibinfo{author}{{Kutyrev} A},
  \bibinfo{author}{{Lim} G}, \bibinfo{author}{{S{\'a}nchez-Ram{\'\i}rez} R},
  \bibinfo{author}{{Veilleux} S}, \bibinfo{author}{{Wieringa} MH} and
  \bibinfo{author}{{Yoon} Y} (\bibinfo{year}{2017}), \bibinfo{month}{Nov.}
\bibinfo{title}{{The X-ray counterpart to the gravitational-wave event
  GW170817}}.
\bibinfo{journal}{{\em \nat}} \bibinfo{volume}{551} (\bibinfo{number}{7678}):
  \bibinfo{pages}{71--74}. \bibinfo{doi}{\doi{10.1038/nature24290}}.
\eprint{1710.05433}.

\bibtype{Article}%
\bibitem[{Troja} et al.(2022)]{troja22}
\bibinfo{author}{{Troja} E}, \bibinfo{author}{{Fryer} CL},
  \bibinfo{author}{{O'Connor} B}, \bibinfo{author}{{Ryan} G},
  \bibinfo{author}{{Dichiara} S}, \bibinfo{author}{{Kumar} A},
  \bibinfo{author}{{Ito} N}, \bibinfo{author}{{Gupta} R},
  \bibinfo{author}{{Wollaeger} R}, \bibinfo{author}{{Norris} JP},
  \bibinfo{author}{{Kawai} N}, \bibinfo{author}{{Butler} N},
  \bibinfo{author}{{Aryan} A}, \bibinfo{author}{{Misra} K},
  \bibinfo{author}{{Hosokawa} R}, \bibinfo{author}{{Murata} KL},
  \bibinfo{author}{{Niwano} M}, \bibinfo{author}{{Pandey} SB},
  \bibinfo{author}{{Kutyrev} A}, \bibinfo{author}{{van Eerten} HJ},
  \bibinfo{author}{{Chase} EA}, \bibinfo{author}{{Hu} YD},
  \bibinfo{author}{{Caballero-Garcia} MD} and  \bibinfo{author}{{Castro-Tirado}
  AJ} (\bibinfo{year}{2022}), \bibinfo{month}{Sep.}
\bibinfo{title}{{A long gamma-ray burst from a merger of compact objects}}.
\bibinfo{journal}{{\em arXiv e-prints}} ,
  \bibinfo{eid}{arXiv:2209.03363}\eprint{2209.03363}.

\bibtype{Article}%
\bibitem[{Valenti} et al.(2017)]{Valenti+17}
\bibinfo{author}{{Valenti} S}, \bibinfo{author}{{Sand} DJ},
  \bibinfo{author}{{Yang} S}, \bibinfo{author}{{Cappellaro} E},
  \bibinfo{author}{{Tartaglia} L}, \bibinfo{author}{{Corsi} A},
  \bibinfo{author}{{Jha} SW}, \bibinfo{author}{{Reichart} DE},
  \bibinfo{author}{{Haislip} J} and  \bibinfo{author}{{Kouprianov} V}
  (\bibinfo{year}{2017}), \bibinfo{month}{Oct.}
\bibinfo{title}{{The Discovery of the Electromagnetic Counterpart of GW170817:
  Kilonova AT 2017gfo/DLT17ck}}.
\bibinfo{journal}{{\em ApJl}} \bibinfo{volume}{848} (\bibinfo{number}{2}),
  \bibinfo{eid}{L24}. \bibinfo{doi}{\doi{10.3847/2041-8213/aa8edf}}.
\eprint{1710.05854}.

\bibtype{Article}%
\bibitem[{van Dalen} et al.(2024)]{vandalen24}
\bibinfo{author}{{van Dalen} JND}, \bibinfo{author}{{Levan} AJ},
  \bibinfo{author}{{Jonker} PG}, \bibinfo{author}{{Malesani} DB},
  \bibinfo{author}{{Izzo} L}, \bibinfo{author}{{Sarin} N},
  \bibinfo{author}{{Quirola-V{\'a}squez} J}, \bibinfo{author}{{Mata
  S{\'a}nchez} D}, \bibinfo{author}{{de Ugarte Postigo} A},
  \bibinfo{author}{{van Hoof} APC}, \bibinfo{author}{{Torres} MAP},
  \bibinfo{author}{{Schulze} S}, \bibinfo{author}{{Littlefair} SP},
  \bibinfo{author}{{Chrimes} A}, \bibinfo{author}{{Ravasio} ME},
  \bibinfo{author}{{Bauer} FE}, \bibinfo{author}{{Martin-Carrillo} A},
  \bibinfo{author}{{Fraser} M}, \bibinfo{author}{{van der Horst} AJ},
  \bibinfo{author}{{Jakobsson} P}, \bibinfo{author}{{O'Brien} P},
  \bibinfo{author}{{De Pasquale} M}, \bibinfo{author}{{Pugliese} G},
  \bibinfo{author}{{Sollerman} J}, \bibinfo{author}{{Tanvir} NR},
  \bibinfo{author}{{Zafar} T}, \bibinfo{author}{{Anderson} JP},
  \bibinfo{author}{{Galbany} L}, \bibinfo{author}{{Gal-Yam} A},
  \bibinfo{author}{{Gromadzki} M}, \bibinfo{author}{{Muller-Bravo} TE},
  \bibinfo{author}{{Ragosta} F} and  \bibinfo{author}{{Terwel} JH}
  (\bibinfo{year}{2024}), \bibinfo{month}{Sep.}
\bibinfo{title}{{The Einstein Probe transient EP240414a: Linking Fast X-ray
  Transients, Gamma-ray Bursts and Luminous Fast Blue Optical Transients}}.
\bibinfo{journal}{{\em arXiv e-prints}} ,
  \bibinfo{eid}{arXiv:2409.19056}\bibinfo{doi}{\doi{10.48550/arXiv.2409.19056}}.
\eprint{2409.19056}.

\bibtype{Article}%
\bibitem[{van Paradijs} et al.(1997)]{vanparadijs97}
\bibinfo{author}{{van Paradijs} J}, \bibinfo{author}{{Groot} PJ},
  \bibinfo{author}{{Galama} T}, \bibinfo{author}{{Kouveliotou} C},
  \bibinfo{author}{{Strom} RG}, \bibinfo{author}{{Telting} J},
  \bibinfo{author}{{Rutten} RGM}, \bibinfo{author}{{Fishman} GJ},
  \bibinfo{author}{{Meegan} CA}, \bibinfo{author}{{Pettini} M},
  \bibinfo{author}{{Tanvir} N}, \bibinfo{author}{{Bloom} J},
  \bibinfo{author}{{Pedersen} H}, \bibinfo{author}{{N{\o}rdgaard-Nielsen} HU},
  \bibinfo{author}{{Linden-V{\o}rnle} M}, \bibinfo{author}{{Melnick} J},
  \bibinfo{author}{{van der Steene} G}, \bibinfo{author}{{Bremer} M},
  \bibinfo{author}{{Naber} R}, \bibinfo{author}{{Heise} J},
  \bibinfo{author}{{in't Zand} J}, \bibinfo{author}{{Costa} E},
  \bibinfo{author}{{Feroci} M}, \bibinfo{author}{{Piro} L},
  \bibinfo{author}{{Frontera} F}, \bibinfo{author}{{Zavattini} G},
  \bibinfo{author}{{Nicastro} L}, \bibinfo{author}{{Palazzi} E},
  \bibinfo{author}{{Bennett} K}, \bibinfo{author}{{Hanlon} L} and
  \bibinfo{author}{{Parmar} A} (\bibinfo{year}{1997}), \bibinfo{month}{Apr.}
\bibinfo{title}{{Transient optical emission from the error box of the
  {$\gamma$}-ray burst of 28 February 1997}}.
\bibinfo{journal}{{\em \nat}} \bibinfo{volume}{386}: \bibinfo{pages}{686--689}.
  \bibinfo{doi}{\doi{10.1038/386686a0}}.

\bibtype{Article}%
\bibitem[{Waxman}(2006)]{waxman06}
\bibinfo{author}{{Waxman} E} (\bibinfo{year}{2006}), \bibinfo{month}{Dec.}
\bibinfo{title}{{Gamma-ray bursts and collisionless shocks}}.
\bibinfo{journal}{{\em Plasma Physics and Controlled Fusion}}
  \bibinfo{volume}{48} (\bibinfo{number}{12B}): \bibinfo{pages}{B137--B151}.
  \bibinfo{doi}{\doi{10.1088/0741-3335/48/12B/S14}}.
\eprint{astro-ph/0607353}.

\bibtype{Inproceedings}%
\bibitem[{White} et al.(2021)]{gamow}
\bibinfo{author}{{White} NE}, \bibinfo{author}{{Bauer} FE},
  \bibinfo{author}{{Baumgartner} W}, \bibinfo{author}{{Bautz} M},
  \bibinfo{author}{{Berger} E}, \bibinfo{author}{{Cenko} B},
  \bibinfo{author}{{Chang} TC}, \bibinfo{author}{{Falcone} A},
  \bibinfo{author}{{Fausey} H}, \bibinfo{author}{{Feldman} C},
  \bibinfo{author}{{Fox} D}, \bibinfo{author}{{Fox} O},
  \bibinfo{author}{{Fruchter} A}, \bibinfo{author}{{Fryer} C},
  \bibinfo{author}{{Ghirlanda} G}, \bibinfo{author}{{Gorski} K},
  \bibinfo{author}{{Grant} C}, \bibinfo{author}{{Guiriec} S},
  \bibinfo{author}{{Hart} M}, \bibinfo{author}{{Hartmann} D},
  \bibinfo{author}{{Hennawi} J}, \bibinfo{author}{{Kann} DA},
  \bibinfo{author}{{Kaplan} D}, \bibinfo{author}{{Kennea} J},
  \bibinfo{author}{{Kocevski} D}, \bibinfo{author}{{Kouveliotou} C},
  \bibinfo{author}{{Lawrence} C}, \bibinfo{author}{{Levan} AJ},
  \bibinfo{author}{{Lidz} A}, \bibinfo{author}{{Lien} A},
  \bibinfo{author}{{Littenberg} TB}, \bibinfo{author}{{Mas-Ribas} L},
  \bibinfo{author}{{Moss} M}, \bibinfo{author}{{O'Brien} P},
  \bibinfo{author}{{O'Meara} J}, \bibinfo{author}{{Palmer} DM},
  \bibinfo{author}{{Pasham} D}, \bibinfo{author}{{Racusin} J},
  \bibinfo{author}{{Remillard} R}, \bibinfo{author}{{Roberts} OJ},
  \bibinfo{author}{{Roming} P}, \bibinfo{author}{{Rud} M},
  \bibinfo{author}{{Salvaterra} R}, \bibinfo{author}{{Sambruna} R},
  \bibinfo{author}{{Seiffert} M}, \bibinfo{author}{{Sun} G},
  \bibinfo{author}{{Tanvir} NR}, \bibinfo{author}{{Terrile} R},
  \bibinfo{author}{{Thomas} N}, \bibinfo{author}{{van der Horst} A},
  \bibinfo{author}{{Verstrand} WT}, \bibinfo{author}{{Willems} P},
  \bibinfo{author}{{Wilson-Hodge} C}, \bibinfo{author}{{Young} ET},
  \bibinfo{author}{{Amati} L}, \bibinfo{author}{{Bozzo} E},
  \bibinfo{author}{{Karczewski} O{\L}}, \bibinfo{author}{{Hernandez-Monteagudo}
  C}, \bibinfo{author}{{Rebolo Lopez} R}, \bibinfo{author}{{Genova-Santos} R},
  \bibinfo{author}{{Martin} A}, \bibinfo{author}{{Granot} J},
  \bibinfo{author}{{Bemiamini} P}, \bibinfo{author}{{Gil} R} and
  \bibinfo{author}{{Burns} E} (\bibinfo{year}{2021}), \bibinfo{month}{Aug.},
  \bibinfo{title}{{The Gamow Explorer: a Gamma-Ray Burst Observatory to study
  the high redshift universe and enable multi-messenger astrophysics}},
  \bibinfo{editor}{{Siegmund} OH}, (Ed.), \bibinfo{booktitle}{UV, X-Ray, and
  Gamma-Ray Space Instrumentation for Astronomy XXII}, \bibinfo{series}{Society
  of Photo-Optical Instrumentation Engineers (SPIE) Conference Series},
  \bibinfo{volume}{11821}, pp. \bibinfo{pages}{1182109}, \eprint{2111.06497}.

\bibtype{Article}%
\bibitem[{Yang} et al.(2022)]{yang22}
\bibinfo{author}{{Yang} J}, \bibinfo{author}{{Ai} S}, \bibinfo{author}{{Zhang}
  BB}, \bibinfo{author}{{Zhang} B}, \bibinfo{author}{{Liu} ZK},
  \bibinfo{author}{{Wang} XI}, \bibinfo{author}{{Yang} YH},
  \bibinfo{author}{{Yin} YH}, \bibinfo{author}{{Li} Y} and
  \bibinfo{author}{{L{\"u}} HJ} (\bibinfo{year}{2022}), \bibinfo{month}{Apr.}
\bibinfo{title}{{A long-duration gamma-ray burst with a peculiar origin}}.
\bibinfo{journal}{{\em arXiv e-prints}} ,
  \bibinfo{eid}{arXiv:2204.12771}\eprint{2204.12771}.

\bibtype{Article}%
\bibitem[{Yonetoku} et al.(2004)]{yonetoku04}
\bibinfo{author}{{Yonetoku} D}, \bibinfo{author}{{Murakami} T},
  \bibinfo{author}{{Nakamura} T}, \bibinfo{author}{{Yamazaki} R},
  \bibinfo{author}{{Inoue} AK} and  \bibinfo{author}{{Ioka} K}
  (\bibinfo{year}{2004}), \bibinfo{month}{Jul.}
\bibinfo{title}{{Gamma-Ray Burst Formation Rate Inferred from the Spectral Peak
  Energy-Peak Luminosity Relation}}.
\bibinfo{journal}{{\em ApJ}} \bibinfo{volume}{609} (\bibinfo{number}{2}):
  \bibinfo{pages}{935--951}. \bibinfo{doi}{\doi{10.1086/421285}}.
\eprint{astro-ph/0309217}.

\bibtype{Inproceedings}%
\bibitem[{Yonetoku} et al.(2020)]{hizgundam}
\bibinfo{author}{{Yonetoku} D}, \bibinfo{author}{{Mihara} T},
  \bibinfo{author}{{Doi} A}, \bibinfo{author}{{Sakamoto} T},
  \bibinfo{author}{{Tsumura} K}, \bibinfo{author}{{Ioka} K},
  \bibinfo{author}{{Amaya} Y}, \bibinfo{author}{{Arimoto} M},
  \bibinfo{author}{{Enoto} T}, \bibinfo{author}{{Fujii} T},
  \bibinfo{author}{{Goto} H}, \bibinfo{author}{{Gunji} S},
  \bibinfo{author}{{Hiraga} J}, \bibinfo{author}{{Ikeda} H},
  \bibinfo{author}{{Kawai} N}, \bibinfo{author}{{Kurosawa} S},
  \bibinfo{author}{{Li} J}, \bibinfo{author}{{Maeda} Y},
  \bibinfo{author}{{Mitsuishi} I}, \bibinfo{author}{{Murakami} T},
  \bibinfo{author}{{Nakagawa} Y}, \bibinfo{author}{{Ogino} N},
  \bibinfo{author}{{Ohno} M}, \bibinfo{author}{{Sawano} T},
  \bibinfo{author}{{Sei} K}, \bibinfo{author}{{Serino} M},
  \bibinfo{author}{{Sugita} S}, \bibinfo{author}{{Tamagawa} T},
  \bibinfo{author}{{Tamura} K}, \bibinfo{author}{{Tanaka} T},
  \bibinfo{author}{{Tanimori} T}, \bibinfo{author}{{Tashiro} MS},
  \bibinfo{author}{{Tomida} H}, \bibinfo{author}{{Wang} H},
  \bibinfo{author}{{Yamaguchi} T}, \bibinfo{author}{{Yamamoto} A},
  \bibinfo{author}{{Yamaoka} K}, \bibinfo{author}{{Yamauchi} M},
  \bibinfo{author}{{Yatsu} Y}, \bibinfo{author}{{Yoshida} A},
  \bibinfo{author}{{Yuhi} D}, \bibinfo{author}{{Akitaya} H},
  \bibinfo{author}{{Fukui} A}, \bibinfo{author}{{Ita} Y},
  \bibinfo{author}{{Kaneda} H}, \bibinfo{author}{{Kawabata} K},
  \bibinfo{author}{{Kawata} Y}, \bibinfo{author}{{Kurimata} M},
  \bibinfo{author}{{Matsumoto} T}, \bibinfo{author}{{Matsuura} S},
  \bibinfo{author}{{Miyasaka} A}, \bibinfo{author}{{Motohara} K},
  \bibinfo{author}{{Narita} N}, \bibinfo{author}{{Noda} H},
  \bibinfo{author}{{Ohashi} A}, \bibinfo{author}{{Okita} H},
  \bibinfo{author}{{Sano} K}, \bibinfo{author}{{Tanaka} M},
  \bibinfo{author}{{Urata} Y}, \bibinfo{author}{{Wada} T},
  \bibinfo{author}{{Yamaguchi} H}, \bibinfo{author}{{Yanagisawa} K},
  \bibinfo{author}{{Yoshida} M}, \bibinfo{author}{{Asano} K},
  \bibinfo{author}{{Inayoshi} K}, \bibinfo{author}{{Inoue} S},
  \bibinfo{author}{{Ito} H}, \bibinfo{author}{{Izumiura} H},
  \bibinfo{author}{{Kawanaka} N}, \bibinfo{author}{{Kinugawa} T},
  \bibinfo{author}{{Kisaka} S}, \bibinfo{author}{{Kiuchi} K},
  \bibinfo{author}{{Matsumoto} J}, \bibinfo{author}{{Mizuta} A},
  \bibinfo{author}{{Murase} K}, \bibinfo{author}{{Nagakura} H},
  \bibinfo{author}{{Nagataki} S}, \bibinfo{author}{{Nakada} Y},
  \bibinfo{author}{{Nakamura} T}, \bibinfo{author}{{Niino} Y},
  \bibinfo{author}{{Suwa} Y}, \bibinfo{author}{{Takahashi} K},
  \bibinfo{author}{{Tanaka} T}, \bibinfo{author}{{Toma} K},
  \bibinfo{author}{{Totani} T}, \bibinfo{author}{{Yamazaki} R} and
  \bibinfo{author}{{Yokoyama} J} (\bibinfo{year}{2020}), \bibinfo{month}{Dec.},
  \bibinfo{title}{{High-redshift gamma-ray burst for unraveling the Dark Ages
  Mission: HiZ-GUNDAM}}, \bibinfo{editor}{{den Herder} JWA},
  \bibinfo{editor}{{Nikzad} S} and  \bibinfo{editor}{{Nakazawa} K}, (Eds.),
  \bibinfo{booktitle}{Space Telescopes and Instrumentation 2020: Ultraviolet to
  Gamma Ray}, \bibinfo{series}{Society of Photo-Optical Instrumentation
  Engineers (SPIE) Conference Series}, \bibinfo{volume}{11444}, pp.
  \bibinfo{pages}{114442Z}.

\bibtype{Article}%
\bibitem[{Zanin} et al.(2016)]{zanin16}
\bibinfo{author}{{Zanin} R}, \bibinfo{author}{{Fern{\'a}ndez-Barral} A},
  \bibinfo{author}{{de O{\~n}a Wilhelmi} E}, \bibinfo{author}{{Aharonian} F},
  \bibinfo{author}{{Blanch} O}, \bibinfo{author}{{Bosch-Ramon} V} and
  \bibinfo{author}{{Galindo} D} (\bibinfo{year}{2016}), \bibinfo{month}{Nov.}
\bibinfo{title}{{Gamma rays detected from Cygnus X-1 with likely jet origin}}.
\bibinfo{journal}{{\em A\&A}} \bibinfo{volume}{596}, \bibinfo{eid}{A55}.
  \bibinfo{doi}{\doi{10.1051/0004-6361/201628917}}.
\eprint{1605.05914}.

\bibtype{Article}%
\bibitem[{Zhang} and {Zhang}(2014)]{zhang14}
\bibinfo{author}{{Zhang} B} and  \bibinfo{author}{{Zhang} B}
  (\bibinfo{year}{2014}), \bibinfo{month}{Feb.}
\bibinfo{title}{{Gamma-Ray Burst Prompt Emission Light Curves and Power Density
  Spectra in the ICMART Model}}.
\bibinfo{journal}{{\em ApJ}} \bibinfo{volume}{782} (\bibinfo{number}{2}),
  \bibinfo{eid}{92}. \bibinfo{doi}{\doi{10.1088/0004-637X/782/2/92}}.
\eprint{1312.7701}.

\bibtype{Article}%
\bibitem[{Zhang} et al.(2006)]{zhang06}
\bibinfo{author}{{Zhang} B}, \bibinfo{author}{{Fan} YZ},
  \bibinfo{author}{{Dyks} J}, \bibinfo{author}{{Kobayashi} S},
  \bibinfo{author}{{M{\'e}sz{\'a}ros} P}, \bibinfo{author}{{Burrows} DN},
  \bibinfo{author}{{Nousek} JA} and  \bibinfo{author}{{Gehrels} N}
  (\bibinfo{year}{2006}), \bibinfo{month}{May}.
\bibinfo{title}{{Physical Processes Shaping Gamma-Ray Burst X-Ray Afterglow
  Light Curves: Theoretical Implications from the Swift X-Ray Telescope
  Observations}}.
\bibinfo{journal}{{\em ApJ}} \bibinfo{volume}{642} (\bibinfo{number}{1}):
  \bibinfo{pages}{354--370}. \bibinfo{doi}{\doi{10.1086/500723}}.
\eprint{astro-ph/0508321}.

\bibtype{Article}%
\bibitem[{Zhang} et al.(2009)]{zhang09}
\bibinfo{author}{{Zhang} B}, \bibinfo{author}{{Zhang} BB},
  \bibinfo{author}{{Virgili} FJ}, \bibinfo{author}{{Liang} EW},
  \bibinfo{author}{{Kann} DA}, \bibinfo{author}{{Wu} XF},
  \bibinfo{author}{{Proga} D}, \bibinfo{author}{{Lv} HJ},
  \bibinfo{author}{{Toma} K}, \bibinfo{author}{{M{\'e}sz{\'a}ros} P},
  \bibinfo{author}{{Burrows} DN}, \bibinfo{author}{{Roming} PWA} and
  \bibinfo{author}{{Gehrels} N} (\bibinfo{year}{2009}), \bibinfo{month}{Oct.}
\bibinfo{title}{{Discerning the Physical Origins of Cosmological Gamma-ray
  Bursts Based on Multiple Observational Criteria: The Cases of z = 6.7 GRB
  080913, z = 8.2 GRB 090423, and Some Short/Hard GRBs}}.
\bibinfo{journal}{{\em ApJ}} \bibinfo{volume}{703} (\bibinfo{number}{2}):
  \bibinfo{pages}{1696--1724}.
  \bibinfo{doi}{\doi{10.1088/0004-637X/703/2/1696}}.
\eprint{0902.2419}.

\end{thebibliography*}

\end{document}